\documentclass[twocolumn,reprint,superscriptaddress,amsmath,amssymb,prl,aps]{revtex4-2}

\usepackage{graphicx}
\usepackage{bm}
\usepackage{bbm}
\usepackage{hyperref}
\usepackage{braket}
\usepackage{soul}
\usepackage{color}
\usepackage{orcidlink}
\usepackage{tikz}
\usetikzlibrary{quantikz2}

\hypersetup{
    colorlinks=true,
    linkcolor=blue,
    filecolor=magenta,      
    urlcolor=blue,
    citecolor=blue
}

\begin{document}

\title{Imaginary Time Spectral Transforms for Excited State Preparation}

\author{D.~A.~Millar\,\orcidlink{0000-0003-3713-8997}}
\email{declan.millar@ibm.com}
\affiliation{IBM Quantum, IBM Research Europe, Hursley, Winchester, SO21 2JN, United Kingdom}
\affiliation{School of Physics and Astronomy, University of Southampton, Southampton SO17 1BJ, UK}
\author{L.~W.~Anderson\,\orcidlink{0000-0003-0269-3237}}
\affiliation{IBM Quantum, IBM Research Europe, Hursley, Winchester, SO21 2JN, United Kingdom}
\author{E.~Altamura\,\orcidlink{0000-0001-6973-1897}}
\affiliation{The Hartree Centre, STFC, Sci-Tech Daresbury, Warrington WA4 4AD, United Kingdom }
\affiliation{Yusuf Hamied Department of Chemistry, University of Cambridge, Lensfield Road, Cambridge CB2 1EW, United Kingdom }
\author{O.~Wallis\,\orcidlink{0009-0002-7323-2059}}
\affiliation{The Hartree Centre, STFC, Sci-Tech Daresbury, Warrington WA4 4AD, United Kingdom }
\author{M.~E.~Sahin\,\orcidlink{0000-0002-5996-0407}}
\affiliation{The Hartree Centre, STFC, Sci-Tech Daresbury, Warrington WA4 4AD, United Kingdom }
\author{J.~Crain\,\orcidlink{0000-0001-8672-9158}}
\affiliation{IBM Research Europe, The Hartree Centre, Sci-Tech Daresbury, Warrington WA4 4AD, UK}
\affiliation{Clarendon Laboratory, University of Oxford, Oxford OX1 3PU, UK}
\author{S.~J.~Thomson\,\orcidlink{0000-0001-9065-9842}}
\email{steven.thomson@ed.ac.uk}
\affiliation{SUPA, School of Physics and Astronomy, University of Edinburgh, Edinburgh EH9 3FD, UK}
\affiliation{IBM Quantum, IBM Research Europe, Hursley, Winchester, SO21 2JN, United Kingdom}

\date{\today}

\begin{abstract}
Excited states of many-body quantum systems play a key role in a wide range of physical and chemical phenomena. Despite this, there is a notable lack of scalable algorithms capable of preparing highly excited eigenstates. To address this challenge, we introduce a general approach that directly targets eigenstates near a chosen energy, applicable to both classical and quantum simulation frameworks. Our approach combines the shift-invert mechanism with imaginary time evolution, enabling the construction of excited states of large many-body quantum systems. We demonstrate the technique classically by computing mid-spectrum eigenstates of disordered spin chains with up to $L=128$ sites. Based on this, we propose a hybrid scheme compatible with near-term quantum hardware.
\end{abstract}

\maketitle

\emph{Introduction} - The excited state manifold of quantum systems is important for a variety of physical and chemical properties.
In condensed matter physics, excited states govern key phenomena such as information transport, critical behavior, and thermodynamics in various applications, including molecular magnets, optically addressable spins, quantum dots, and trapped ions~\cite{mishra2021observation, bera2022emergent, bayliss2020optically, katcko2021encoding}. Knowledge of the excited state structure can even allow construction of thermal Gibbs states at low temperatures~\cite{Cocchiarella+25}.
In chemistry, molecular excitations can alter reactivities, redox properties, and proton transfer barriers~\cite{jankowska2021modern,pfau2024accurate}. They can even activate reaction pathways and isomerization transitions that are inaccessible in the ground state~\cite{balzani2014photochemistry}, with potentially transformative consequences for drug discovery and design, as well as affect bonding, geometry, chemical properties, and spectroscopic signatures~\cite{dral2021molecular, westermayr2020machine}. 
Similarly, in electronic structure, the excited state of a molecular species often exhibits distinct chemical properties compared to its ground state, influencing bonding, geometry, and spectroscopic signatures~\cite{zewail1992chemical, crespo2004ultrafast, crisenza2020synthetic}.
Broad classes of materials underpinning optoelectronics and quantum devices, such as those in light-emitting diodes and photovoltaic cells, all rely on harnessing excited state attributes to confer functionality~\cite{stoneham2003exploiting}.
Excited states offer a vastly expanded molecular design space, playing a pivotal and growing role in various applications that rely upon light-matter interactions or charge injection~\cite{yam2023using}.

Despite the importance of excited states, a full understanding of their structure has remained elusive. Unlike ground states, for which many efficient and widely used algorithms exist (including quantum Monte Carlo~\cite{Hammond+94,Foulkes+01} or the density matrix renormalization group (DMRG)~\cite{Schollwock+11,Bridgeman+17}) there are few ways to compute excited states outside of full exact diagonalization.
Constructing low-lying excited states iteratively from the ground state is often feasible, but constructing highly excited states in this manner poses a significant challenge, often requiring the computation of hundreds or thousands of unwanted states before obtaining the desired one. The ability to directly compute highly excited eigenstates---without requiring the explicit construction of all lower-energy states---is highly desirable for many fundamental problems, including those related to eigenstate thermalization and many-body localization~\cite{Alet+18,Abanin+19,Sierant+25}. Classical tensor network techniques exist for directly computing excited states of large systems in various limits, such as strong disorder~\cite{Khemani+16,Pollmann+16, Yu+17} or weakly entangled states such as quantum many-body scars~\cite{Zhang+23}. Despite impressive results, these algorithms are ultimately hindered by the inability of classical computers to efficiently capture highly entangled states. There is a clear need for a robust algorithm which can be deployed on both classical and quantum computers, allowing us to overcome these limitations and study highly entangled excited states---this is the central challenge we address in this work.

\begin{figure*}[ht!]
    \includegraphics[width=\textwidth]{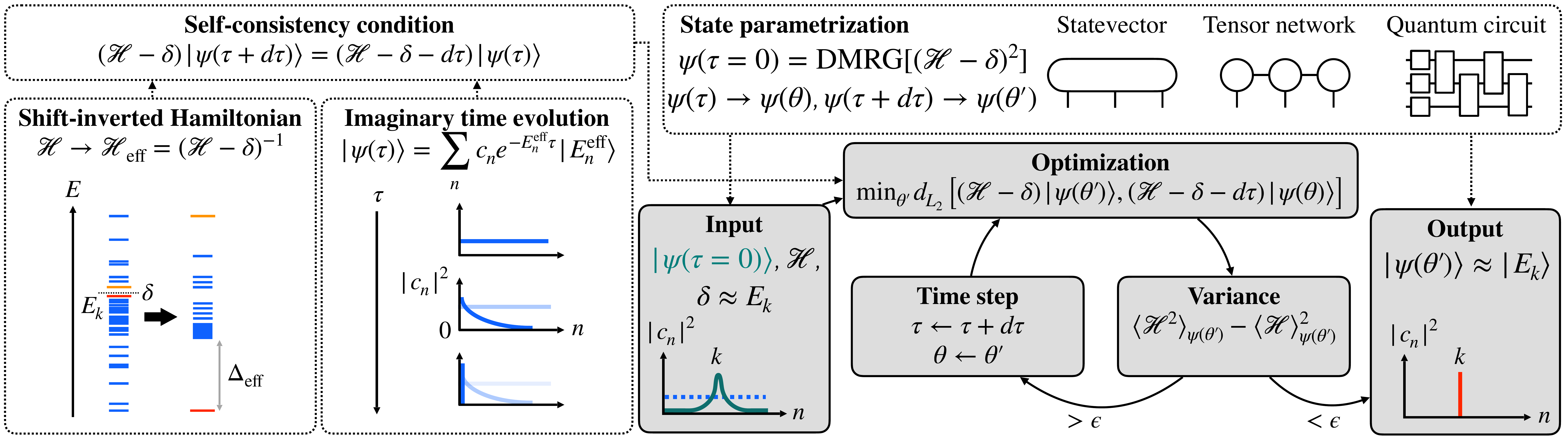}
    \caption{Schematic of the algorithm. The method combines shift-invert and imaginary time evolution: the shift-invert process maps eigenstates of an initial Hamiltonian $\mathcal{H}$ near a target energy $\delta$ to the extremal eigenstates of an effective Hamiltonian $\mathcal{H}_{\mathrm{eff}}$, which are then determined via imaginary time evolution. To avoid explicit inversion of $(\mathcal{H}-\delta)$, we derive a self-consistency condition for an infinitesimal timestep $\textrm{d}\tau$, which we solve variationally. We evolve the state iteratively until we reach a target variance. To provide a `warm start' to the algorithm, we initialize it using an approximation to the ground state of $(\mathcal{H}-\delta)^2$.}
    \label{fig1}
\end{figure*}

\emph{Model} - We demonstrate our technique using the disordered Heisenberg model, which is given by:
\begin{align}
    \mathcal{H} = J \sum_i {\bf S}_i \cdot {\bf S}_{i+1} + \sum_i h_i S^z_i
    \label{eq.hamiltonian}
\end{align}
where $h_i \in [-W,W]$ is the on-site disorder taken from a uniform distribution, and we set $J=1$ as our unit of energy. We restrict our analysis to $W>0$, as in the limit of $W \to 0$, the energy spectrum is highly degenerate. In the presence of degeneracies, our method is likely to converge to a superposition of degenerate states. Other methods (including DMRG) have similar limitations.

\emph{Method} - Our algorithm, sketched in Fig.~\ref{fig1}, combines two core techniques: spectral transforms and imaginary time evolution. Spectral transforms involve modifying the Hamiltonian such that an eigenstate at some target energy $\delta$ becomes an extremal eigenstate of a new effective Hamiltonian $\mathcal{H}_{\textrm{eff}}$. We may then use any reliable ground state search technique to find the ground state of $\pm\mathcal{H}_{\textrm{eff}}$. One example is \emph{spectral folding} $\mathcal{H} \to \mathcal{H}_{\textrm{eff}} = (\mathcal{H}-\delta)^2$, which has been used in the context of quantum chemistry~\cite{CadiTazi+24}. Throughout, we abbreviate constant terms like $\mathbbm{1} \delta \to \delta$ for clarity. This transform typically results in a large density of states in the vicinity of the target state, which can be challenging for variational algorithms (results shown in~\cite{SM}). This problem can be avoided by instead using the \emph{shift-invert technique}~\cite{Pietracaprina+18}:
\begin{align}
    \mathcal{H} \to \mathcal{H}_{\textrm{eff}} = (\mathcal{H} - \delta)^{-1}.
\end{align}
The ground state of $\mathcal{H}_{\textrm{eff}}$ will now be the closest eigenstate of $\mathcal{H}$ with energy less than $\delta$, while the maximally excited state of $\mathcal{H}_{\mathrm{eff}}$ will be the closest state with energy greater than $\delta$. Note that if $\delta$ is an exact eigenvalue, the shifted Hamiltonian is singular. This is typically not a problem in practice, as we are extremely unlikely to pick an exact eigenvalue by accident. We can find the highest excited state of $\mathcal{H}_{\textrm{eff}}$ by solving for the ground state of $-\mathcal{H}_{\mathrm{eff}}$. Since the shift-invert transform corresponds to a resolvent of the Hamiltonian, the procedure may also be viewed as an energy-selective spectral transformation.

The key obstacle in applying the shift-invert method is that direct inversion of the Hamiltonian is costly, with naive methods scaling like $\mathcal{O}(N_{H}^3)$ where $N_H$ is the dimension of the Hilbert space, which is itself typically exponentially large in the system size. Moreover, even if $\mathcal{H}_{\mathrm{eff}}$ can be computed, it is unlikely to be sparse, making it complicated to directly implement in frameworks such as tensor networks or on near-term quantum hardware. Sophisticated techniques have been developed to avoid having to invert the Hamiltonian, often involving solving an equivalent implicit problem. These are, however, often specific to particular classical implementations~\cite{Yu+17,Pietracaprina+18} and are not convenient for direct implementation on quantum hardware. Here, we avoid explicit inversion of the Hamiltonian in a way that does not depend on any particular implementation framework, either quantum or classical, by combining the shift-invert procedure with \emph{imaginary time evolution}.

Imaginary time evolution is a well-established technique with guarantees for obtaining ground states of quantum systems~\cite{AltlandSimons}. It involves preparing the system in an arbitrary initial state, presumed to be far from the ground state of the chosen Hamiltonian -- but with non-zero overlap -- and evolving in imaginary time $\tau$. This evolution is generated by an operator of the form $\textrm{e}^{-\mathcal H \tau} $, where $\mathcal{H}$ is the microscopic Hamiltonian and $\tau$ represents imaginary time. The ultimate goal is to prepare the state $\ket{\psi(\tau)}/ \| \ket{\psi(\tau)}  \|_2$, where $\ket{\psi(\tau)} = \textrm{e}^{-\mathcal H \tau} \ket{\psi(0)}$, which in the limit of $\tau \to \infty$ will be the zero-temperature ground state of $\mathcal{H}$. Our crucial modification to this technique is that we shall perform the imaginary time evolution with respect to the shift-inverted Hamiltonian $\mathcal{H}_\textrm{eff}$, allowing us to use imaginary time evolution to target excited states.

The (un-normalized) time-evolved state is given by:
\begin{align}
\ket{\psi(\tau)} = \textrm{e}^{-\mathcal H_\textrm{eff} \tau} \ket{\psi(0)} = \textrm{e}^{-(\mathcal{H} - \delta)^{-1} \tau} \ket{\psi(0)}.
\end{align}
We start by making an infinitesimal imaginary time step of size $\textrm{d} \tau$ and expand the evolution operator to obtain:
\begin{align}
|\psi(\textrm{d}\tau + \tau) \rangle
\approx (1 - (\mathcal{H} - \delta)^{-1} \textrm{d}\tau) | \psi(\tau) \rangle.
\end{align}
We then multiply both sides by $\mathcal{H}_{\textrm{shift}}=(\mathcal{H} - \delta)$ to obtain
\begin{align}
\mathcal{H}_{\textrm{shift}} |\psi(\textrm{d}\tau + \tau) \rangle = (\mathcal{H}_{\textrm{shift}}- \textrm{d}\tau) | \psi(\tau) \rangle.
\label{eq.variational}
\end{align}
This gives us a self-consistency condition that can be solved variationally to evolve the state in (imaginary) time $\ket{\psi(\tau)} \to\ket{\psi(\tau + \textrm{d} \tau)}$. At each time step, we evolve the state by minimizing the Hilbert-Schmidt distance. Other cost functions can be used, some of which are simpler to measure on a quantum computer, however we found the Hilbert-Schmidt distance to be significantly more numerically stable than all alternatives we tested. 
\begin{align}
D &= \| \mathcal{H}_{\textrm{shift}} \ket{\psi(\tau + \textrm{d} \tau)} - (\mathcal{H}_{\textrm{shift}}- \textrm{d}\tau)\ket{\psi(\tau)} \|_2.
\label{eq.timestep}
\end{align}

We have thus traded the problematic matrix inverse for a variational update step which can be performed classically or on near-term quantum hardware~\cite{Cincio+18,Cao+24}. The iterative nature of the process ensures that as long as $\mathrm{d}\tau$ is small, the convergence properties of the algorithm remain favorable~\cite{Puig+25}, while also inheriting the performance and accuracy guarantees of conventional imaginary time evolution.

Evaluating the cost function in Eq.~(\ref{eq.timestep}) on a quantum computer requires estimating
$\langle\psi(\tau+\mathrm d\tau)|\mathcal H_{\mathrm{shift}}^2|\psi(\tau+\mathrm d\tau)\rangle$,
$\langle\psi(\tau)|(\mathcal H_{\mathrm{shift}}-\mathrm d\tau)^2|\psi(\tau)\rangle$,
and
$\mathrm{Re}\!\left[\langle\psi(\tau+\mathrm d\tau)|\mathcal H_{\mathrm{shift}}(\mathcal H_{\mathrm{shift}}-\mathrm d\tau)|\psi(\tau)\rangle\right]$.
The first two terms are expectation values of positive semidefinite operators and can be estimated using standard measurement routines.
The cross term can be written as
\begin{align}
(\mathcal H-\delta)(\mathcal H-\delta-\mathrm d\tau)=\sum_i \alpha_i P_i,
\end{align}
where $P_i$ are Pauli strings.
This reduces the overlap term to a sum of quantities of the form $\mathrm{Re}[\langle \psi(\tau+\mathrm d\tau)|P_i|\psi(\tau)\rangle]$, each of which can be measured using a modified Hadamard test~\cite{SM}. For local Hamiltonians like Eq.~\ref{eq.hamiltonian}, this expansion generally contains $\mathcal O(L^2)$ Pauli strings, although mutually commuting terms can be grouped to reduce the number of measurement bases. Consequently, for local Hamiltonians the number of grouped observables scales polynomially with system size, as does the sampling cost per variational update.
Estimating each grouped observable to additive precision $\epsilon$ with failure probability $\eta$ requires $\mathcal O(\log(1/\eta)/\epsilon^2)$ shots. This can be further reduced to $\mathcal O(\log(1/\eta)/\epsilon)$ if amplitude amplification is used, albeit at the cost of deeper circuits~\cite{brassard2000quantum, berry2014exponential}. 
In this context, the circuit depth would be controlled by the ansatz chosen to represent the state, and would not grow with the total imaginary time $\tau$ as in conventional (real-time) Trotter evolution. Instead, increasing $\tau$ corresponds to performing additional variational updates of a fixed circuit structure, more akin to the McLachlan variational approach to time evolution.
Further details regarding measurement protocols, compilation cost, and noise considerations are discussed in~\cite{SM}.

We now have all of the essential ingredients for our algorithm (Fig.~\ref{fig1}). i) First, we use standard techniques to obtain a loose approximation to the ground state of the folded Hamiltonian $(\mathcal{H}-\delta)^2$. This could be obtained on quantum hardware using, for example, the variational quantum eigensolver (VQE)~\cite{Peruzzo+14,Tilly+22}. Here, we compute it classically using DMRG (typically with initial bond dimension $\chi_0 \leq 4$). The state we obtain will in general not be an exact eigenstate, but will serve as a good initial choice for $\ket{\psi(\tau=0)}$. This procedure ensures that our initial state has non-zero overlap with eigenstates near the target energy, mitigating against the `orthogonality catastrophe' that would cause randomly chosen states to have a vanishingly small overlap with the target state. ii) We then take $\ket{\psi(\tau=0)}$ -- which could be expressed as a shallow quantum circuit~\cite{Ran20,Malz+24,mpstocircuit2025,Robertson+25} -- and use it as the input state for the shift-inverted imaginary time evolution.
iii) We evolve the system in time by repeatedly applying Eq.~\ref{eq.timestep}. As $\tau \to \infty$, we asymptotically approach an eigenstate with energy close to $\delta$. The precise final state will depend on the overlap of the initial state with states in the vicinity of the target energy. Closeness to the desired energy can be verified by computing $E(\tau) = \braket{\psi(\tau)|\mathcal{H}|\psi(\tau)}$ and comparing with the target energy. We verify closeness to an eigenstate by computing the variance $\mathcal{\sigma}(\tau) = \braket{\psi(\tau)|(\mathcal{H}-E(\tau))^2|\psi(\tau)}$, which vanishes for an exact eigenstate.

\emph{Results} - We present a proof of concept implementation of the algorithm using a Matrix Product State (MPS) representation, based on \texttt{ITensors.jl}~\cite{ITensor}. Details of the implementation are given in~\cite{SM}. The use of matrix product states is not essential to the algorithm: it can equally be implemented directly on statevectors~\cite{SM}, directly on quantum circuits, or directly on any other desired framework. While MPS methods are not efficient for highly entangled states, there generically exist states in the bulk of the spectrum with moderate bond dimension that we expect to capture with this approach~\cite{Rai+24}. 
To ensure accuracy, we update the size of the timestep $\textrm{d}\tau$ such that the Taylor expansion $\exp(- \mathcal{H}_{\textrm{eff}} \phantom{.} \textrm{d} \tau) \ket{\psi(\tau)} \approx (1 - \mathcal{H}_{\textrm{eff}} \phantom{.} \textrm{d} \tau + \mathcal{O}(\textrm{d} \tau^2))\ket{\psi(\tau)}$ is valid. We do this by computing $E(\tau)$ and choosing $\textrm{d} \tau_{\textrm{min}} \leq \textrm{d} \tau \ll 2 (E(\tau) - \delta)$, imposing the lower limit $\textrm{d} \tau_{\textrm{min}} = 10^{-3}$ to prevent slowdown when $E(\tau) \approx \delta$.

\begin{figure}[t!]
    \centering
    \includegraphics[width=\linewidth]{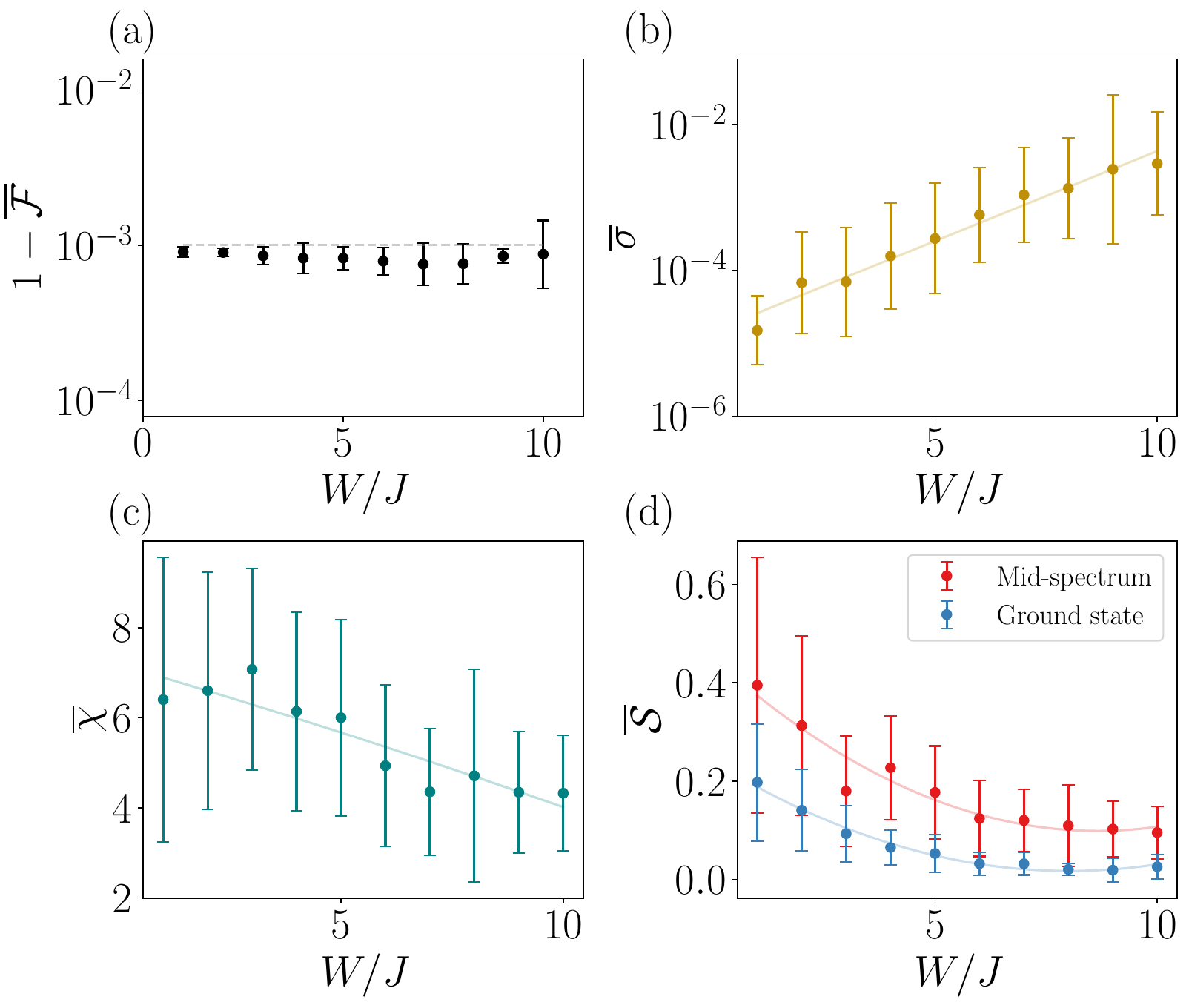}
    \caption{Results for system size $L=12$ for disorder strengths $W \in [1,10]$, averaged over $N_s \in [15,30]$ disorder realizations for each point, depending on the disorder strength. Overlines denote the disorder average, and error bars indicate the standard deviation. a) The deviation from unit fidelity $(1-\overline{\mathcal{F}})$ of mid-spectrum states. b) The final variance $\overline{\sigma}$. c) The final bond dimension $\overline{\chi}$. d) Comparison of the bond-averaged von Neumann entanglement entropy of the mid-spectrum states (red) and the corresponding ground states (blue). Solid lines in panels (b), (c) and (d) are guides to the eye.}
    \label{fig2}
\end{figure}

First, we demonstrate the results of the method for a small, exactly solvable system of size $L=12$, for a variety of disorder strengths. We evolve the initial state until the fidelity with any exact eigenstate $\mathcal{F} = \textrm{max}[\{|\braket{\psi(\tau)|\phi_i}|^2 : i = 1,...,2^L \}]$ is at least $\mathcal{F}=0.999$ or the variance decreases below $\sigma^* = 10^{-6}$. We also compute the von Neumann entanglement entropy $\mathcal{S} = -\textrm{Tr}[\rho_A \log \rho_A] = -\textrm{Tr}[\rho_B \log \rho_B]$ where subsystems $A$ and $B$ are formed by a bipartition at a given bond, and we average $\mathcal{S}$ over all bonds. We set a target energy of $\delta=0$, which is typically a mid-spectrum state, representing a challenging area of the spectrum where we can expect high entanglement and a high density of states. The results are shown in Fig.~\ref{fig2}, where the overline indicates the average over disorder realizations. These results enable us to i) demonstrate that the method is capable of recovering the exact solution to high accuracy, and ii) show that the mid-spectrum states we obtain are always more highly entangled than the corresponding ground states (obtained with DMRG), giving us confidence that we are finding representative mid-spectrum states rather than exceptionally weakly entangled states. We see that the algorithm converges to high accuracy regardless of the disorder strength. As the disorder strength decreases, the entanglement in the mid-spectrum states grows much faster than in the corresponding ground states. The more highly entangled the target state is, the larger the bond dimension we require, and the more computationally costly it is to solve Eq.~\ref{eq.timestep} for each time step due to the rapidly increasing number of parameters. Ultimately, for highly entangled states in large systems, the bond dimension growth will become prohibitive on classical hardware, eventually necessitating the use of quantum computers to construct such states using this algorithm. Even classically, it may be advantageous to represent states as tailored quantum circuits, which can have far fewer parameters than their MPS representation~\cite{Lin+21}. Interestingly, we see no clear indication of the many-body localization (MBL) transition close to $W/J \approx 4$~\cite{Luitz+15,Alet+18,Abanin+19,Thomson23}, although the entanglement entropy appears to flatten off above this point, and convergence becomes significantly slower below it. It would be interesting to perform a systematic study of large systems to investigate whether or not a clear signature of an MBL phase transition can be found~\cite{Doggen+18,Suntajs+20,Sels+21,Sels+23,Sierant+25}.

\begin{figure}
    \centering
    \includegraphics[width=\linewidth]{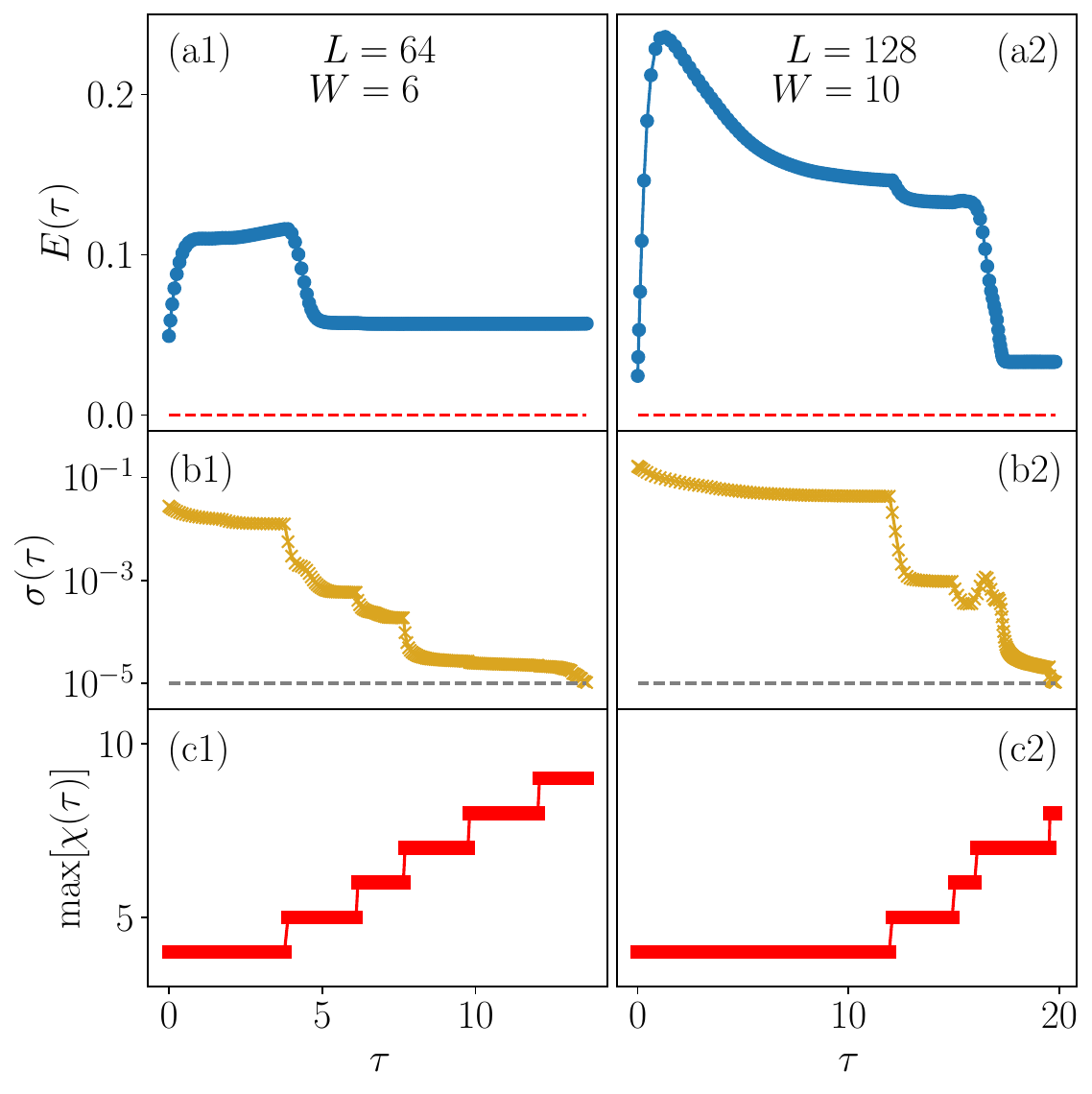}
    \caption{Imaginary time evolution of a) the energy $E(\tau)$, b) the variance $\sigma(\tau)$ and c) the maximum bond dimension. Columns show system size $L=64$ with disorder strength $W/J=6$ (left) and $L=128$ with $W/J=10$ (right). The dashed red line in (a) shows the target energy, $\delta=0$, and the dashed grey line in (b) shows the variance threshold, $\sigma^{*} = 1 \times 10^{-5}$.}
    \label{fig3}
\end{figure}

We now show that in the regime of moderate-to-strong disorder where the entanglement growth is inhibited, we can classically construct low-variance states of large systems. We show representative results for both $L=64$ with $W/J=6$ and $L=128$, $W/J=10$. We use a demanding cutoff of $\sigma^{*} < 1 \times 10^{-5}$, which is much lower than the average variance required to reach an eigenstate for $L=12$ at $W/J=6$ (see Fig.~\ref{fig2}). When comparing different system sizes, it may be fairer to scale the variance by system size~\cite{Devakul+17}, however we opt not to do so here in order to present our results as transparently as possible. The results are shown in Fig.~\ref{fig3}. Using DMRG to extract the extremal eigenvalues of the Hamiltonian, we find that with respect to the target energy $\delta=0$, the final states shown have a relative energy error of $\Delta E = (E(\tau \to \infty) - \delta)/(E_{\textrm{max}}-E_{\textrm{min}})\approx 10^{-4}$ and $\Delta E \approx 10^{-5}$ respectively. In this regime, we would expect a very high density of states with level spacings which may be smaller than can be resolved by machine precision. Starting in a low bond dimension MPS partially mitigates against this, as we do not expect this very sparse initial state will have significant overlap with many contiguous eigenstates within a given energy window. Further details are given in the Supplemental Material~\cite{SM}, along with additional results for system sizes $L \in [48, 64]$ with $W/J=6$, and $L=128$ with $W/J \in [6,10]$. Our strategy of starting in an initial MPS with a small bond dimension and allowing $\chi(\tau)$ to grow gradually may take more timesteps than directly starting with an MPS of large bond dimension, but it is typically significantly faster in terms of elapsed real time. This also suggests a potential hybrid quantum/classical approach, where the initial evolution is performed classically until the bond dimension begins to grow too large. The low bond dimension MPS can then be translated into a shallow quantum circuit~\cite{Ran20,Malz+24,mpstocircuit2025,Robertson+25}, and the final portion of the evolution performed on a quantum computer.

\emph{Discussion \& Conclusion} - In this work, we have described a novel algorithm for the preparation of excited eigenstates (and low-variance states) of many-body quantum systems. We have shown results obtained numerically on classical hardware, and proposed a hybrid implementation compatible with near-term quantum computers. We have showcased the method using a common condensed matter model, however the technique itself is Hamiltonian-agnostic, requiring no prior knowledge of the problem structure, and will work for a wide variety of many-body quantum systems, including applications in quantum chemistry. Nor does it depend on the simulation method used: the core algorithm can be applied to matrix product states, statevectors, quantum circuits, or any other framework in which a quantum state can be represented.

Our results demonstrate the feasibility of using the combination of the shift-invert method and imaginary time evolution to construct weakly entangled excited eigenstates of large quantum systems on classical hardware. We have demonstrated the method using classical tensor network techniques, however, as with all classical methods, the computational cost will grow rapidly as the system becomes more entangled.
Generic mid-spectrum eigenstates are expected to exhibit volume-law entanglement, rendering tensor-network approaches inefficient in the worst case. A quantum circuit can capture such states without incurring exponential classical memory or bond-dimension overhead, so that the principal costs shift to state preparation and measurement. In this sense, our approach differs from algorithms specific to tensor networks, such as DMRG-X or SIMPS. Our method directly targets eigenstates near a chosen energy via a spectral transformation and is not tied to any specific representation of the state, making it naturally compatible with quantum circuit-based implementations in regimes where classical descriptions become inefficient.

In this setting one would variationally optimise the parameters of a parametrised quantum circuit, but otherwise the key steps of the algorithm would be precisely the same. We leave this as a goal for future work, as our focus in this manuscript has been to develop and benchmark the algorithm itself, allowing us to illustrate the fundamental mechanism underlying the technique without encountering noise or precision issues. Nonetheless, such issues are important for practical implementations, and further details regarding scaling and resource estimation may be found in~\cite{SM}. One particularly promising near-term hybrid approach would be to initially run the algorithm classically, then map the resulting MPS onto a quantum circuit using established techniques~\cite{Ran20,Malz+24,mpstocircuit2025,Robertson+25} which could then be used as a `warm start' initial configuration for the algorithm on quantum hardware. Alternatively, for translationally invariant systems, one could implement the algorithm using the infinite MPS (iMPS)~\cite{Vidal07} or infinite PEPS (iPEPS)~\cite{Jordan+08} frameworks, where only a single unit cell is considered. Similar strategies exploiting translation invariance can be employed for quantum circuits. This would significantly lower the computational cost, as only a single unit cell would need to be optimized, potentially allowing construction of highly excited, highly entangled states in one and two dimensions.

Our focus here has been on highly excited states, however the method is very general and can also be used to target lower-lying excited states. For example, one could imagine finding the ground state energy of a gapped system using, e.g., DMRG~\cite{Schollwock+11,Bridgeman+17}, VQE~\cite{Peruzzo+14,Tilly+22} or sample-based quantum diagonalization (SQD)~\cite{Kanno+23,Robledo+24,Barison+24}, and then constructing the first excited state using our method, enabling an accurate determination of the energy gap. Likewise, given an approximate location of the bandgap in a material, our technique could be used to construct the eigenstates immediately above and below the bandgap, along with their eigenvalues, enabling the bandgap itself to be precisely measured. In fact, we need not restrict ourselves to excited states: we can also use this method to directly target ground states. The main reason to do so is that if $\delta$ can be chosen to be sufficiently close to the ground state energy, the resulting $\mathcal{H}_{\textrm{eff}}$ will have a larger energy gap than the initial Hamiltonian~\cite{SM}. This may make it easier for variational methods to converge in systems where the initial Hamiltonian is gapless or otherwise has a challenging energy landscape. Classically efficient lower bounds for ground state energies may be used to provide good choices for $\delta$ in this case~\cite{Anderson51,Baumgratz+12,Eisert23}. 

Similar variational methods have been widely used in many proposed near-term quantum algorithms~\cite{Yuan+19,Cerezo+21}, including in the case of imaginary time evolution~\cite{McArdle+19,Motta+20}. However, they have not to our knowledge been used in conjunction with the shift-invert technique. To that end, while this algorithm will likely be more widely used on classical hardware in the near term, we anticipate that its true long-term value will be in constructing highly entangled states directly on quantum hardware.

\begin{acknowledgments}
    \emph{Acknowledgements} - We acknowledge helpful discussions with L.~Henaut, C.~Willby, N.~Mariella, T.~Murphy and O.~T.~Brown. SJT thanks J.~Eisert for pointing out the existence of efficiently computable lower bounds for ground state energies. All code and data for this work are available at~\cite{data}. This work was supported by the Hartree National Centre for Digital Innovation, a UK Government-funded collaboration between STFC and IBM, as well as the Engineering and Physical Sciences Research Council (grant no. EP/Z533518/1). All numerical results were obtained on the University of Edinburgh School of Physics \& Astronomy Compute Cluster. 
IBM, the IBM logo, and \href{ibm.com}{www.ibm.com} are trademarks of International Business Machines Corp., registered in many jurisdictions worldwide. Other product and service names might be trademarks of IBM or other companies. The current list of IBM trademarks is available at \href{https://www.ibm.com/legal/copytrade}{www.ibm.com/legal/copytrade}.
\end{acknowledgments}

\emph{Author Contribution Statement} - SJT and DAM contributed equally to this work. SJT developed the core algorithm. SJT and DAM jointly wrote the MPS code and performed HPC simulations. LWA and DAM made key contributions to ensuring feasibility on quantum hardware. ES assisted with running HPC simulations. All authors contributed to writing the final manuscript.

\bibliography{refs}

\end{document}


\title{Supplemental Material to ``Imaginary Time Spectral Transforms for Excited State Preparation"}

\author{D.~A.~Millar\,\orcidlink{0000-0003-3713-8997}}
\email{declan.millar@ibm.com}
\affiliation{IBM Quantum, IBM Research Europe, Hursley, Winchester, SO21 2JN, United Kingdom}
\affiliation{School of Physics and Astronomy, University of Southampton, Southampton SO17 1BJ, UK}
\author{L.~W.~Anderson\,\orcidlink{0000-0003-0269-3237}}
\affiliation{IBM Quantum, IBM Research Europe, Hursley, Winchester, SO21 2JN, United Kingdom}
\author{E.~Altamura\,\orcidlink{0000-0001-6973-1897}}
\affiliation{The Hartree Centre, STFC, Sci-Tech Daresbury, Warrington WA4 4AD, United Kingdom }
\affiliation{Yusuf Hamied Department of Chemistry, University of Cambridge, Lensfield Road, Cambridge CB2 1EW, United Kingdom }
\author{O.~Wallis\,\orcidlink{0009-0002-7323-2059}}
\affiliation{The Hartree Centre, STFC, Sci-Tech Daresbury, Warrington WA4 4AD, United Kingdom }
\author{M.~E.~Sahin\,\orcidlink{0000-0002-5996-0407}}
\affiliation{The Hartree Centre, STFC, Sci-Tech Daresbury, Warrington WA4 4AD, United Kingdom }
\author{J.~Crain\,\orcidlink{0000-0001-8672-9158}}
\affiliation{IBM Research Europe, The Hartree Centre, Sci-Tech Daresbury, Warrington WA4 4AD, UK}
\affiliation{Clarendon Laboratory, University of Oxford, Oxford OX1 3PU, UK}
\author{S.~J.~Thomson\,\orcidlink{0000-0001-9065-9842}}
\email{steven.thomson@ed.ac.uk}
\affiliation{SUPA, School of Physics and Astronomy, University of Edinburgh, Edinburgh EH9 3FD, UK}
\affiliation{IBM Quantum, IBM Research Europe, Hursley, Winchester, SO21 2JN, United Kingdom}

\date{\today}

\begin{abstract}
This Supplementary Material contains additional information about the implementation of our algorithm on quantum hardware, state vector simulations to demonstrate an alternative implementation of the algorithm and its utility for constructing ground states, extensive further numerical results for a variety of system sizes and disorder strengths, and full details on all aspects of the tensor network implementation described in the main text.
\end{abstract}

\maketitle
\onecolumngrid
\section{Implementation on a Quantum Computer}
\label{sec.hadamard}

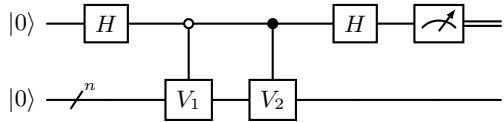
\begin{figure}[t]
\begin{quantikz}
 \lstick{$\ket{0}$} & \gate{H} & \octrl{1} & \ctrl{1} & \gate{H} & \meter{} & \setwiretype{c}\\
 \lstick{$\ket{0}$} & \qwbundle{n} & \gate{V_1} & \gate{V_2} & & &
\end{quantikz}
\caption{The circuit for the modified Hadamard test, required to measure the quantities shown in Eq.~6 of the main text. This allows reconstruction of overlaps of the form $\braket{\psi|\phi}$ where $\ket{\psi} = V_1 \ket{0}$, $\ket{\phi} = V_2 \ket{0}$, and $V_1$ and $V_2$ are unitary operators.}
\label{fig.hadamard}
\end{figure}

Let us define the (non-normalized) states $\ket{a} = (\mathcal{H} - \delta) |\psi(\textrm{d}\tau + \tau) \rangle$ and $\ket{b} = ((\mathcal{H} - \delta)- \textrm{d}\tau)\ket{\psi(\tau)}$, such that the Hilbert-Schmidt distance can be written as:
\begin{align}
    D &= \sqrt{\braket{a|a} + \braket{b|b} -2 \textrm{Re}[\braket{a|b}]}.
    \label{eq.hilbert_schmidt}
\end{align}
The first two terms are expectation values of positive semidefinite operators (i.e., squares of Hermitian operators) which can be straightforwardly measured on a quantum computer. Note that as the states $\ket{a}$ and $\ket{b}$ are not normalized, the first two terms do not immediately simplify to the identity. To measure the third term, we decompose the product of Hermitian operators as a sum over Pauli strings $P_i$:
\begin{align}
(\mathcal{H} - \delta)(\mathcal{H} - \delta- \textrm{d}\tau) = \sum_i \alpha_i P_i,
\label{eq.pauli_strings}
\end{align}
where the $\alpha_i$ coefficients can be read off by expanding the brackets on the left-hand side of Eq.~\ref{eq.pauli_strings} and collecting together like terms. For local Hamiltonians, this expansion generally contains a number of Pauli strings that scales polynomially with system size. This scaling is quadratic for nearest-neighbor models , as considered here. In addition, mutually commuting terms can be grouped to further reduce the measurement cost (i.e still polynomial, but with a smaller prefactor).
Here, we group Pauli strings into sets that can be measured within a common tensor-product measurement basis (i.e., qubit-wise commuting groups). The quantity to measure becomes:
\begin{align}
    \text{Re}[\braket{a|b}] &= \bra{\psi(\textrm{d}\tau + \tau)} (\mathcal{H} - \delta)(\mathcal{H} - \delta- \textrm{d}\tau) \ket{\psi(\tau)} \nonumber\\
    & = \sum_i \alpha_i \text{Re}[\bra{\psi(\textrm{d}\tau + \tau)} P_i \ket{\psi(\tau)}] \nonumber\\
    & = \sum_i \alpha_i \text{Re}[\bra{0} U(\textrm{d}\tau + \tau) P_i  U(\tau) \ket{0}],
    \label{eq.hadamard_test}
\end{align}
where $U(\tau)$ is a parametrized quantum circuit to prepare the time-evolved state. 

For each Pauli string $P_i$, we wish to measure
\begin{align}
    \text{Re}[\bra{0} U(\textrm{d}\tau + \tau) P_i  U(\tau) \ket{0}]
    = \text{Re}[\braket{\psi_1|\psi_2}],
\end{align}
where 
\begin{align}
    U(\tau + d\tau)\ket{0} &\equiv V_1\ket{0} = \ket{\psi_1} \text{ and}\\
    P_i U(\tau)\ket{0} &\equiv V_2 \ket{0} = \ket{\psi_2}.
\end{align}

This can be measured efficiently on quantum hardware using the \emph{modified Hadamard test} (Fig.~\ref{fig.hadamard}) as follows.

We begin with a $1$-qubit ancilla register and $n$-qubit target register in the state $\ket{0}\ket{0^n}$. We apply a Hadamard gate to the ancilla to prepare the state $\ket{+}\ket{0}$. We then apply $V_1$ (controlled on the ancilla being in the $\ket{0}$ state) and $V_2$ (controlled on the ancilla being in the $\ket{1}$ state) to the target register. Finally, we apply a second Hadamard to the ancilla. The corresponding circuit diagram is shown in Fig.~\ref{fig.hadamard}. This prepares the state
\begin{align}
(H\otimes I ) \frac{1}{\sqrt{2}}\left( \ket{0}\ket{\psi_1} + \ket{1}\ket{\psi_2} \right)\nonumber = \frac{1}{2}\left(\ket{0}(\ket{\psi_1} + \ket{\psi_2}) +  \ket{1}(\ket{\psi_1} - \ket{\psi_2})  \right).
\end{align}
The probability of measuring $0$ is then given by
\begin{align}
p(0)=\frac{1+\text{Re}[\braket{\psi_1|\psi_2}]}{2}.
\label{eq.p0}
\end{align}
By the Chernoff-Hoeffding inequality~\cite{Chernoff52, Hoeffding63}, estimating
$\mathrm{Re}[\langle \psi_1 | \psi_2\rangle]$
to additive precision $\epsilon$ with failure probability at most $\eta$
requires
\begin{equation}
    N \ge \frac{2}{\epsilon^2}\log\!\left(\frac{2}{\eta}\right)
\end{equation}
shots.
Using quantum amplitude estimation, this precision dependence can in principle be improved from $\mathcal O(1/\epsilon^2)$ to $\mathcal O(1/\epsilon)$ in query complexity, at the cost of deeper circuits~\cite{brassard2000quantum, berry2014exponential}.

\section{Scaling, Compilation Cost, and Noise Considerations}

In this section, we discuss how the computational cost of the algorithm scales with system size for different state representations, and comment on compilation and noise considerations in the quantum setting. Our focus is to make explicit which parts of the cost are intrinsic to the algorithm and which are a consequence of the chosen classical representation.

The key point is that the method is defined by the variational self-consistency condition,
\begin{align}
    \mathcal{H}_{\textrm{shift}}|\psi(\tau + \textrm{d}\tau)\rangle
    = (\mathcal{H}_{\textrm{shift}}-\textrm{d}\tau)|\psi(\tau)\rangle,
    \label{eq.sm_self_consistency_repeat}
\end{align}
rather than by an explicit implementation of $(\mathcal{H}-\delta)^{-1}$ as an operator. This is important both for scaling and for noise: we never need to compile an inverse Hamiltonian, and the quantum subroutines reduce to estimating expectation values and overlaps appearing in the Hilbert--Schmidt distance, Eq.~\ref{eq.hilbert_schmidt}.

\subsection{Classical scaling: state vectors and MPS}

When representing the state as a full state vector, the Hilbert space dimension grows as $2^L$. Thus even storing $\ket{\psi}$ requires exponential memory. Any variational update that accesses the full state vector therefore has an exponential cost,
\begin{align}
    \textrm{cost per update} \sim \mathcal{O}(2^L),
\end{align}
up to polynomial factors associated with the chosen optimization routine. This is the regime used in Sec.~\ref{sec.state_vector_simulations} only as a small-system validation.

For a matrix product state of bond dimension $\chi$, the dominant cost is the contraction of tensor networks required to evaluate the objective function and its updates. For local one-dimensional Hamiltonians, the cost of computing expectation values and related contractions scales polynomially as
\begin{align}
    \textrm{cost per update} \sim \mathcal{O}(L\,\chi^3).
\end{align}
As discussed in Sec.~\ref{sec.state_vector_simulations}, we explored both sequential and parallel tensor updates; parallel updates preserve the same polynomial scaling but may reduce wall-clock time when the multi-core efficiency is sufficient.

We emphasize that classical MPS scaling is ultimately limited by representability. Generic mid-spectrum eigenstates are expected to have volume-law entanglement, implying that representing them with high fidelity requires $\chi$ that grows exponentially with $L$. In this sense, failure to scale classically at fixed $\chi$ reflects the limitations of the MPS representation rather than an intrinsic limitation of the algorithm itself.

\subsection{Quantum resource scaling: measurement and optimization}

On quantum hardware the state is represented directly by a parametrized circuit $U(\theta)\ket{0}$, and the variational step is performed by estimating the terms in the Hilbert--Schmidt distance,
\begin{align}
    D = \| \ket{a} - \ket{b} \|_2
    = \sqrt{\braket{a|a} + \braket{b|b} - 2\,\textrm{Re}[\braket{a|b}]},
\end{align}
with $\ket{a}$ and $\ket{b}$ defined in Sec.~\ref{sec.hadamard}.
The first two terms are expectation values of positive semidefinite operators and can be estimated by standard measurement routines. The cross term can be written as a sum over Pauli strings,
\begin{align}
    (\mathcal{H}-\delta)(\mathcal{H}-\delta-\textrm{d}\tau) = \sum_i \alpha_i P_i,
\end{align}
leading to
\begin{align}
    \textrm{Re}[\braket{a|b}] = \sum_i \alpha_i\,\textrm{Re}\left[\bra{\psi(\tau+\textrm{d}\tau)}P_i\ket{\psi(\tau)}\right],
\end{align}
where each overlap term can be measured using the modified Hadamard test shown in Fig.~\ref{fig.hadamard}. This expansion contains a number of Pauli strings that scales polynomially with system size for local Hamiltonians (quadratically for nearest-neighbor models). Mutually commuting terms can be grouped, and the measurements can be performed non-uniformly according to $|\alpha_i|$ to reduce the overall shot cost. In practice, the number of such measurement groups (i.e., sets of Pauli strings measurable in a common basis) is determined by the interaction locality and geometry of the Hamiltonian, and remains manageable for the local models considered here.

The modified Hadamard test yields an estimator for each grouped observable with additive precision $\epsilon$ using $O(\log(1/\eta)/\epsilon^2)$ shots at failure probability $\eta$, as discussed following Eq.~\ref{eq.p0}. The total shot cost per variational update therefore depends on the number of resulting measurement groups after commuting Pauli terms are combined.

In addition to this sampling cost, there is a classical optimization overhead: the number of objective evaluations required to update $\theta$ depends on the optimizer and the noise level in the objective estimates. Gradient-based methods can be used by estimating derivatives via the parameter-shift rule~\cite{Wierichs+22}, at the expense of additional circuit evaluations. 

\subsection{Compilation cost and circuit depth}

The circuit depth is set by the chosen ansatz $U(\theta)$ and by the overlap-measurement subroutine. Importantly, the circuit structure can be held fixed throughout the evolution: only the parameters $\theta$ change from step to step. This contrasts with Trotterized time evolution, where circuit depth typically grows with total evolution time. In the present approach, compilation can therefore be viewed as a one-time cost for the ansatz layout (and its controlled version as shown in Fig.~\ref{fig.hadamard}), followed by repeated parameter updates.

We note that, unlike conventional (real or imaginary) time evolution algorithms, the resource cost here does not scale with the total imaginary time $\tau$ through increasing circuit depth. Instead, larger $\tau$ corresponds to additional variational updates of a fixed-depth circuit, so that the relevant scaling is governed by measurement cost and optimizer convergence rather than by time-evolution depth. This formulation places the quantum implementation within the broader class of variational and hybrid imaginary-time algorithms, where evolution is realized through repeated parameter updates driven by measured expectation values rather than by explicit time-evolution circuits~\cite{McArdle+19,Yuan+19,Cerezo+21,Gacon+23}.

In practice, the modified Hadamard test requires controlled application of two unitaries, which in our setting correspond to the state-preparation circuits for $\ket{\psi(\tau+\textrm{d}\tau)}$ and $\ket{\psi(\tau)}$. The overhead associated with controlled operations remains proportional to the ansatz depth rather than growing with the number of imaginary-time steps.

\subsection{Noise and mitigation}

Our manuscript does not claim a near-term quantum advantage; however, it is useful to clarify how noise will affect the procedure. The variational update depends on expectation values and overlaps, so noise appears in the estimators of $\braket{a|a}$, $\braket{b|b}$ and $\textrm{Re}[\braket{a|b}]$. This differs from sequential simulation of imaginary time evolution, where errors accumulate directly through long gate sequences.

The reliance on expectation values and overlap estimation makes the method naturally compatible with standard mitigation strategies. Measurement-error mitigation can be applied to all circuits involved. Zero-noise extrapolation can be applied directly to the circuits used in Fig.~\ref{fig.hadamard} and to the circuits used for measuring $\braket{a|a}$ and $\braket{b|b}$.

We note that noise can still slow convergence by increasing the variance of the objective and gradient estimates, and can bias the optimum found by the classical optimizer. A quantitative analysis necessarily depends on the choice of ansatz, compilation strategy (particularly for the controlled-unitary primitives), and the noise properties of the target device. These considerations define a natural direction for future work in which the algorithm is deployed on specific quantum hardware platforms.

\section{State Vector Simulations}
\label{sec.state_vector_simulations}

In this section, we will first demonstrate that the technique works for small systems where we can compare with exact solutions, and we will compare the convergence properties with conventional imaginary time evolution for ground states. We make use of exact state vectors here, and pass the entire (exponentially large) statevector into the numerical optimization routine. This avoids any complications due to bond dimension, but results in an optimization problem that is exponentially large in the system size. This approach is therefore not scalable, but provides useful intuition as to the performance of the method.

We will begin by testing the method on a (disorder-free) transverse field Ising model given by:
\begin{align}
    \mathcal{H} = -J \sum_i \sigma^{x}_i \sigma^{x}_{i+1} - h \sum_i \sigma^{z}_i + h_x \sum_i \sigma^{x}_i.
\end{align}
where the final term is a weak field of strength $h_x/J=0.05$ used to weakly lift degeneracies in the energy spectrum.

We first compare the convergence of the shift-invert method against conventional imaginary time evolution for the purpose of finding ground states, i.e. we choose $\delta = E_{\textrm{min}} + \epsilon$, and for demonstration purposes, we find $E_{\textrm{min}}$ exactly and choose $\epsilon=0.1$. In both cases, we initialize the system in a Néel state and evolve it in imaginary time with a fixed time step size $\textrm{d}\tau=0.1$. In the shift-invert case, we evolve according to Eq.~6 of the main text, while in the conventional case we repeatedly multiply by the time evolution operator $U = \exp(- \mathcal{H} \phantom{.} \textrm{d} \tau)$ and normalize the resulting state after each step. We compute the fidelity of these states with the exact ground state obtained via exact diagonalization, $\mathcal{F}(\tau) = |\braket{\psi(\tau)|\psi_{ED}}|^2$, and stop the procedure when the fidelity reaches $\mathcal{F}(\tau)=0.999$.

\begin{figure}
    \centering
    \includegraphics[width=0.65\linewidth]{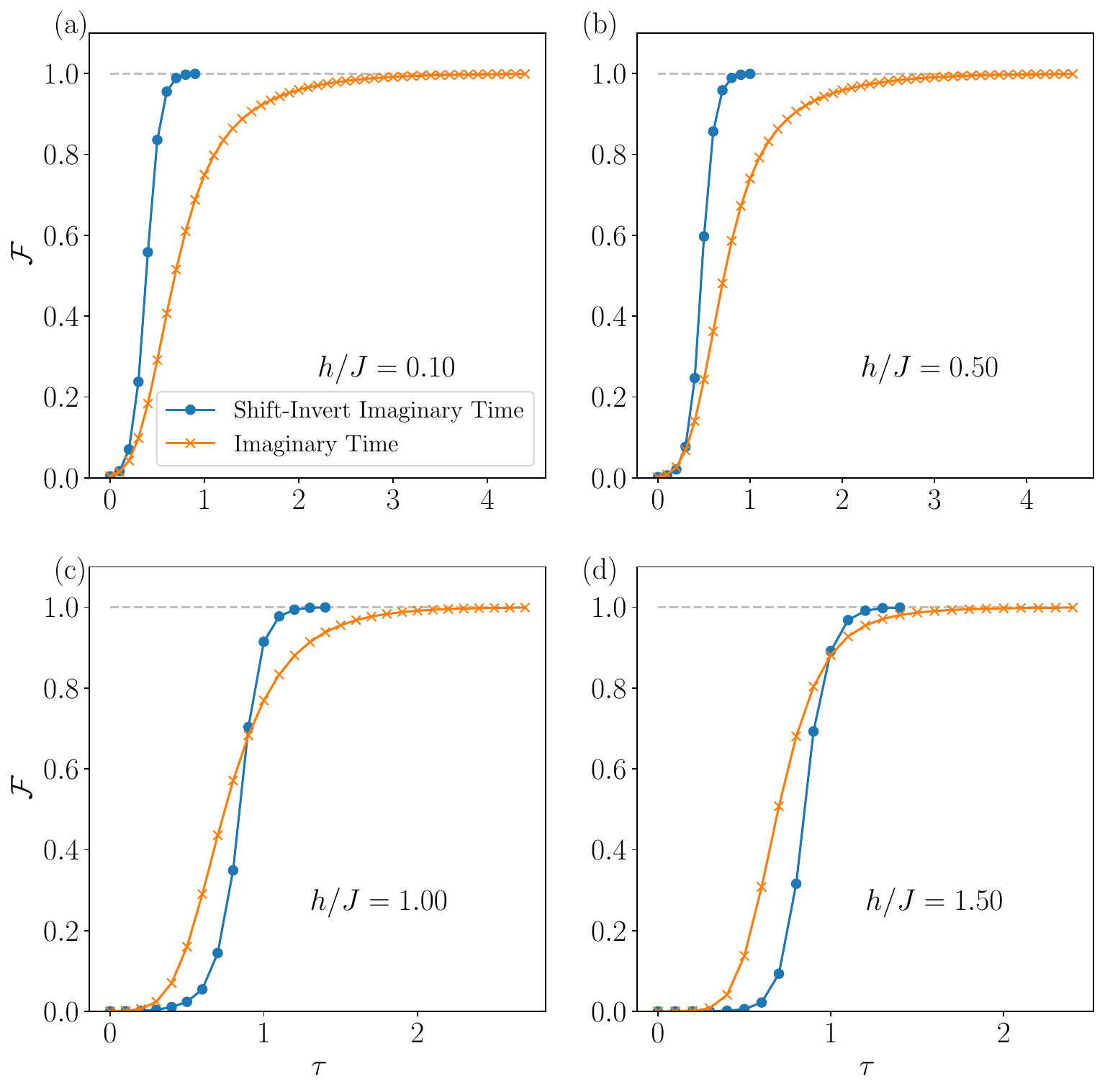}
    \caption{A comparison showing convergence to the ground state of the transverse Ising model. Conventional imaginary time evolution is shown in orange, and the shift-invert imaginary time evolution is shown in blue. All panels show the fidelity $\mathcal{F}$ with the exact ground state versus the number of time steps required to reach a fidelity of $\mathcal{F} = 0.999$. The dashed gray line indicates $\mathcal{F}=1.0$. System size is $L=8$. a) $h/J = 0.1$. b) $h/J=0.5$. c) $h/J=1.0$. d) $h/J=1.5$.}
    \label{fig.comparison_ed}
\end{figure}

The results are shown in Fig.~\ref{fig.comparison_ed} for a small system of size $L=8$, for several different values of $h/J$ in both the ordered and the disordered phase. We find that in both phases, the shift-invert procedure outperforms conventional imaginary time evolution by a significant margin, requiring far fewer timesteps to converge to the target accuracy, although it should be noted that each step of the shift-invert process involves a variational minimization procedure and takes longer to produce than one step of conventional imaginary time evolution. This confirms our expectations that the shift-invert method can in some senses be a more efficient way of finding ground states of complex systems, provided one has access to a good classical estimate of the ground state energy~\cite{Anderson51,Baumgratz+12,Eisert23}. It is worth noting that the performance of this algorithm depends critically on the classical solver used for the variational step. For these exact results, we use the \texttt{SLSQP} solver from \texttt{SciPy}.

\begin{figure}
    \centering
    \includegraphics[width=0.65\linewidth]{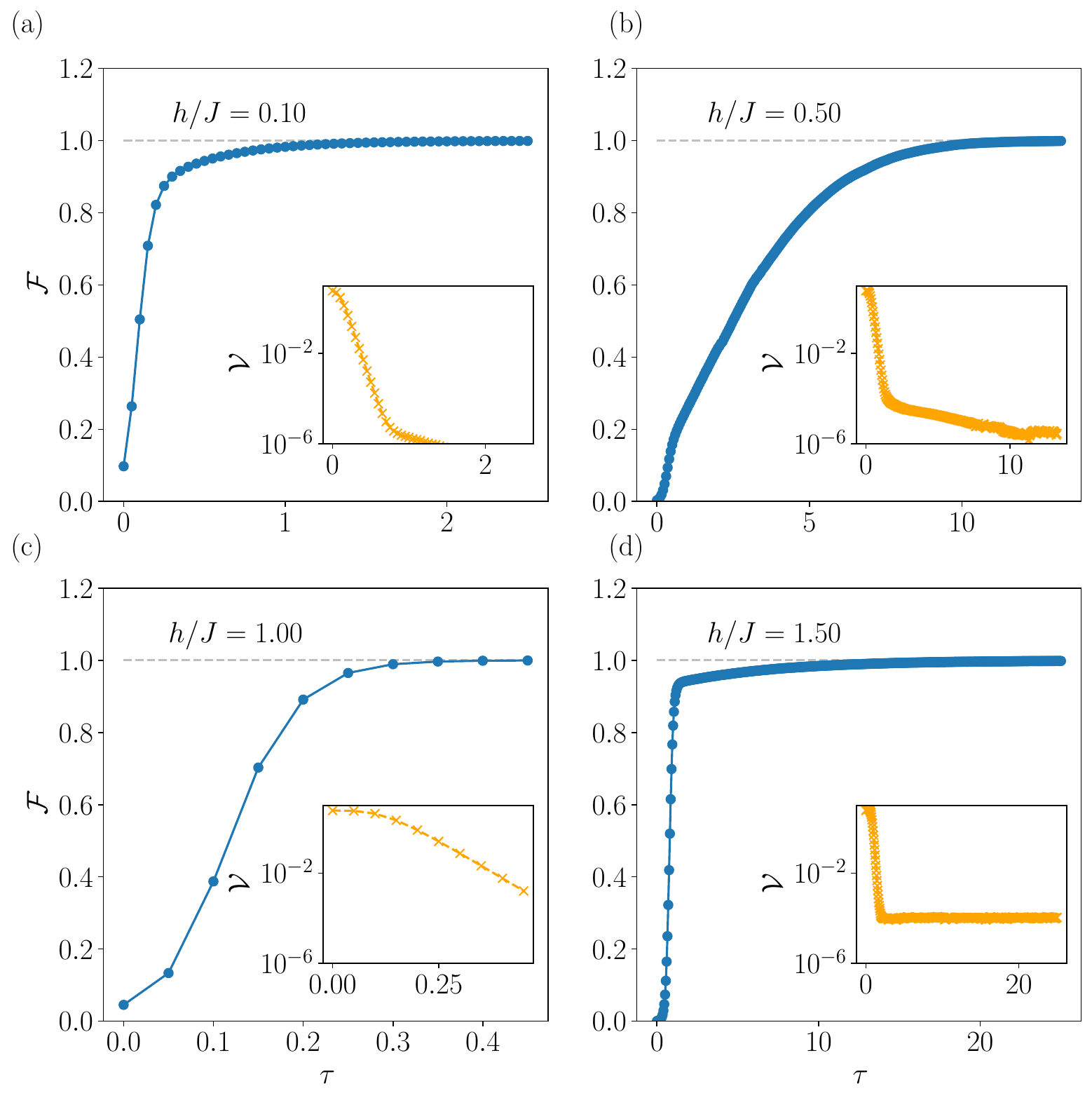}
    \caption{Performance of the shift-invert imaginary time evolution algorithm for finding mid-spectrum states of a transverse Ising model, with system size $L=7$. All panels show the fidelity $\mathcal{F}$ with the exact target state versus the number of time steps. Again, we use a cutoff fidelity of $\mathcal{F}=0.999$. Insets show the convergence of the variance $\mathcal{V}$. a) $h/J = 0.1$. b) $h/J=0.5$. c) $h/J=1.0$. d) $h/J=1.5$.}
    \label{fig.midspectrum_ed}
\end{figure}

We now turn to excited states, and we specify to mid-spectrum states where we choose the target energy to be $\delta = (E_{\textrm{max}} + E_{\textrm{min}})/2 + \epsilon$. Again, we start from a Néel state and evolve according to Eq.~6 of the main text until we reach a fidelity of $\mathcal{F}=0.999$. As shown in Fig.~\ref{fig.midspectrum_ed}, we again find excellent results, confirming that the algorithm performs as expected and is able to smoothly converge to highly excited states. Note that depending on the size of the timestep $\textrm{d}\tau$, the algorithm may converge to a state close, but not precisely equal to, to the target state. This can be understood as essentially a form of Trotter error. The time required to converge depends crucially on both the initial state and how close the target state is to neighboring eigenstates of $\mathcal{H}_{\textrm{eff}}$---naively, we would expect energy differences on the order of $\Delta E$ to require a timescale of $\tau \propto 1/\Delta E$ in order to resolve them. In Fig.~\ref{fig.midspectrum_ed}, we also demonstrate that the variance quickly decays as the fidelity rises, confirming that it is an appropriate metric to use in situations where we do not know the target state and cannot compute a fidelity. It is important to emphasise, however, that there is no one-to-one link between the fidelity and the variance: some samples reach the target fidelity with a variance several orders of magnitude higher than that of other samples. Naively, the variance of a completely random state (not an eigenstate) will depend on both the system size and the disorder bandwidth. For large, strongly disordered systems, the variance of a randomly chosen initial state will be greater than for small, weakly disordered systems, making the task of reaching a constant threshold variance more of a challenge. Nonetheless, as the variance should vanish for an exact eigenstate, we have opted to maintain the target of constant variance and not rescale the variance by system size or normalise the eigenspectra in order to present our results as clearly as possible. If one were to compute instead the variance of the \emph{energy density}, or to normalise the eigenspectrum, the variances reported in the main text would in general be three to four orders of magnitude smaller.

\section{Details of the MPS Implementation}
\label{sec.mps_implementation}

\begin{figure}
    \centering
    \includegraphics[width=0.35\linewidth]{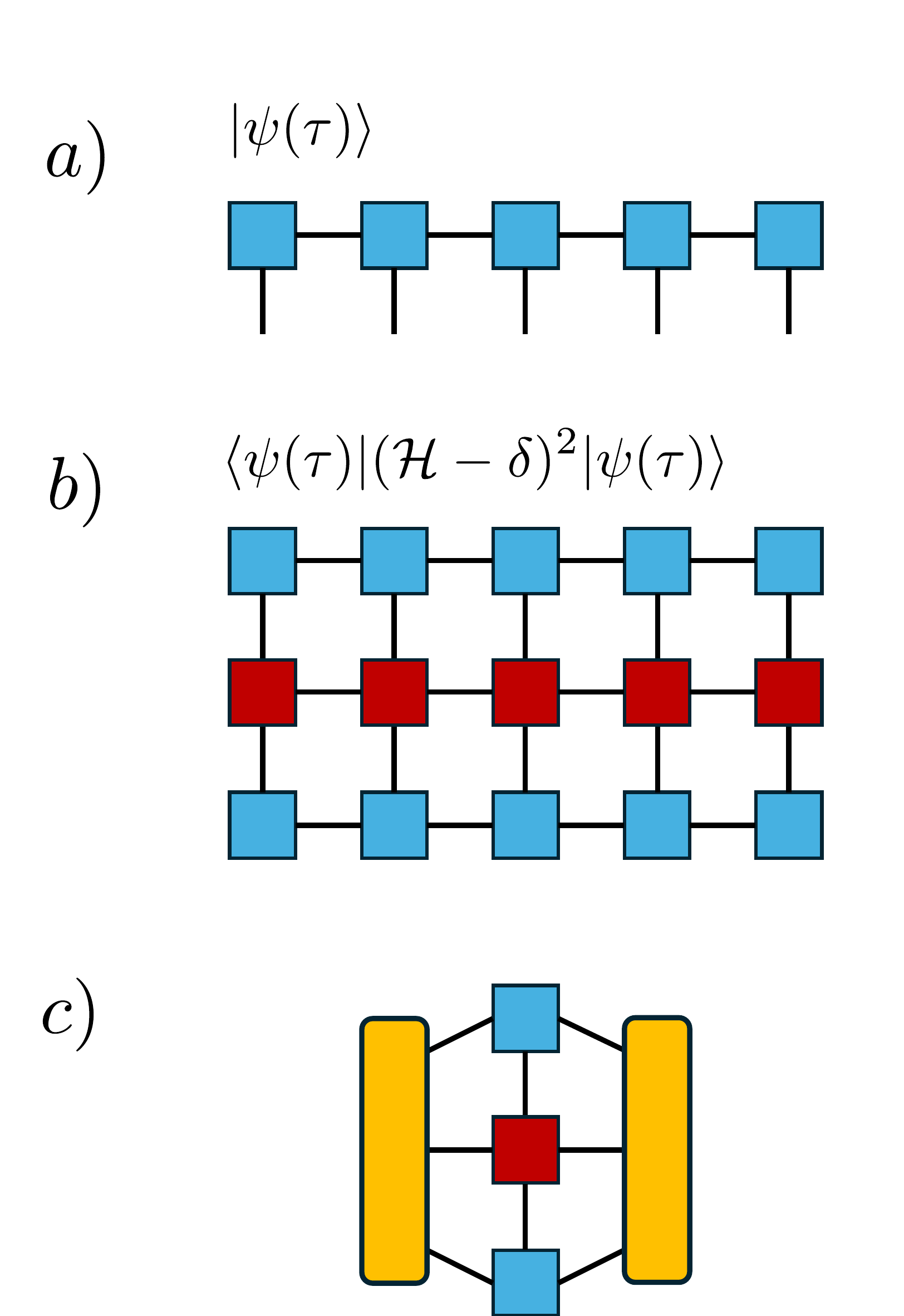}
    \caption{A sketch showing how the MPS is updated at each imaginary time step. a) The initial MPS, for example representing $\ket{\psi(\tau)}$. b) The type of expectation value we need to compute, here showing the term $\braket{a|a} = \langle \psi(\tau) | (\mathcal{H} - \delta)^2 | \psi(\tau) \rangle$ in Eq.~7 of the main text. The red squares represent a matrix product operator (MPO). c) Each lattice site is updated individually by tracing out the other lattice sites surrounding it, then the resulting updated tensor is inserted back into the original MPS. This procedure can be performed in parallel for each lattice site, yielding a potential speedup over sequential evaluation.}
    \label{fig.mps_contractions}
\end{figure}

One of the key benefits of using a matrix product state (MPS) representation for this algorithm is that the problem of variationally solving Eq.~6 in the main text can be broken down into a series of individual optimization problems for each lattice site. A sketch is shown in Fig.~\ref{fig.mps_contractions}. Note that unlike conventional tensor network algorithms, here it is possible to update each MPS tensor in parallel, load the updated tensors back into the original MPS, then normalize the final result. We note that the assumption that each MPS tensor can be optimized individually is not rigorous, and in cases where the convergence is slow or otherwise challenging, it may be better to either use a sequential sweeping technique (as in conventional DMRG) or to contract pairs of neighbouring tensors together, optimize them as a pair, and then separate them again using a singular value decomposition or similar method.
For the majority of our results, we used sequential updates, as we found the CPU efficiency of the parallel update scheme to be anomalously low, however for larger problem sizes and bigger bond dimensions this approach may prove more efficient in the long run.
Normalization can be performed after the update step: imaginary time evolution is inherently not a unitary operation, so normalization must be enforced by hand. We have also explored stochastic update steps, either by selecting a subset of elements in each tensor to be optimized over (reducing the cost of the gradient descent step), or by updating a random sample of tensors at each timestep rather than every single one. In the present work, we switch to stochastic updates when the bond dimension increases above $\chi=12$, however we did not encounter this situation very often. We did not employ the randomly chosen tensor update strategy for the results presented in this work. For very large system sizes and bond dimensions, these strategies are likely to be the only ones which are classically computationally tractable. 

Computing these contractions over all lattice sites bar the one we are updating allows the optimization step to be broken into a series of small optimizations for each lattice site. We made use of two approaches, the \texttt{L-BFGS} gradient descent algorithm with finite difference methods and the \texttt{Newton}/\texttt{NewtonTrustRegion} optimization routines with automatic differentiation to compute gradients. Both were from the Julia \texttt{Optim.jl} library. We did not find significant performance differences between these methods. We note that while automatic differentiation is not possible on quantum hardware, one can instead use methods such as the parameter shift rule~\cite{Wierichs+22} to efficiently compute the required gradients. Classically, it is also possible that performing the tensor contractions on graphics processing units (GPUs) may offer improved speed and efficiency, however this is beyond the scope of the present work.

\section{Determination of the timestep}

To efficiently perform the imaginary time evolution, we must strike a balance between finding a timestep $\textrm{d}\tau$ small enough that the Taylor expansion remains valid, but large enough that the procedure converges in a reasonable number of steps. We can adaptively choose the timestep by examining the Taylor expansion:
\begin{align}
    \textrm{e}^{-(\mathcal{H}-\delta)^{-1} \textrm{d}\tau} \ket{\psi(\tau)} \approx \left[ 1 - (\mathcal{H}-\delta)^{-1} \textrm{d}\tau + \frac12 (\mathcal{H}-\delta)^{-2} (\textrm{d}\tau)^2 + ... \right] \ket{\psi(\tau)}
\end{align}
In order for the term of order $\mathcal{O}(\textrm{d}\tau^2)$ to be negligible, we require:
\begin{align}
    (\mathcal{H}-\delta)^{-1} \textrm{d}\tau \ket{\psi(\tau)} \gg \frac12 (\mathcal{H}-\delta)^{-2} (\textrm{d}\tau)^2) \ket{\psi(\tau)}
\end{align}
which leads to:
\begin{align}
    \textrm{d}\tau \ket{\psi(\tau)} \ll 2(\mathcal{H}-\delta) \ket{\psi(\tau)}
\end{align}
We can multiply both sides by $\bra{\psi(\tau)}$ and make use of $E(\tau) = \braket{\psi(\tau) | \mathcal{H} | \psi(\tau)}$ to obtain the condition:
\begin{align}
    \textrm{d}\tau \ll 2 (E(\tau) - \delta)
\end{align}
We impose a minimum timestep $\textrm{d}\tau_{\textrm{min}}=10^{-3}$ to prevent the algorithm from getting stuck close to the target energy. For example, one could imagine a scenario where a state that is not an eigenstate by chance has an instantaneous energy $E(\tau)$ close to $\delta$: in this case, the size of the timestep would shrink to something close to zero and the algorithm would slow to a crawl despite not yet having converged to an eigenstate. Retaining a minimum timestep size of $\textrm{d}\tau_{\textrm{min}} > 0$ allows us to avoid such scenarios, which occur often when initialising the system in a fully random state.

In addition to the above, we allow the algorithm to adaptively choose the sign of $\textrm{d}\tau$. As discussed in the main text, and shown in Ref.~\cite{Pietracaprina+18}, the eigenstates just above and below the target energy become the extremal eigenstates of $\mathcal{H}_{\textrm{eff}}$. By changing the sign of $\textrm{d}\tau$, we can choose which of the extremal eigenstates we target, i.e. the ground state or the maximally excited state. In the main text, we set the target energy $\delta = 0$, which corresponds to a mid-spectrum excited state. Our initial `warm-up' step where we find a low bond dimension approximation to the ground state of $(\mathcal{H}-\delta)^{2}$ may be close to either the eigenstate just above $\delta$ or just below $\delta$, and \emph{a priori} we have no way of knowing which. By computing the expectation value of the initial state with respect to $\mathcal{H}$, we can learn whether the initial spectral folding step has produced an initial state with energy above or below the target energy $\delta$, and use this to inform our choice of the sign of $\textrm{d}\tau$, i.e. by using $\textrm{sign}(\textrm{d}\tau) = \textrm{sign}(\braket{\psi(0) | \mathcal{H} | \psi(0)})$. This helps to avoid situations where, for example, the initial spectral folding produces something close to the highest excited state of $\mathcal{H}_{\textrm{eff}}$, yet we choose to target the ground state, essentially making no use of the spectral folding step and starting the imaginary time evolution from something very far from the target state. Instead, by noting the sign of the energy of the initial state $\ket{\psi(0)}$ in this case, we could instead immediately see that the algorithm will converge much more quickly to the ground state of $-\mathcal{H}_{\textrm{eff}}$ (note the minus sign), and so we target this state instead. This approach is valid in the case considered in the main text because we expect all mid-spectrum states which are close in energy to behave similarly, and it does not matter whether we converge to the state just above or just below the target state. This approach would not be appropriate if we wished to target a very specific state, e.g. the first excited state, where the sign of $\mathrm{d}\tau$ is crucial in order to obtain the specified target state.

\section{Comparison with Spectral Folding}

In the main text, we claimed that the ground state of the folded Hamiltonian $(\mathcal{H}-\delta)^2$ at small bond dimension was a good initial state for the shift-invert imaginary time evolution, but that using DMRG directly on the folded Hamiltonian would not produce results as accurate as those we obtain from the shift-invert procedure. Here, we show data to support this statement. The fidelity and variance shown here for the folded Hamiltonian were obtained for each disorder realization after the shift-invert procedure was performed, using the same maximum bond dimension as the shift-invert method converged to at the end of the evolution (rather than restricting the DMRG to the small bond dimension used to initialize the shift-invert runs).

\begin{figure}
    \centering
    \includegraphics[width=0.65\linewidth]{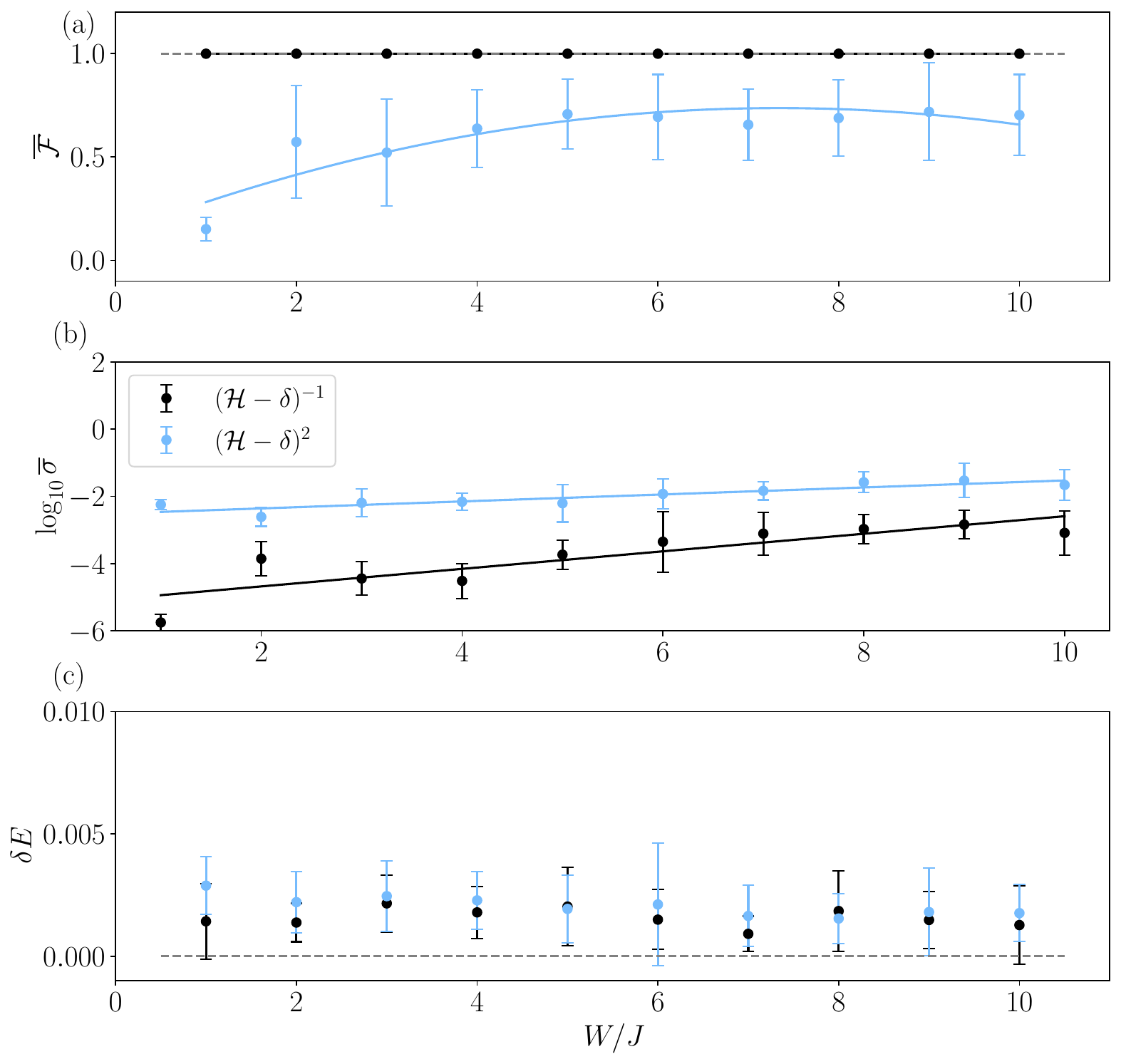}
    \caption{A comparison of the results of our shift-invert imaginary time method with direct use of DMRG on the folded Hamiltonian $(\mathcal{H}-\delta)^2$. a) The disorder-averaged fidelity. b) The logarithm of the disorder-averaged variance. c) The average distance between the energy of the obtained state and the target energy, normalised by the bandwidth of the eigenspectrum. Error bars indicate the standard deviation over disorder realizations, and the solid lines are guides to the eye.}
    \label{fig.folded_comparison}
\end{figure}

\begin{figure}
    \centering
    \includegraphics[width=0.45\linewidth]{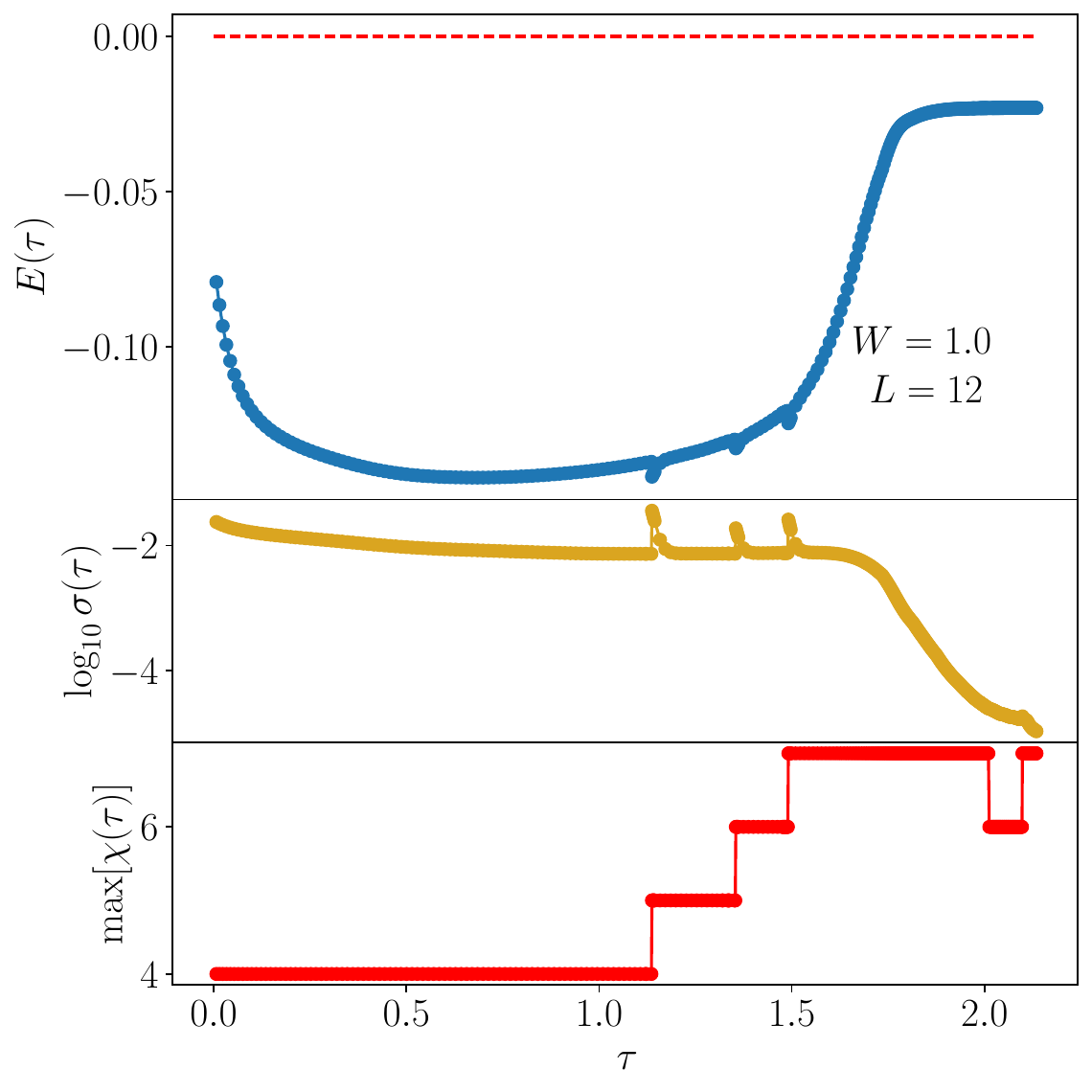}
    \includegraphics[width=0.45\linewidth]{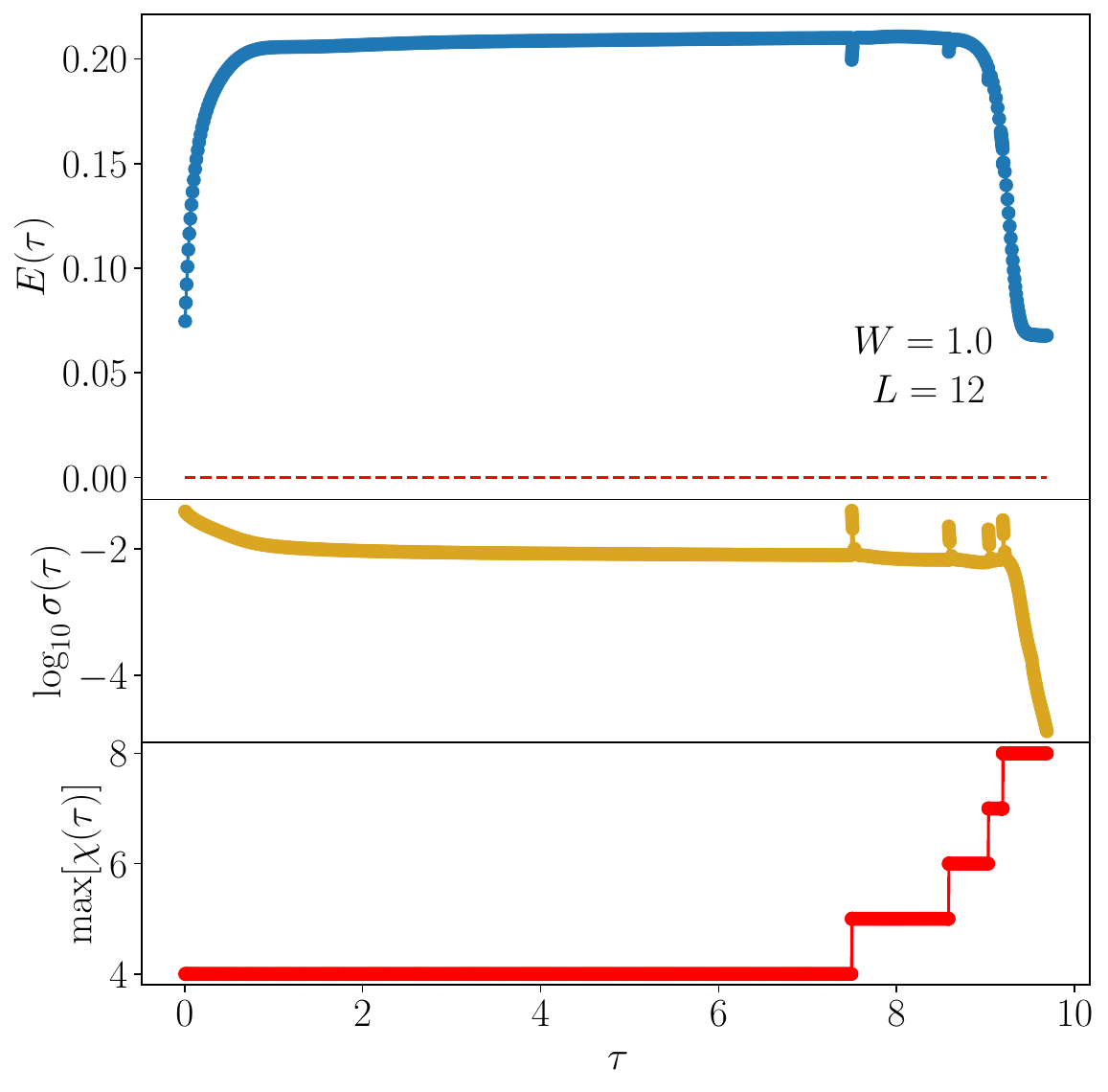}
    \includegraphics[width=0.45\linewidth]{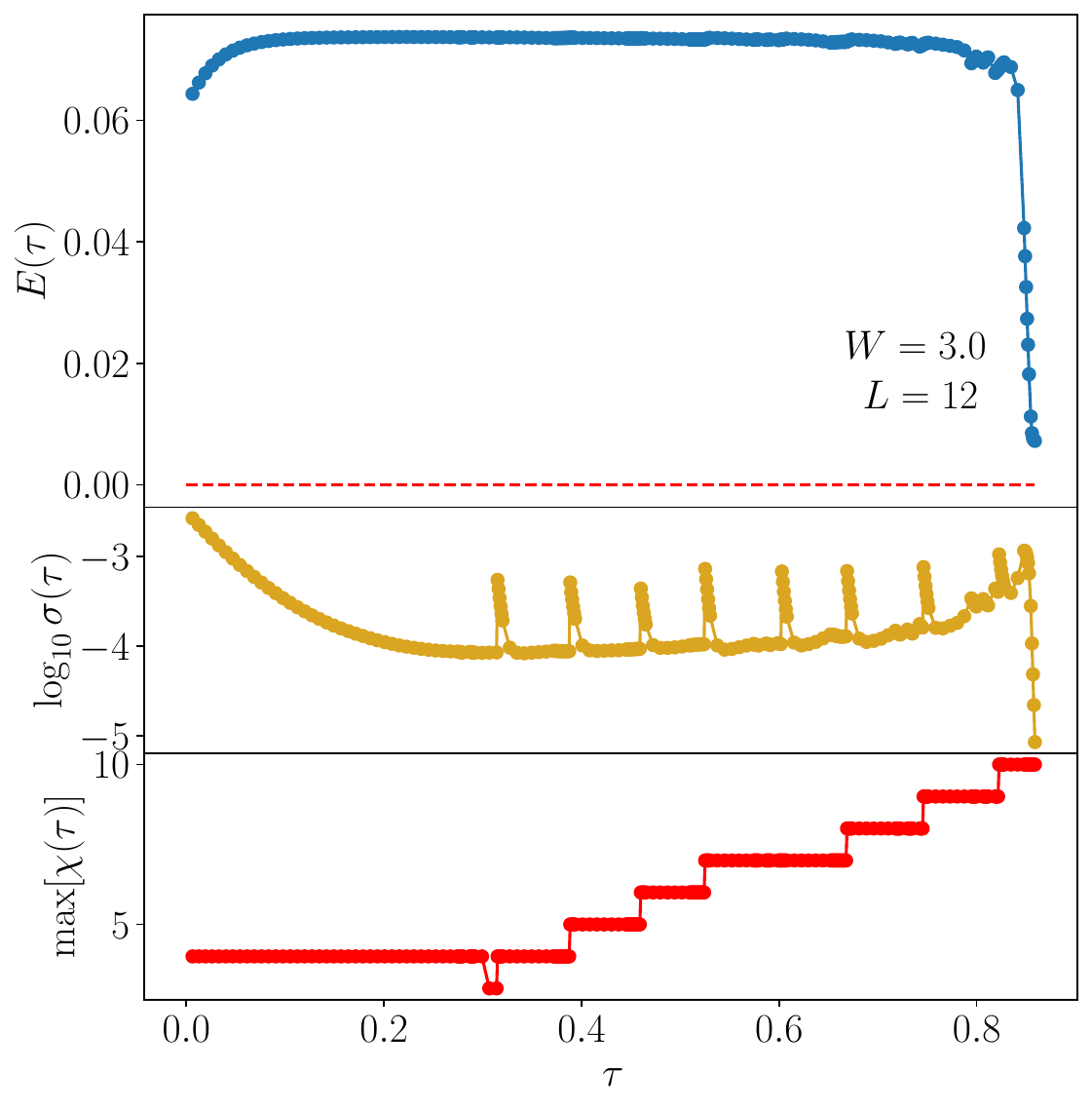}
    \includegraphics[width=0.45\linewidth]{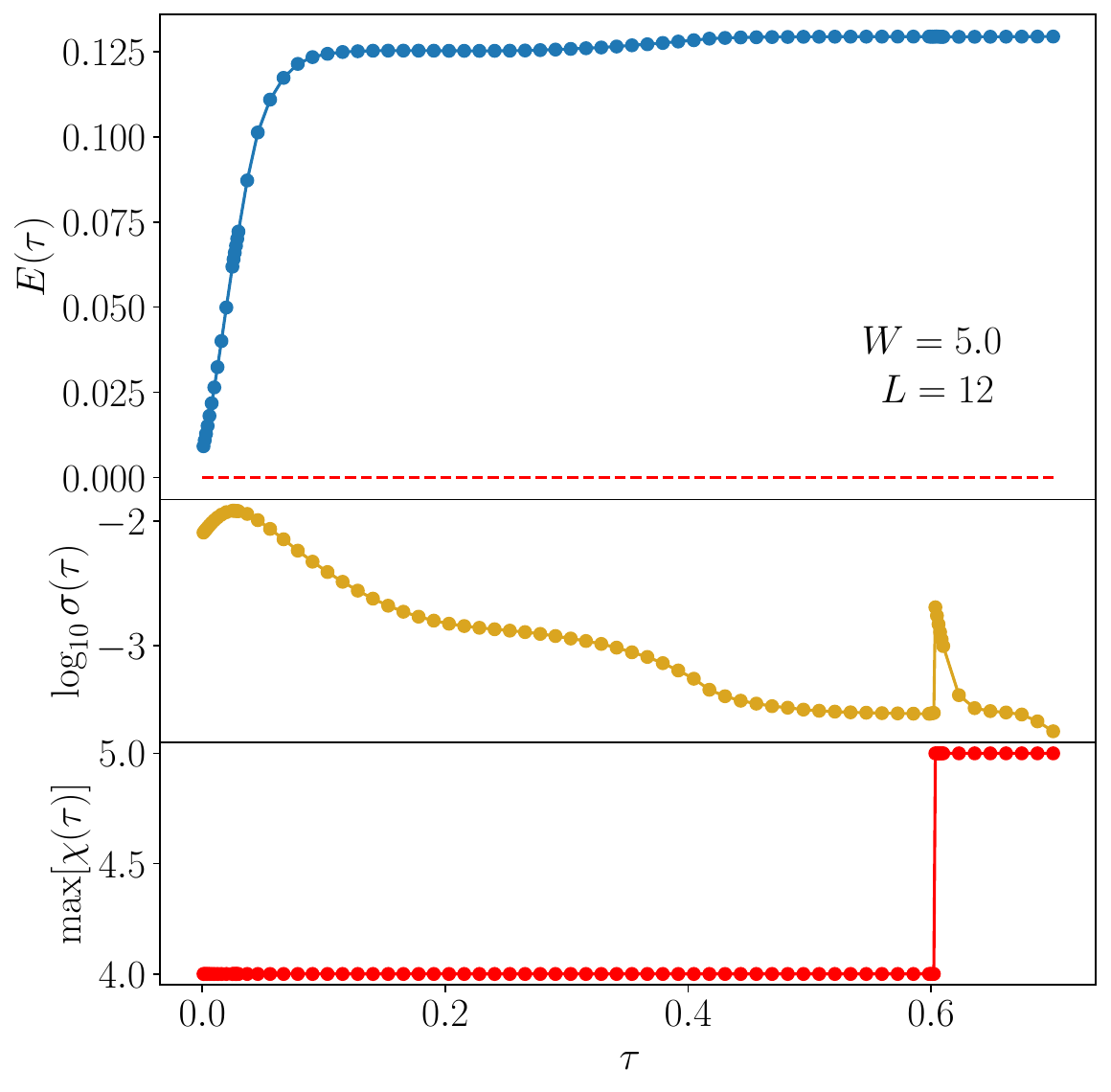}
    \caption{Additional results for $L=12$ and several values of the disorder strength $W$, showing the discontinuities in the variance when the bond dimension is increased by adding a random MPS of bond dimension one. While the end results are eigenstates, the intermediate evolution is noisy and not optimal. This method was used to obtain some of the $L=12$ results, but was not used for larger system sizes.}
    \label{fig.further_results1}
\end{figure}

The results are shown in Fig.~\ref{fig.folded_comparison}. The data here is the same $L=12$ data shown in the main text, as well as in Figs.~\ref{fig.further_results1}, \ref{fig.further_results_12_1}, \ref{fig.further_results_12_2} and \ref{fig.further_results_12_3}. We can see immediately that the spectral folding method produces far worse results, as measured by both the fidelity and the variance. The fidelity is typically around half the value obtained by the shift-invert procedure, with much larger error bars, while the variance is several orders of magnitude larger. Both methods produce states close to the target energy, however in the case of spectral folding these states are superpositions of states clustered around this energy, whereas for the shift-invert procedure the final states are eigenstates. This confirms that the shift-invert imaginary time method outperforms direct DMRG using the folded Hamiltonian, but also clearly demonstrates that this fast, efficient method serves as a suitable preconditioning step to obtain a good `warm start' solution that can be improved upon using our shift-invert procedure.

\section{Bond Dimension Increases}

By contrast with typical tensor network algorithms, here we are variationally updating the individual tensors in a given matrix product state, and as such there is no immediately natural increase of the bond dimension of the sort encountered in e.g. time evolution, where the bond dimension of the MPS would increase when acted on by a time evolution operator or a Hamiltonian.

In our simulations, we monitored the variance and manually increased the bond dimension whenever the instantaneous variance $\sigma(\tau)$ increased above the average of the previous five timesteps, or if the algorithm consistently failed to find an updated MPS $\ket{\psi(\tau + \textrm{d} \tau)}$ with a variance $\sigma(\tau + \textrm{d}\tau) \leq 10 \phantom{.} \sigma(\tau)$, which we took as our criterion for whether to accept or reject a variational update.
We employed two main strategies of increasing the bond dimension in our simulations. The first was the addition of a random MPS with bond dimension one. The second was an approach motivated by more conventional subspace expansion methods~\cite{Hubig+15}, $\ket{\psi'} = (1 + a (\mathcal{H}-\delta)) \ket{\psi}$ where $a = 10^{-3}$ is chosen to be a small constant. While both methods work, the second method was consistently superior, often resulting in sharp decreases in variance whenever the bond dimension was increased. By contrast, the former method tended to result in sharp discontinuities in the variance when the bond dimension was increased, although ultimately the expanded variational space still allowed the algorithm to converge in most scenarios. We did not use this technique for $L>12$, as the convergence was too slow to be practical.

\section{Degeneracies \& Level Spacing}

Imaginary time evolution cannot distinguish between two exactly degenerate states, and in general for two states separated by an energy $\Delta$, we can expect that it will take a timescale $T \propto 1/\Delta$ to be able to resolve them. For weak disorder strengths in particular, where the level spacing can become very small, the method can become stuck in low-variance superpositions of closely separated eigenstates. In these situations, one can either accept the state as likely to be indistinguishable from an eigenstate for many practical purposes, or simply re-run the simulation with a different initial state. The final state obtained from this procedure depends sensitively on the particular initial state one begins with. As we start from matrix product states of low bond dimension, this is loosely equivalent to starting from a sparse state vector with contributions from only a polynomial number of all of the possible states in the Hilbert space. (This is to be contrasted with the conventional starting point for imaginary time evolution, which is often taken to be an infinite temperature state, e.g. a uniform superposition of all possible states, or a Haar-random state which behaves similarly.) If one is unlucky and the initial state contains a strong contribution from several states very close to the target energy, one may have to run the simulation for a very long time in order to distinguish them. Rather than paying this cost, it is usually better to simply restart the simulation with a different initial state.

Similar considerations apply to very large systems, where we can expect the level spacing between adjacent states to be comparable to (or below) machine precision. Again, due to the sparsity of our initial state, we do not anticipate severe issues due to the unlikelihood of the initial state being in a superposition of such closely separated states, however this can be verified \emph{a posteriori} by running the simulation several times, starting from different initial states, and computing the overlaps of the obtained states. If the overlaps are zero or one, the obtained states are orthonormal and are very likely to be eigenstates. If the overlap is between zero and one, the obtained states are not orthonormal and are not eigenstates. At this point, one can continue evolving the states until the desired accuracy is reached.

\section{Further Numerical Results}

\begin{figure}
    \centering
    \includegraphics[width=0.4\linewidth]{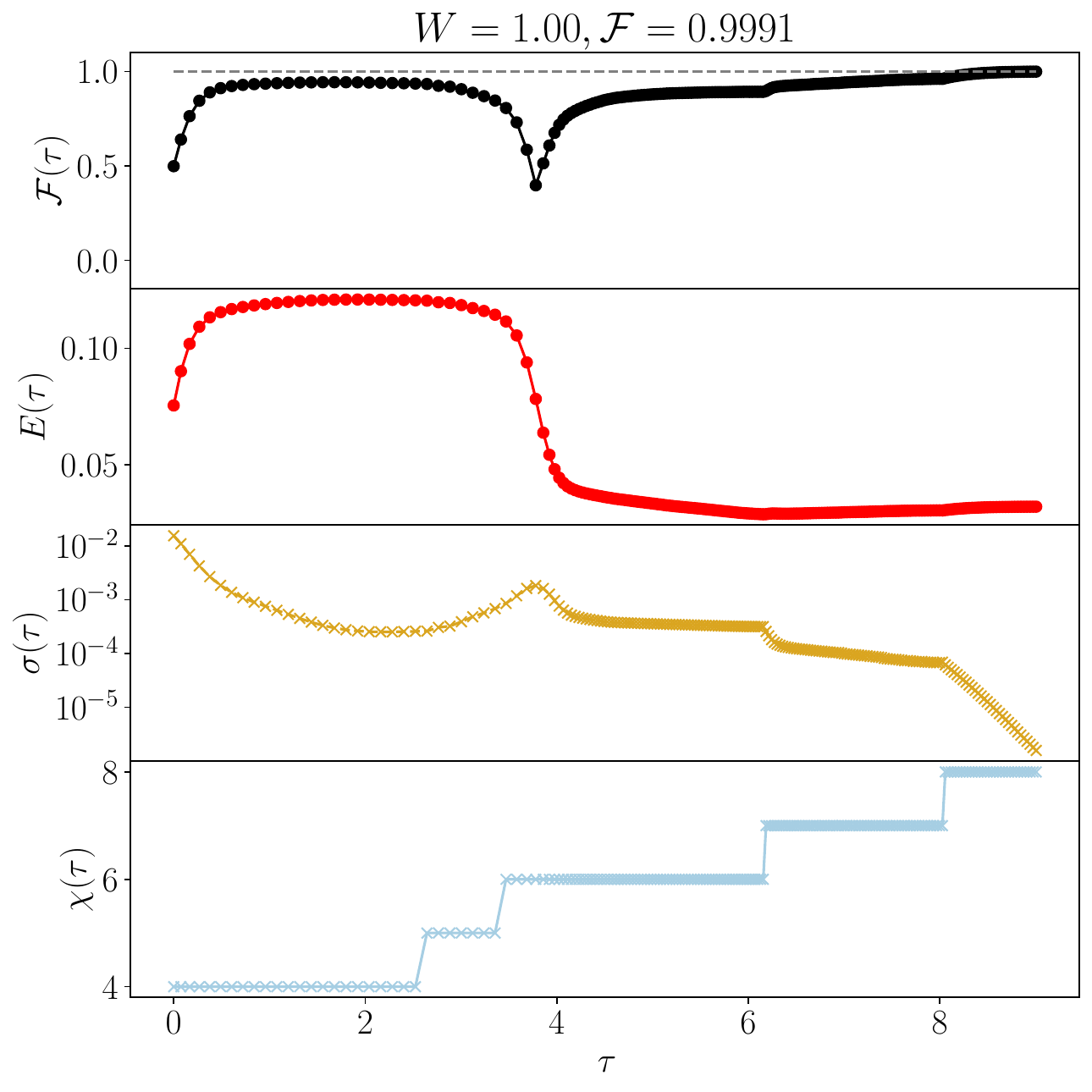}
    \includegraphics[width=0.4\linewidth]{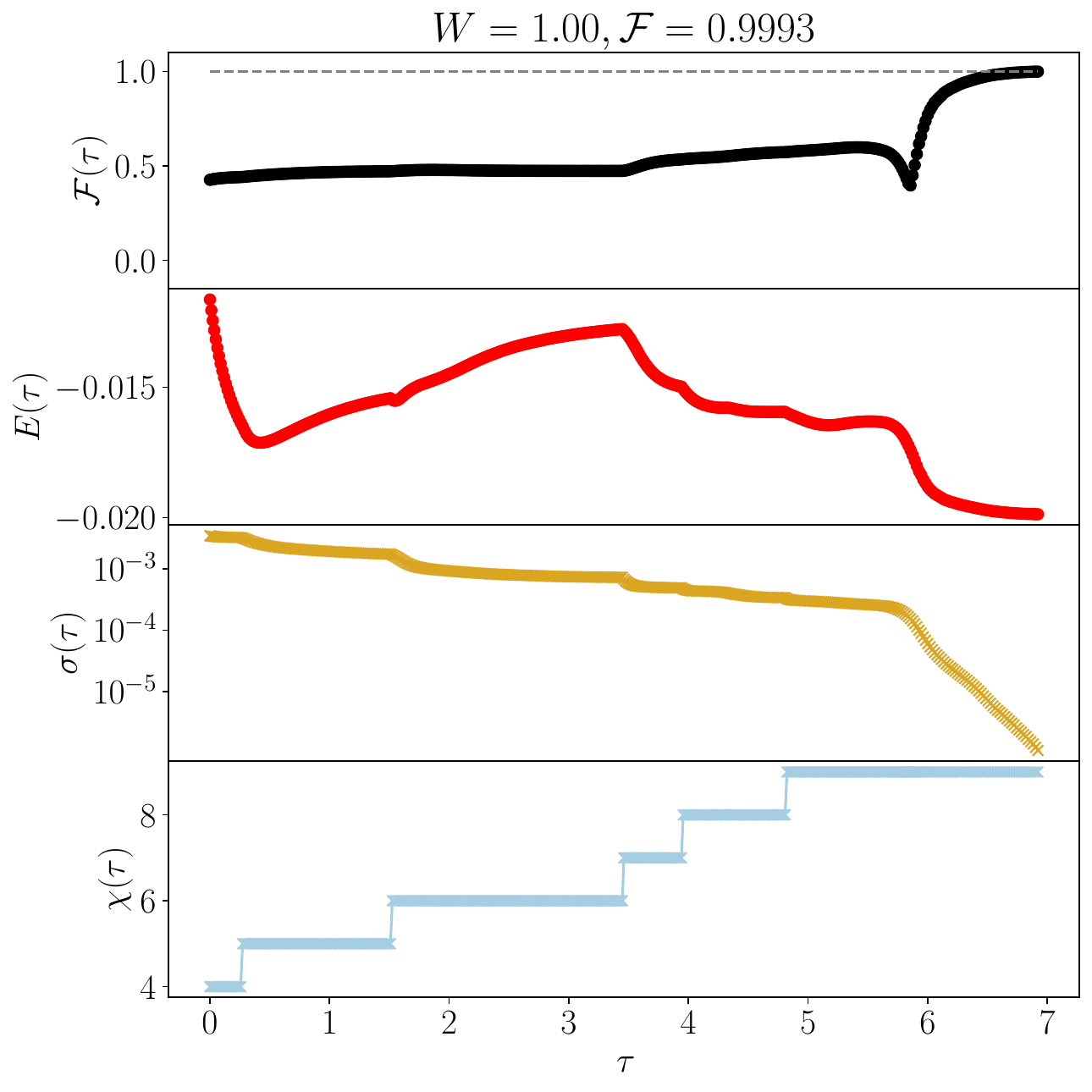}
    \includegraphics[width=0.4\linewidth]{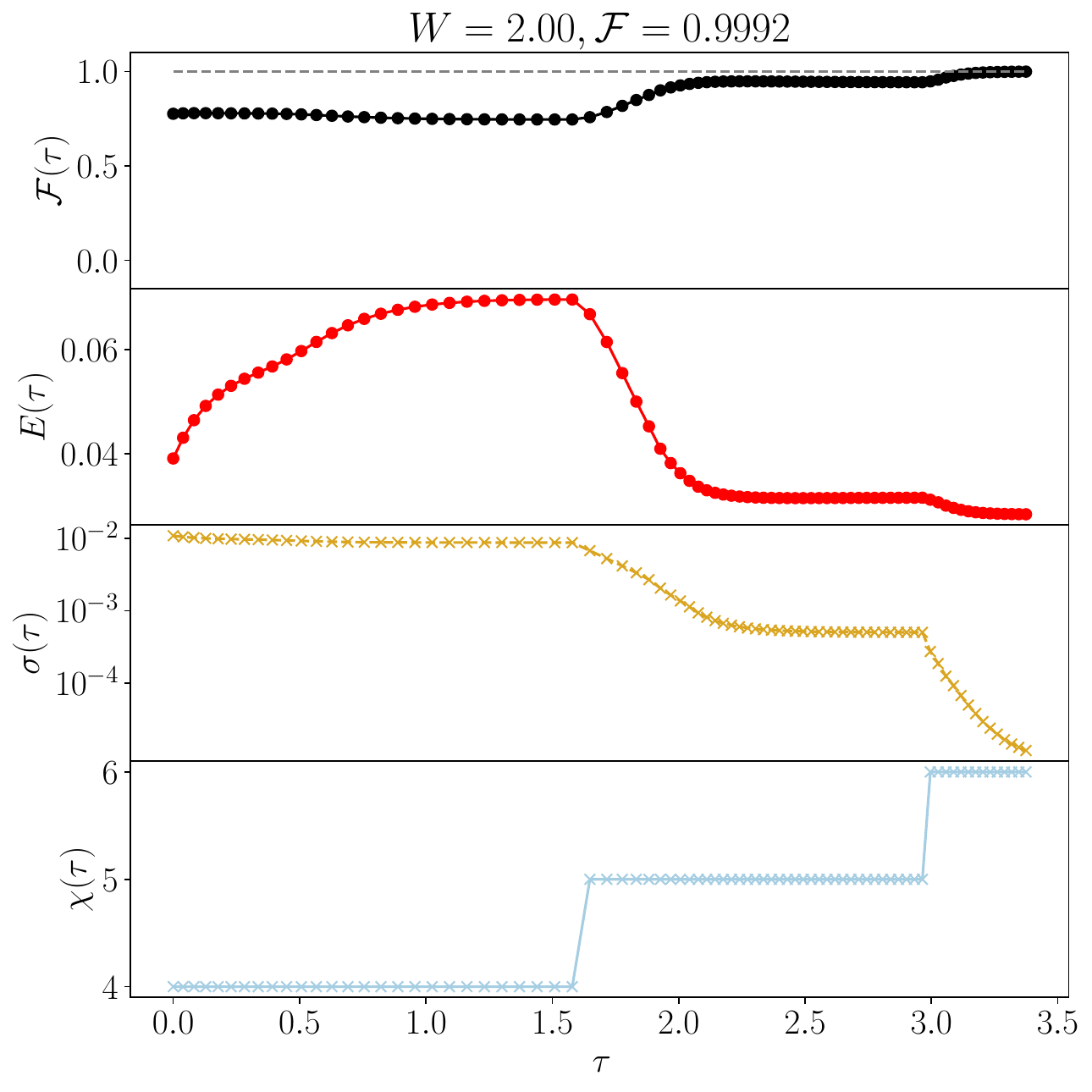}
    \includegraphics[width=0.4\linewidth]{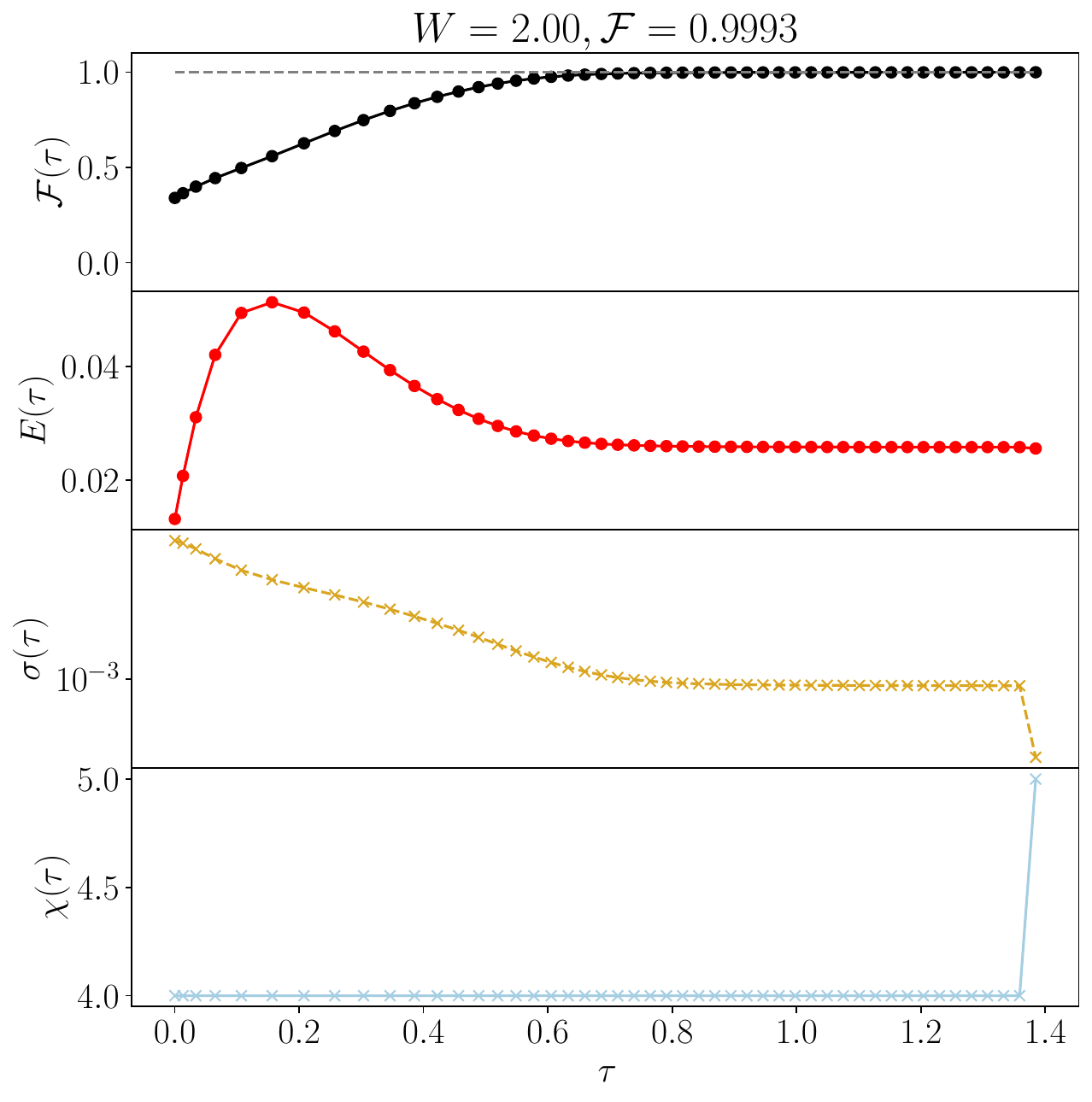}
    \includegraphics[width=0.4\linewidth]{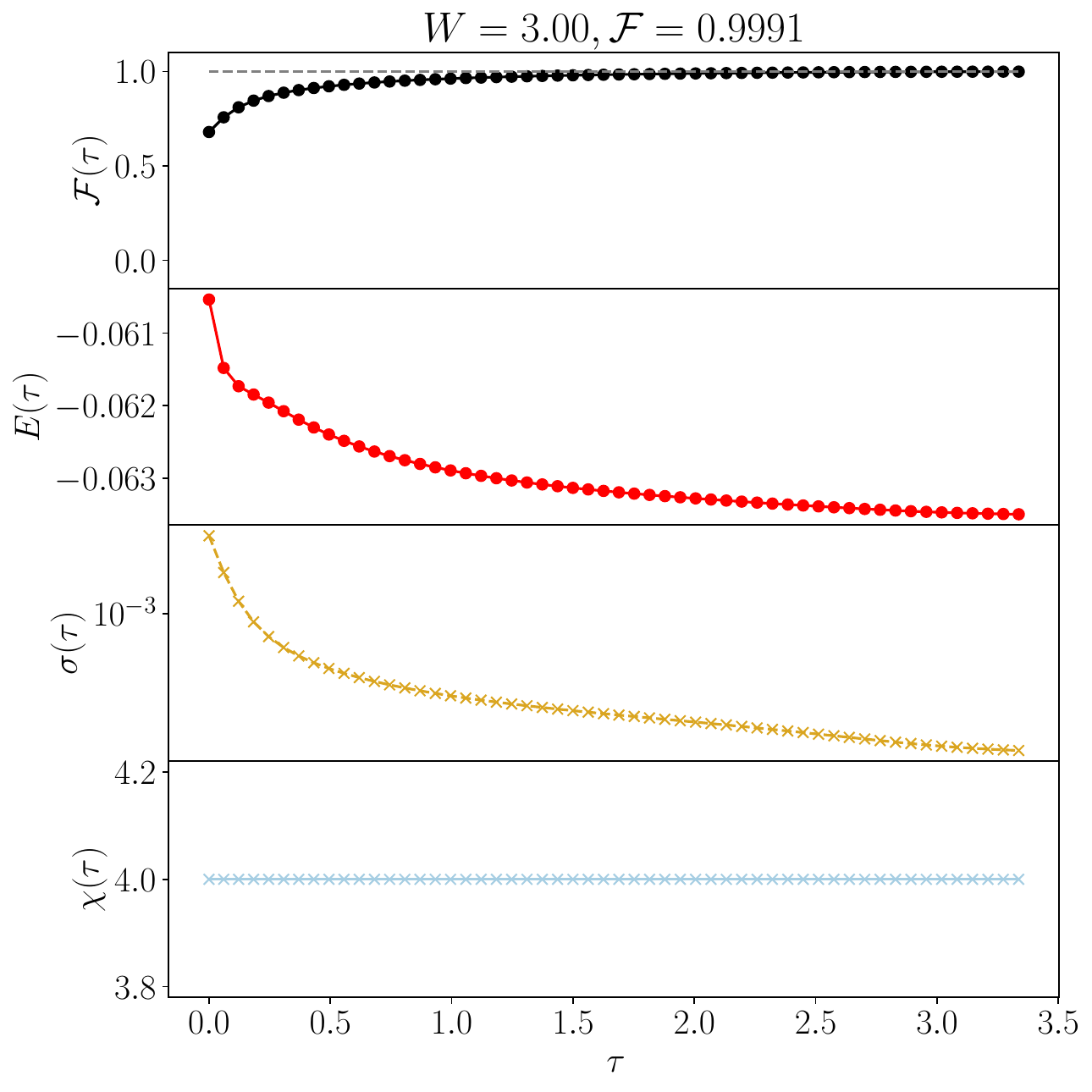}
    \includegraphics[width=0.4\linewidth]{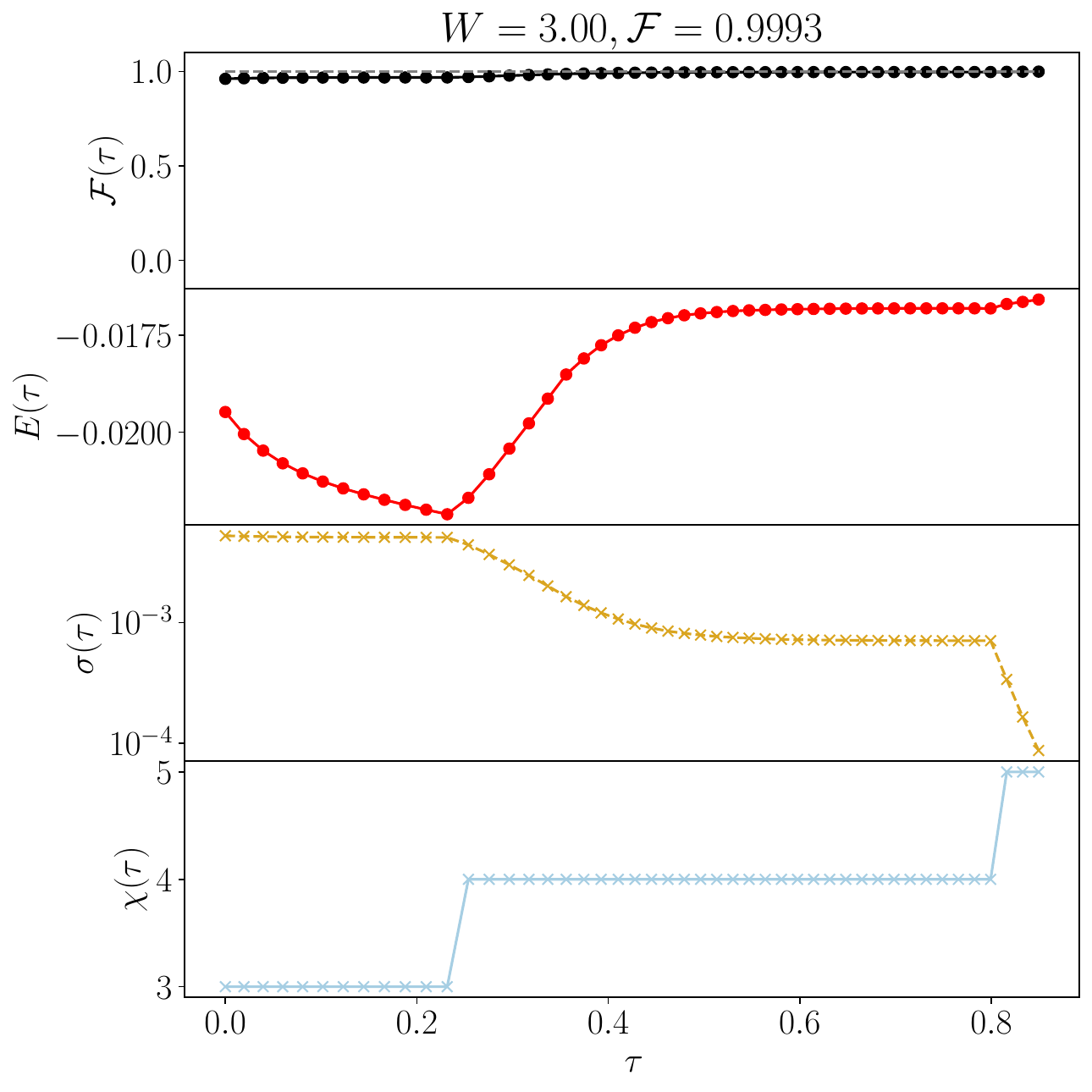}
    \caption{Additional representative results for $L=12$. Panels show the evolution of the fidelity, the energy, the variance, and the bond dimension. The figure label indicates the maximum fidelity achieved at the end of the evolution.}
    \label{fig.further_results_12_1}
\end{figure}

\begin{figure}
    \centering
    \includegraphics[width=0.4\linewidth]{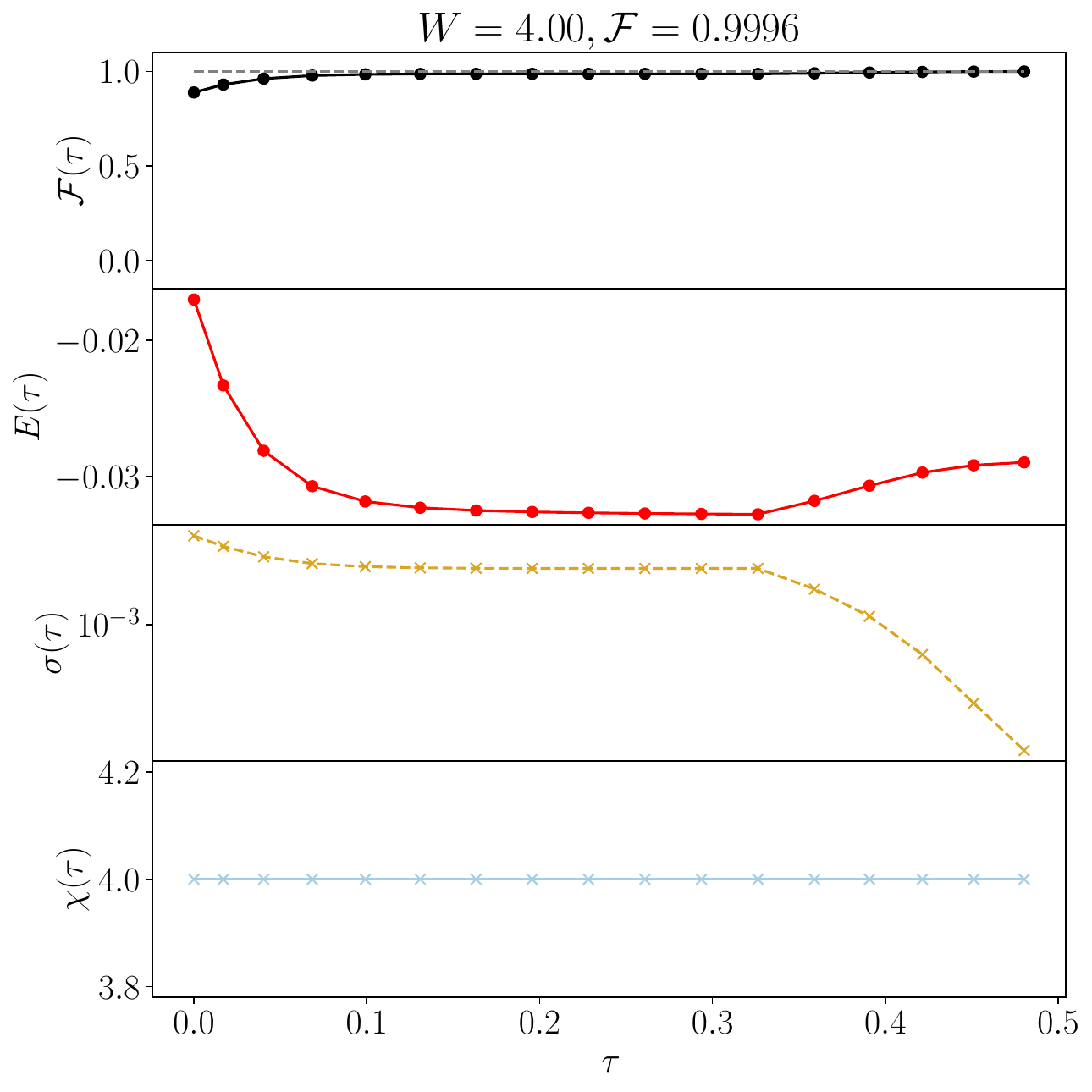}
    \includegraphics[width=0.4\linewidth]{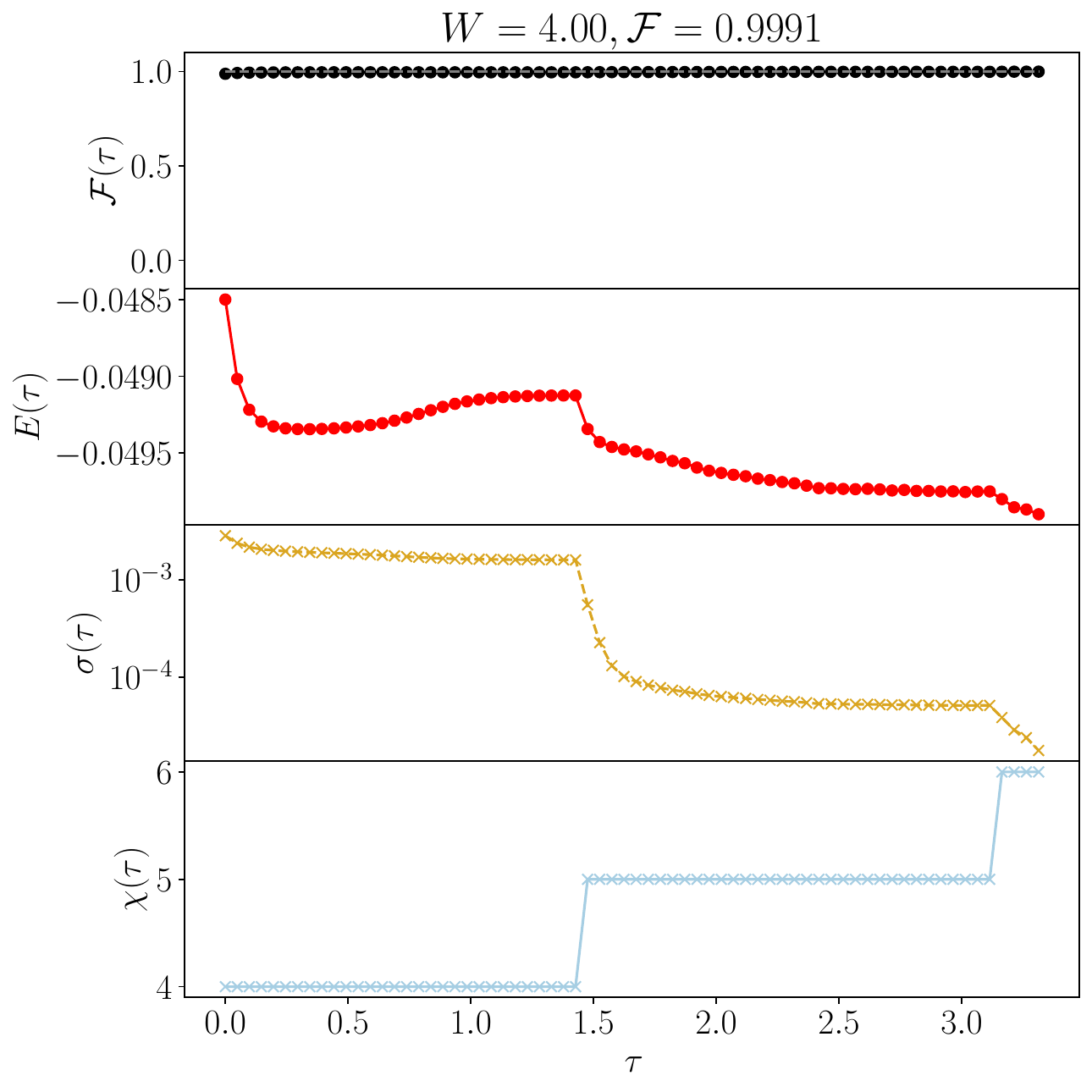}
    \includegraphics[width=0.4\linewidth]{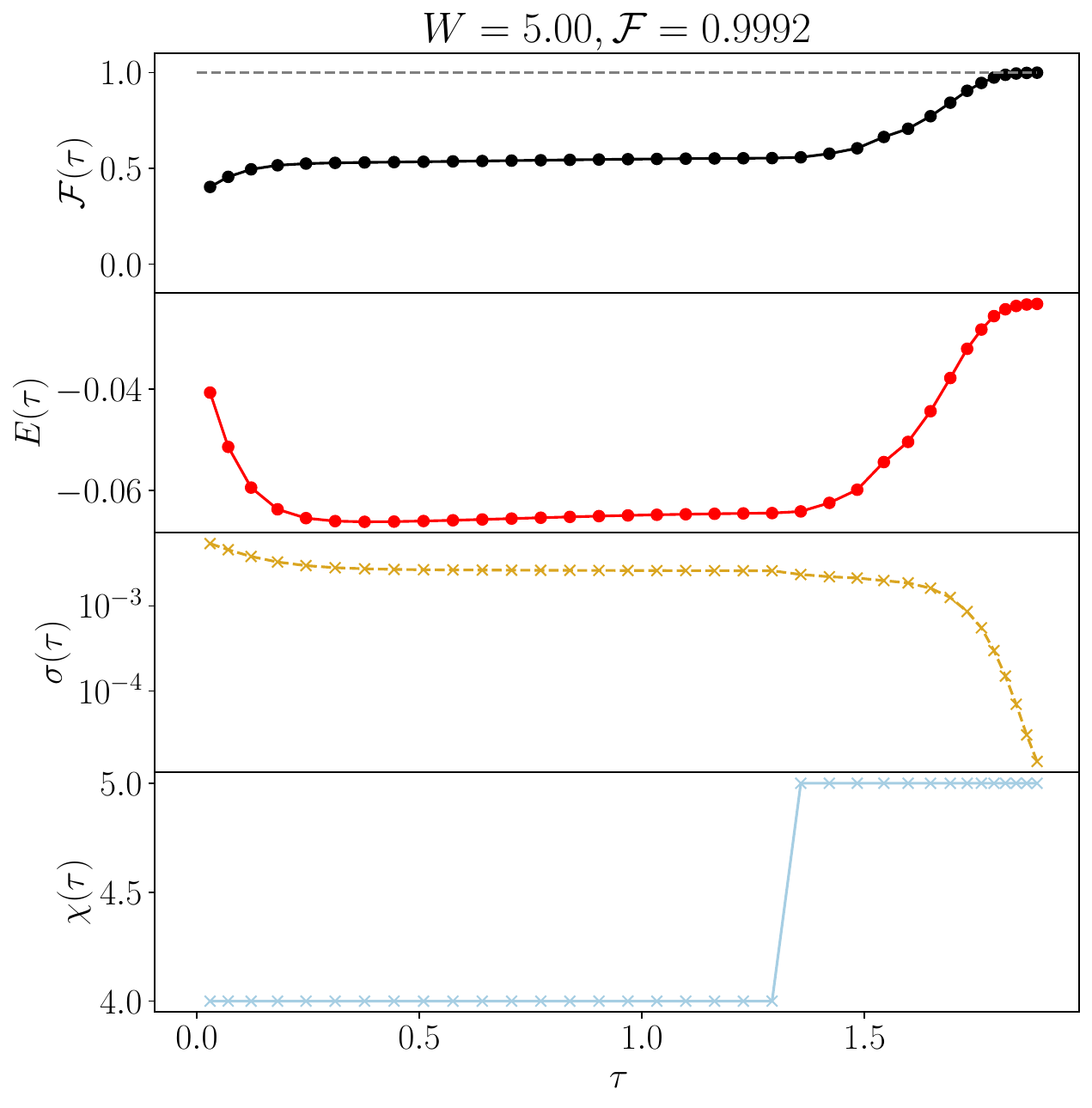}
    \includegraphics[width=0.4\linewidth]{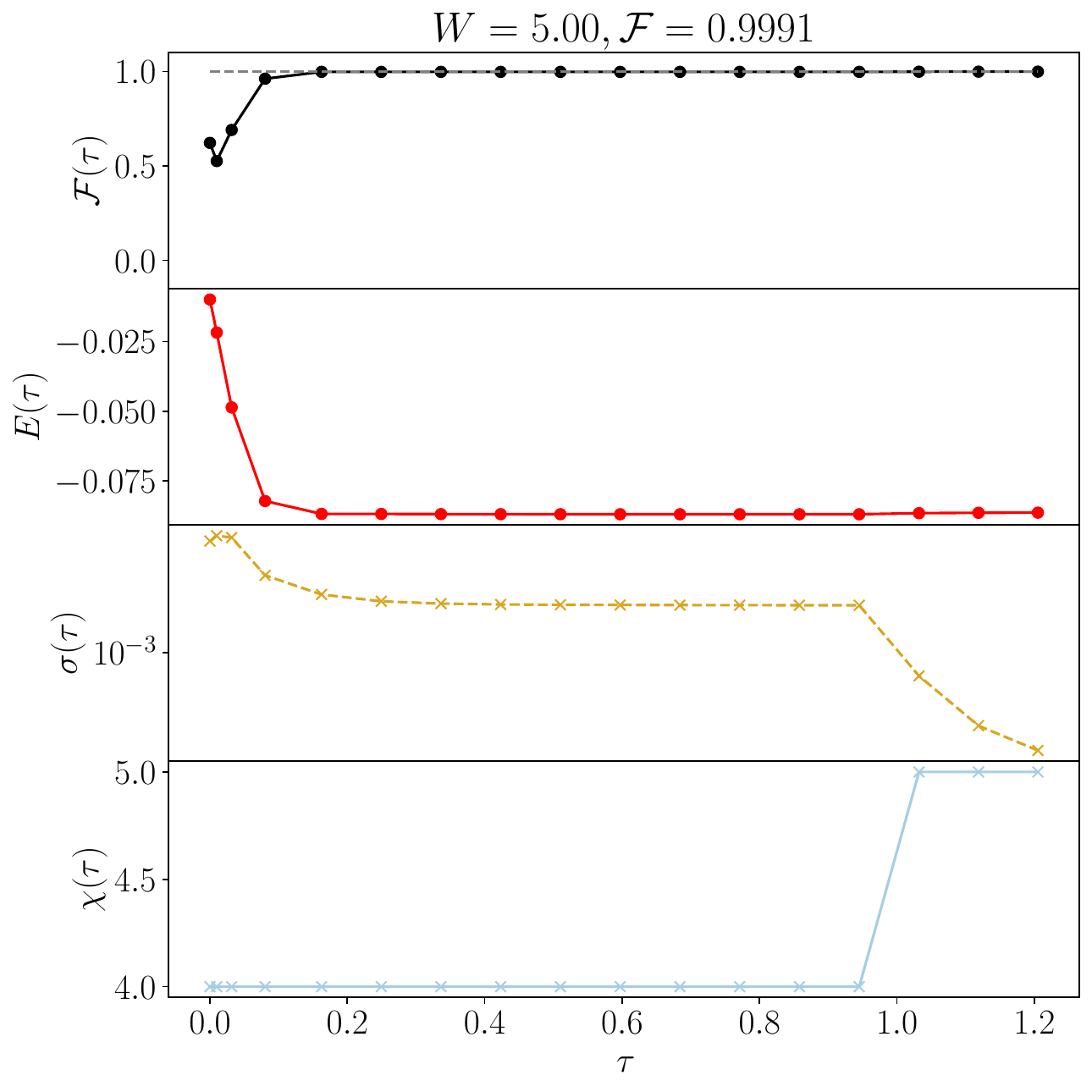}
    \includegraphics[width=0.4\linewidth]{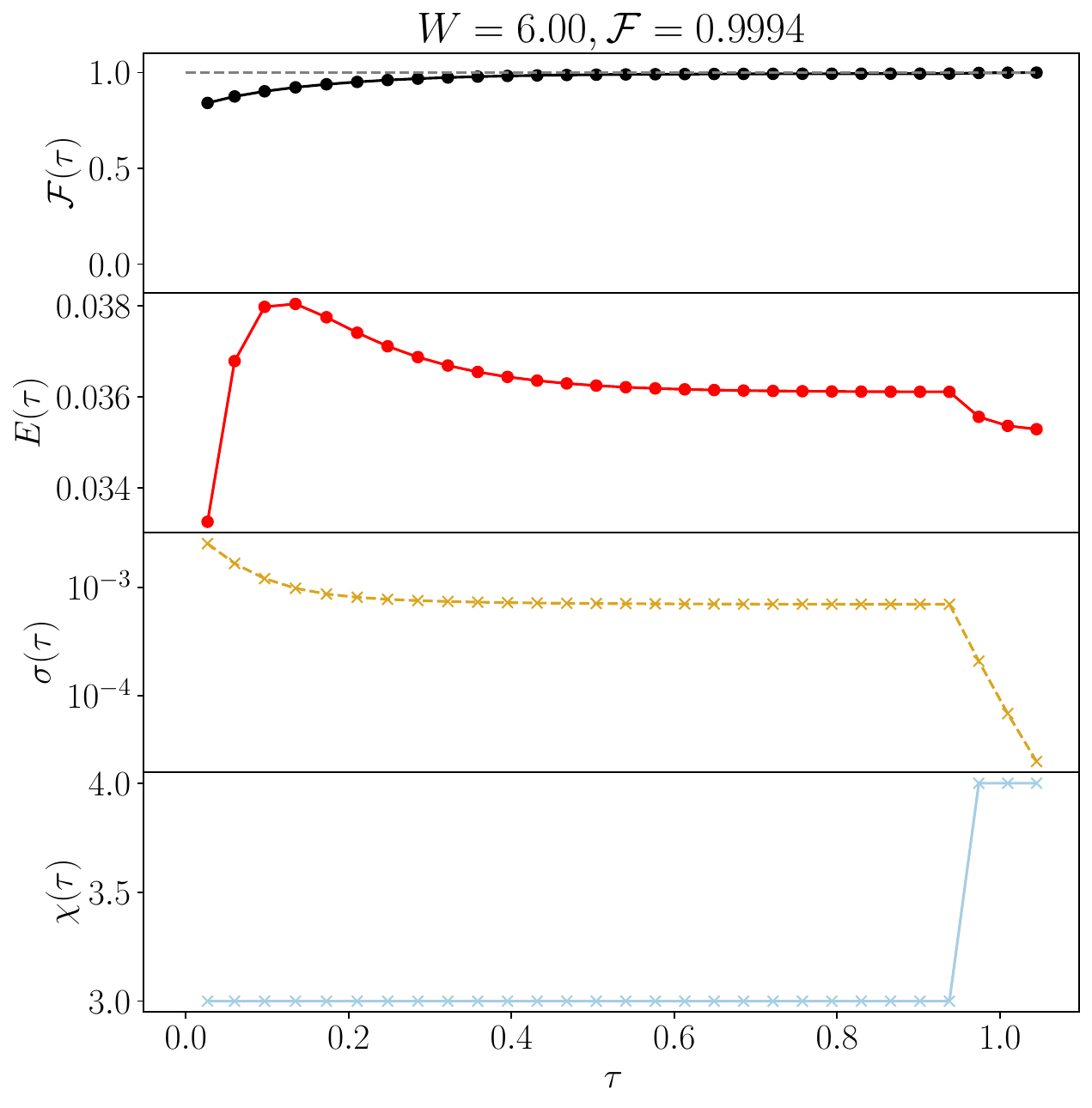}
    \includegraphics[width=0.4\linewidth]{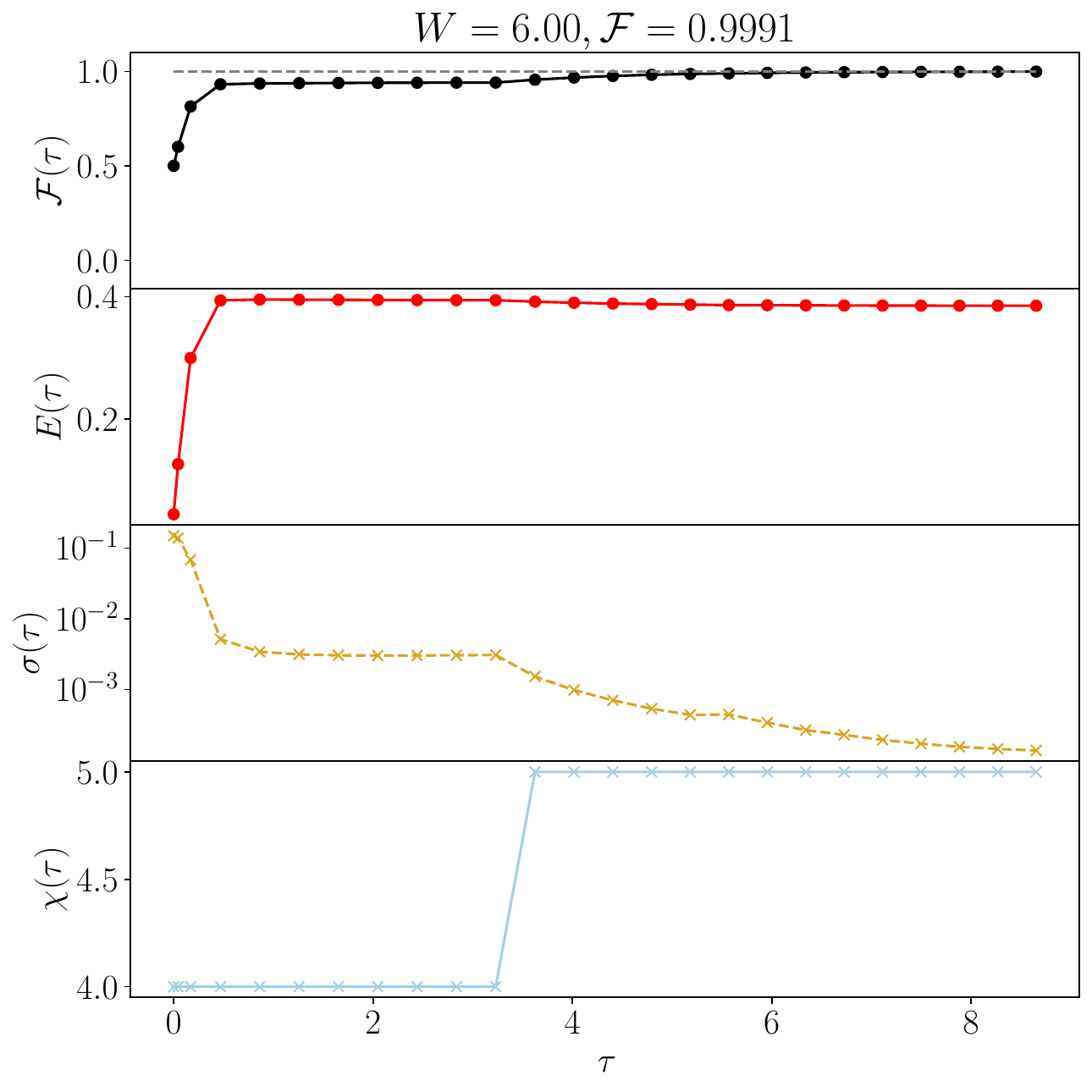}
    \caption{Additional representative results for $L=12$. Panels show the evolution of the fidelity, the energy, the variance, and the bond dimension. The figure label indicates the maximum fidelity achieved at the end of the evolution.}
    \label{fig.further_results_12_2}
\end{figure}

\begin{figure}
    \centering
    \includegraphics[width=0.4\linewidth]{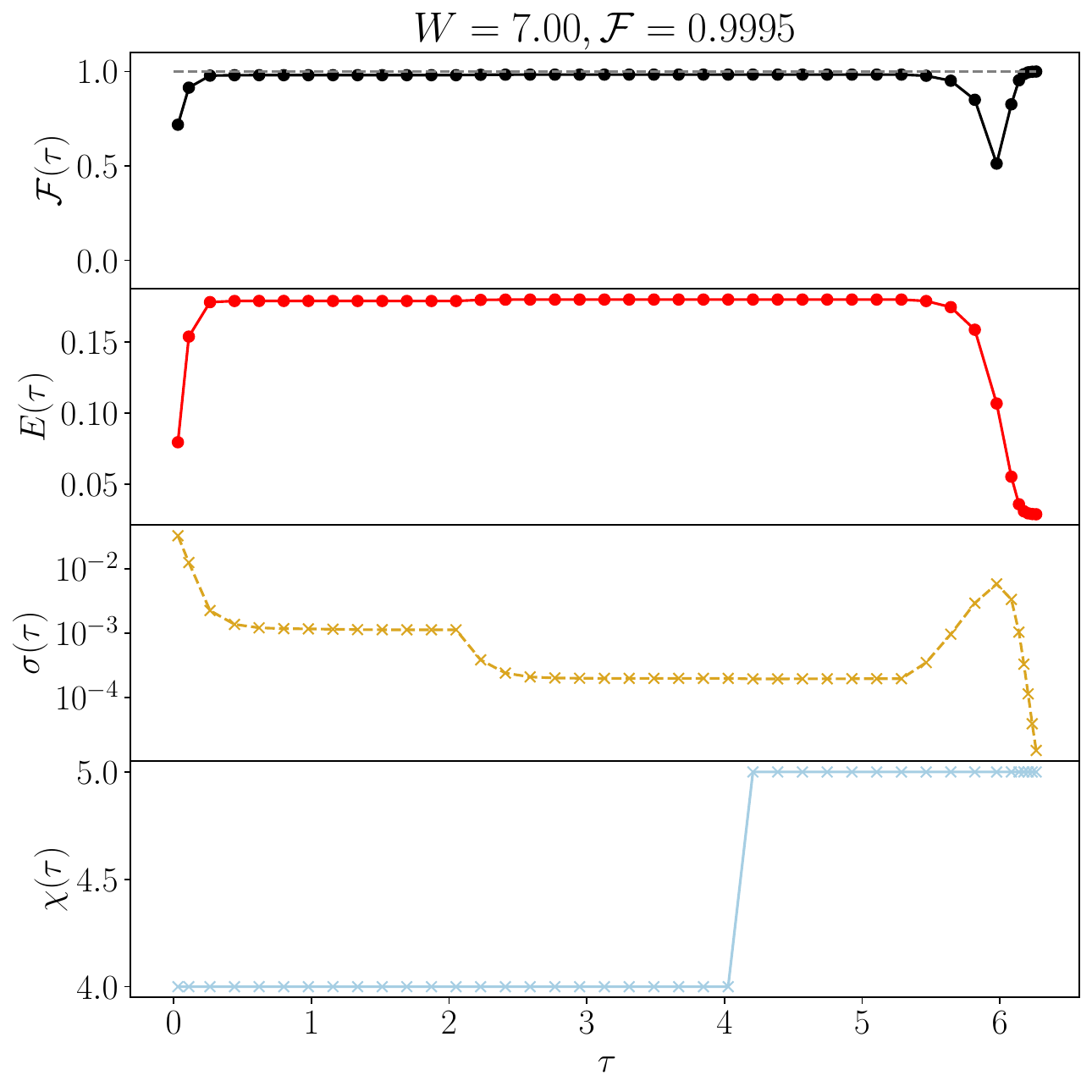}
    \includegraphics[width=0.4\linewidth]{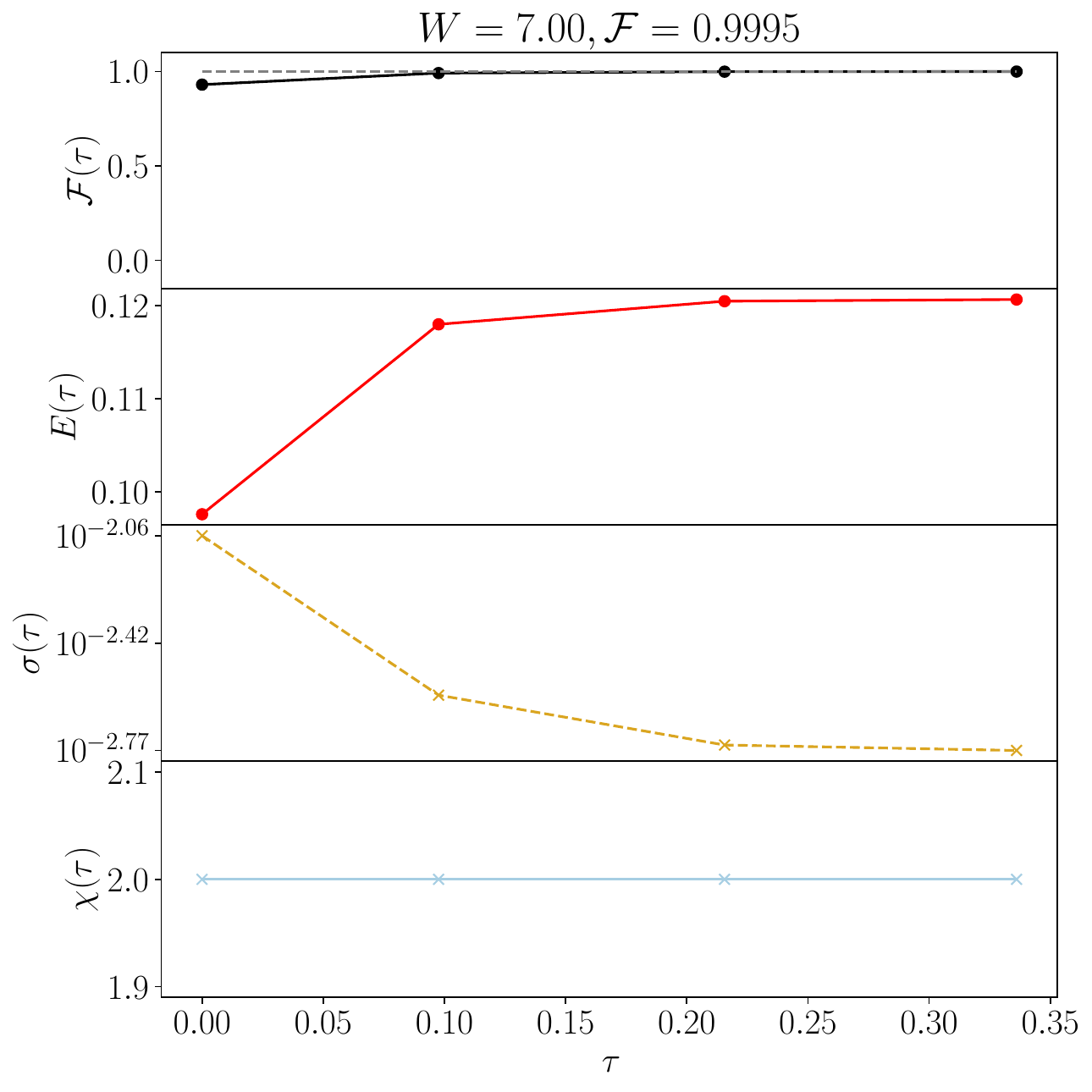}
    \includegraphics[width=0.4\linewidth]{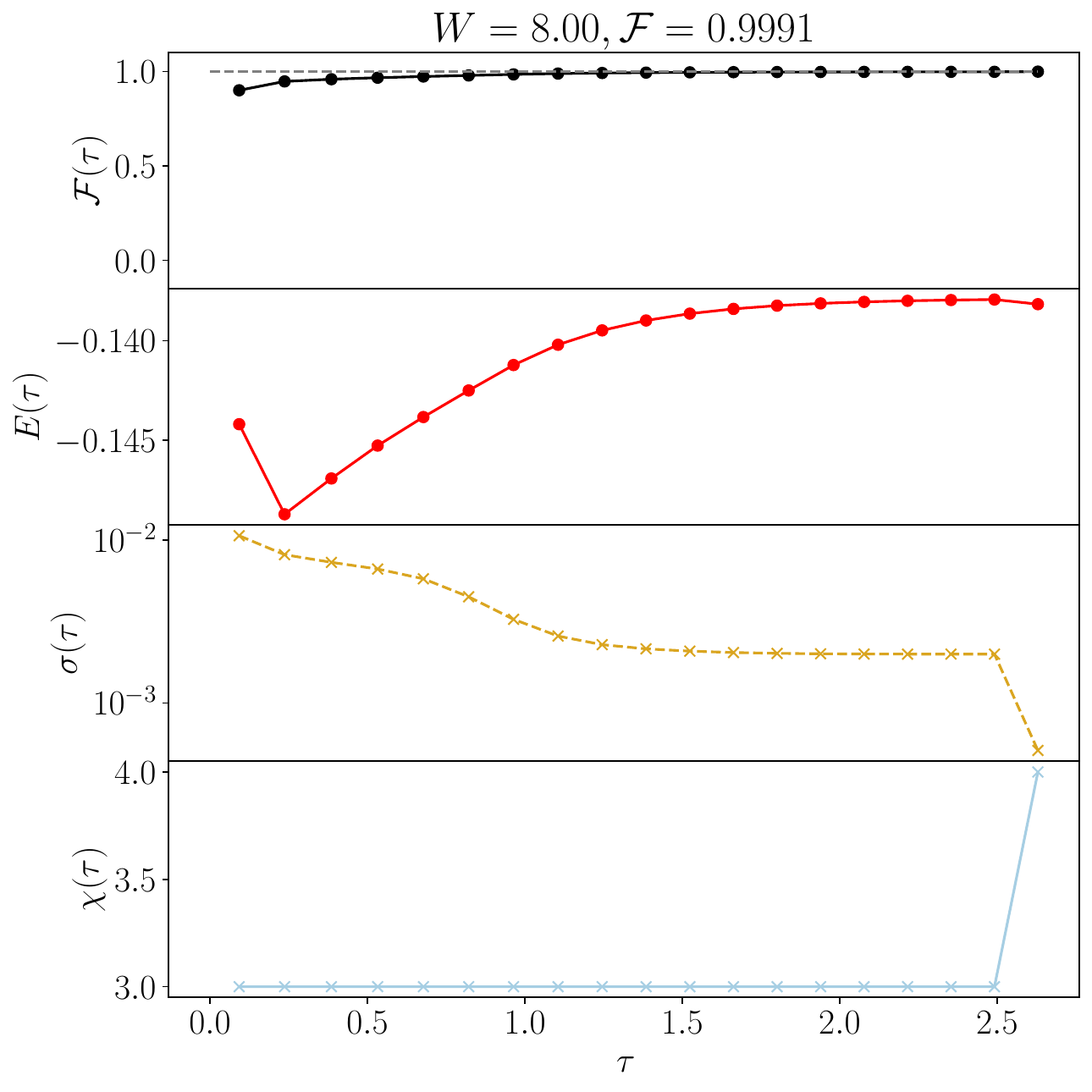}
    \includegraphics[width=0.4\linewidth]{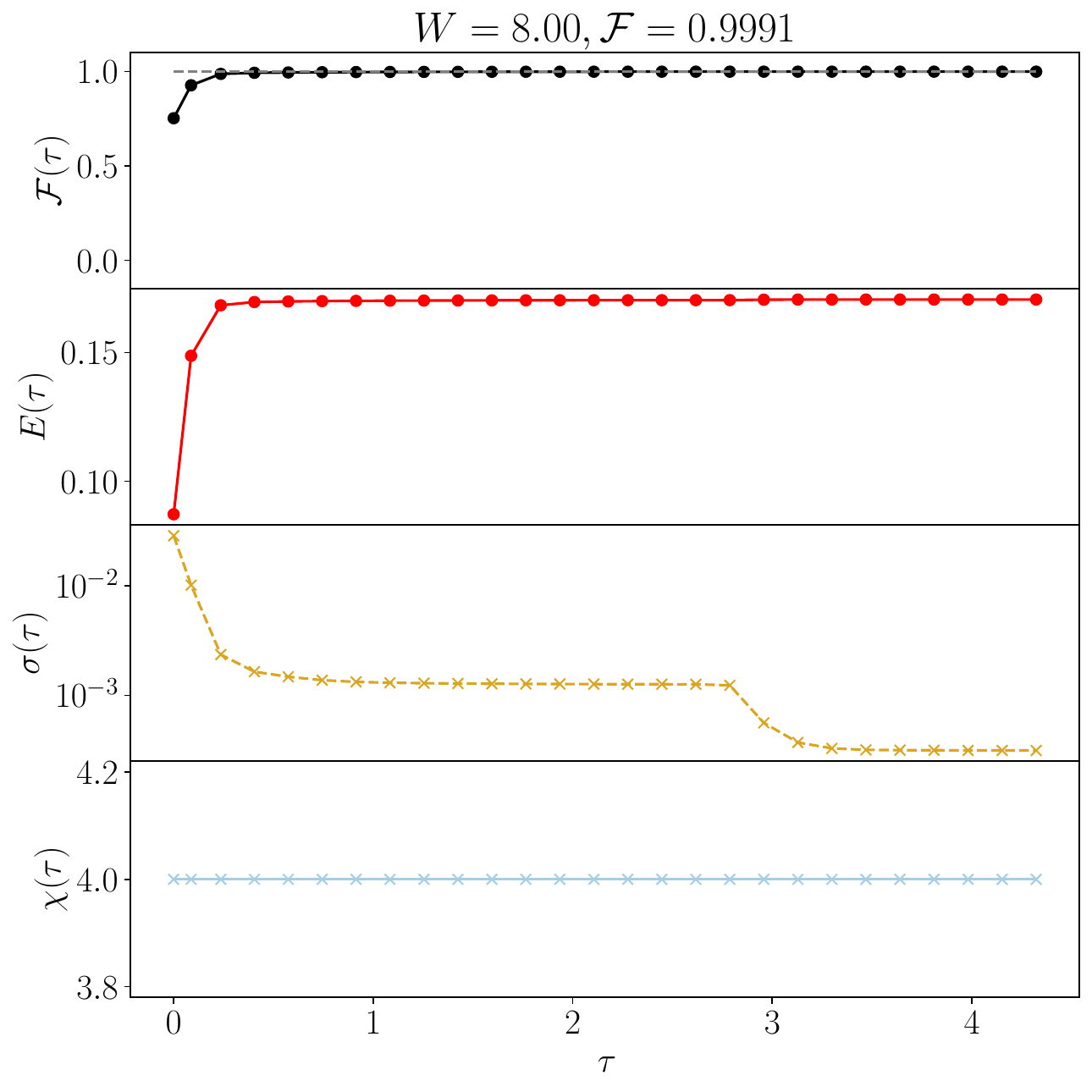}
    \includegraphics[width=0.4\linewidth]{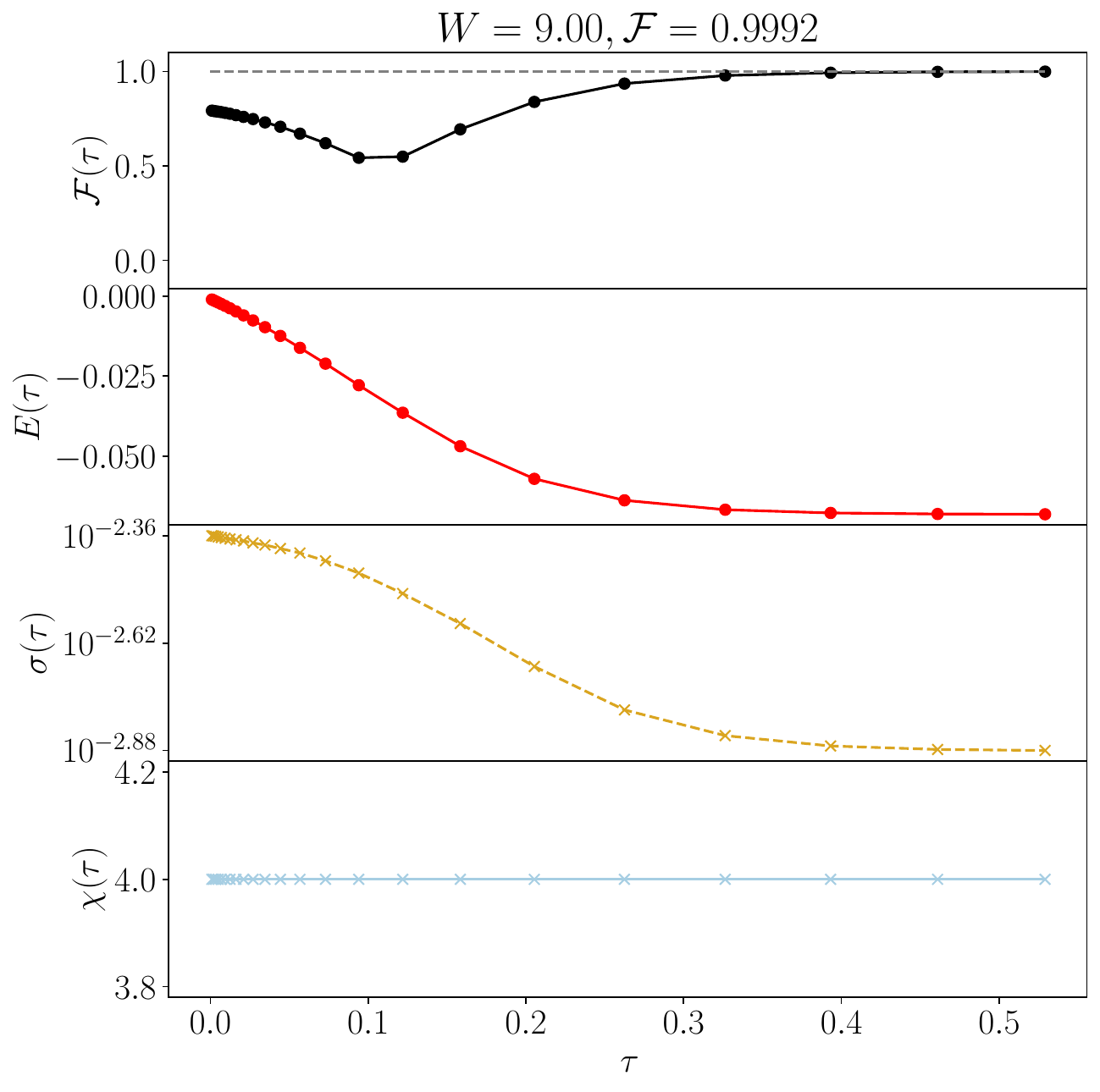}
    \includegraphics[width=0.4\linewidth]{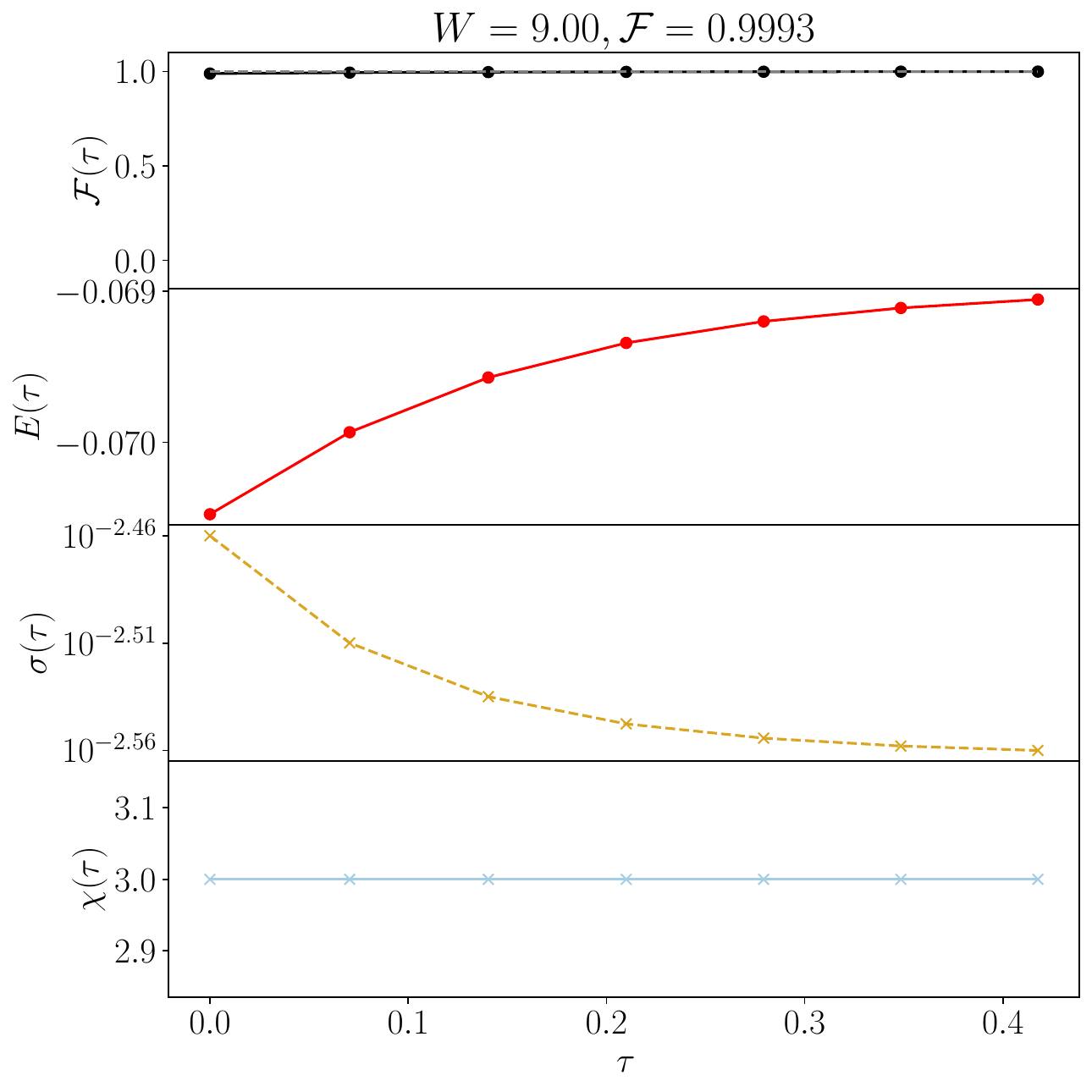}
    \caption{Additional representative results for $L=12$. Panels show the evolution of the fidelity, the energy, the variance, and the bond dimension. The figure label indicates the maximum fidelity achieved at the end of the evolution.}
    \label{fig.further_results_12_3}
\end{figure}

Additional results similar to Fig.~3 of the main text are shown in Figs.~\ref{fig.further_results_12_1}, \ref{fig.further_results_12_2} and \ref{fig.further_results_12_3} for a system size $L=12$ and a variety of disorder strengths. As the system size is sufficiently small, here we also show the evolution of the fidelity with (imaginary) time. (Note that oscillations in the fidelity typically mean that the system is transitioning from having a strong overlap with one eigenstate, to having a strong overlap with another one.)

As expected for a small disordered system, there is quite some variation between different disorder realizations, however there is a general pattern that weaker disorder strengths tend to require larger bond dimensions and/or longer evolution times in order to converge to the target threshold of fidelity $\mathcal{F} \geq 0.999$. There are exceptions to this, where weakly disordered systems can occasionally converge quickly, or where strongly disordered systems may require large bond dimensions due to e.g. anomalous rare regions, or long evolution times due to near-degeneracies in the spectrum that are difficult to resolve. Although strong disorder can help the classical MPS implementation of our algorithm by reducing the entanglement, energy levels in strongly disordered systems follow a Poisson distribution and do not exhibit Wigner-Dyson-like level repulsion---the lack of level repulsion can result in eigenvalues which can be arbitrarily close together, meaning that resolving excited eigenstates at strong disorder is far from trivial, even though their bond dimensions may remain small and classically tractable.

\begin{figure}
    \centering
    \includegraphics[width=0.4\linewidth]{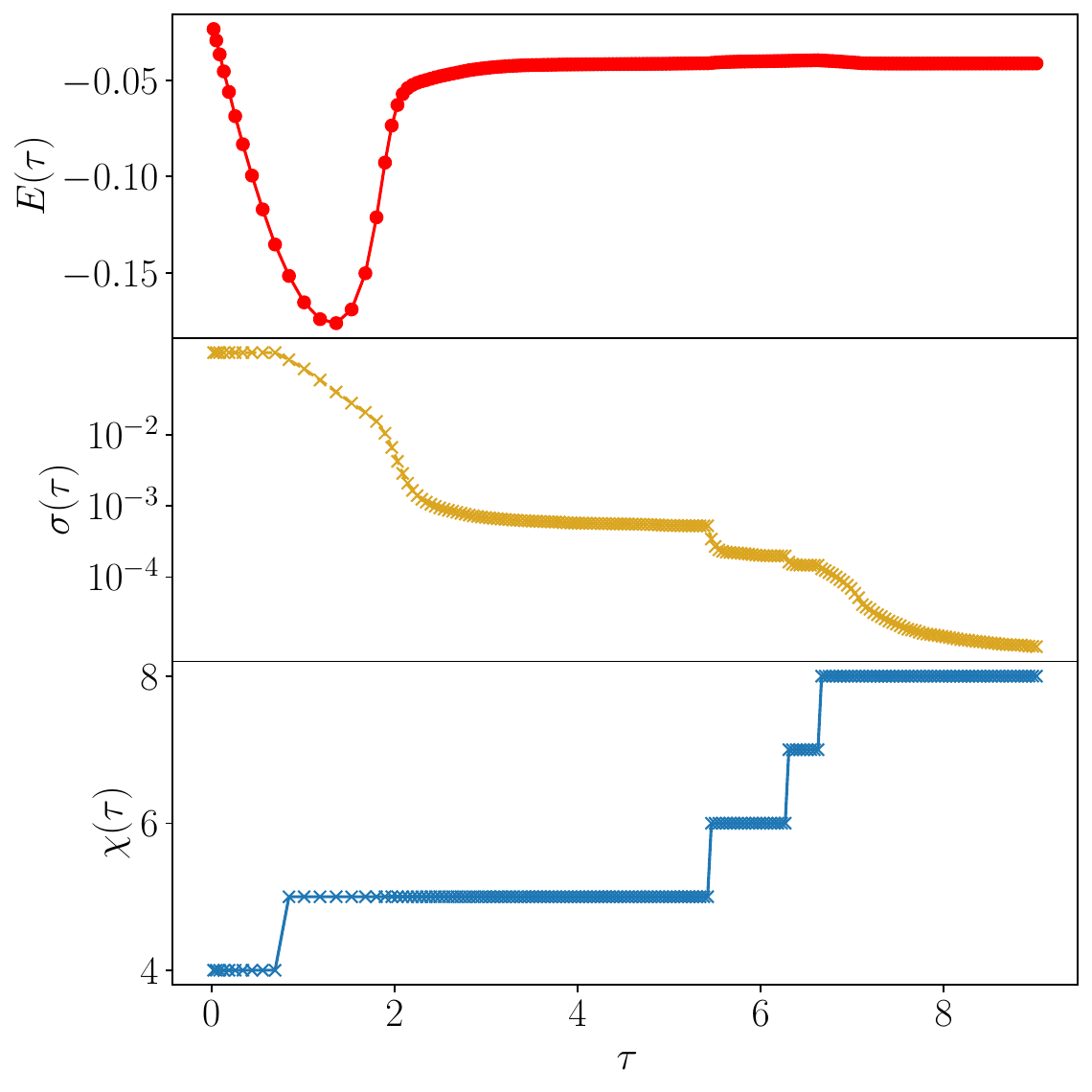}
    \includegraphics[width=0.4\linewidth]{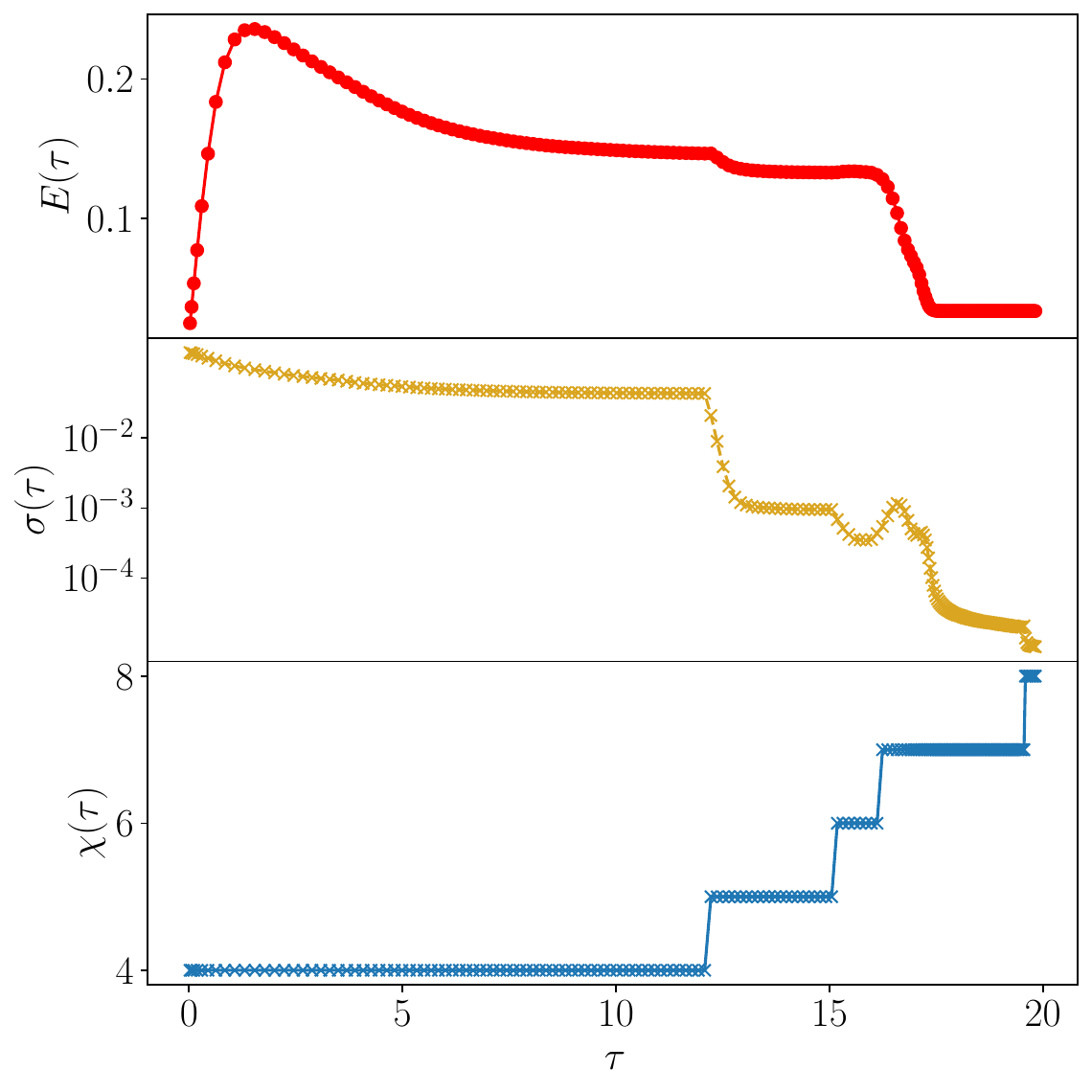}
    \includegraphics[width=0.4\linewidth]{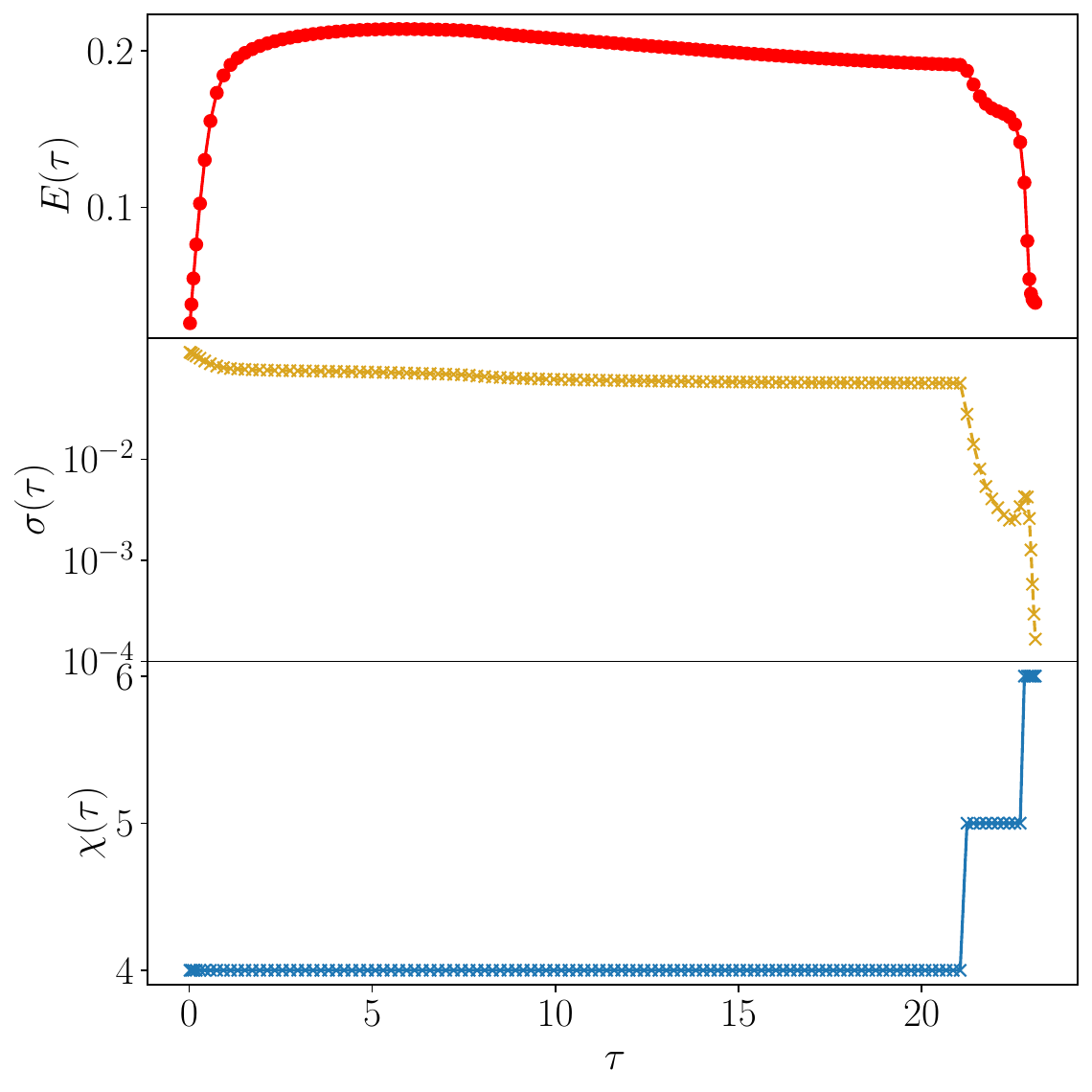}
    \includegraphics[width=0.4\linewidth]{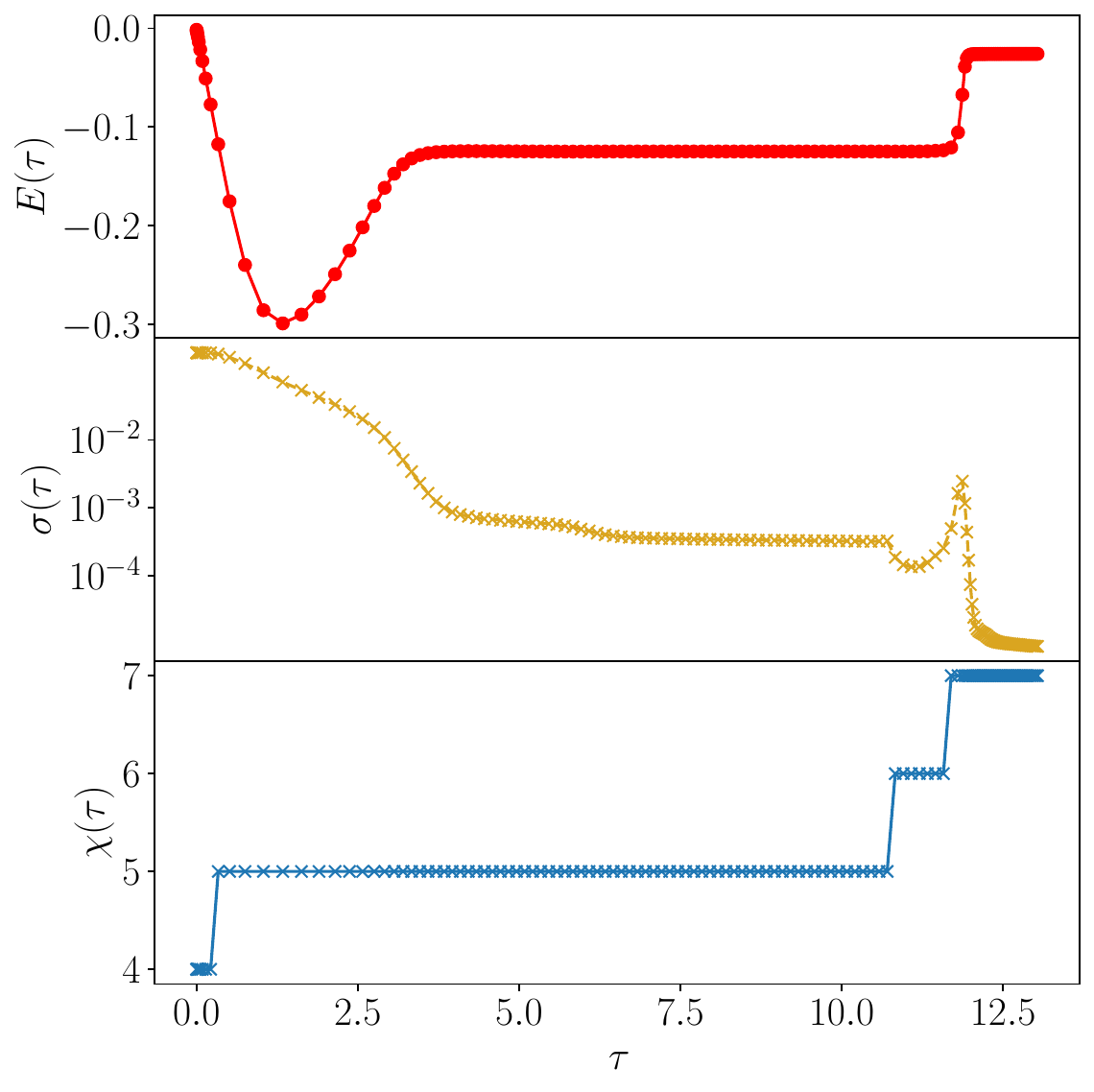}
    \includegraphics[width=0.4\linewidth]{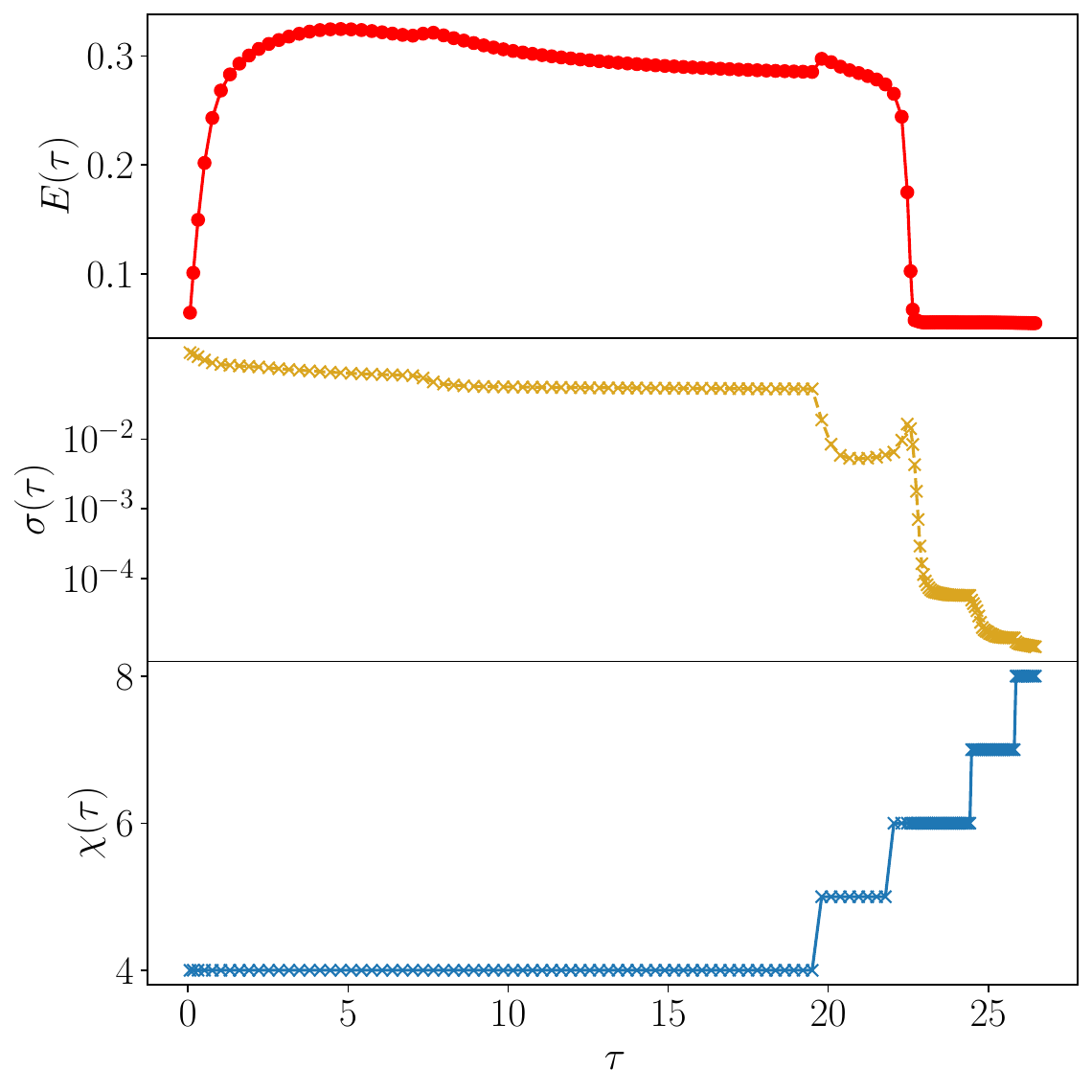}
    \caption{Additional representative results for $L=128$ and $W=10$.}
    \label{fig.further_results_128}
\end{figure}

In addition, we also show additional results for $L=128$ and $W=10$ in Fig.~\ref{fig.further_results_128} and results for $L=48$ and $W=6$ in Fig.~\ref{fig.further_results_48}, demonstrating the robustness of the technique and that it functions for a variety of system sizes and disorder strengths beyond those accessible to numerically exact methods or other state-of-the-art shift-invert implementations. The method remains smooth and well-controlled in all cases.

\begin{figure}
    \centering
    \includegraphics[width=0.4\linewidth]{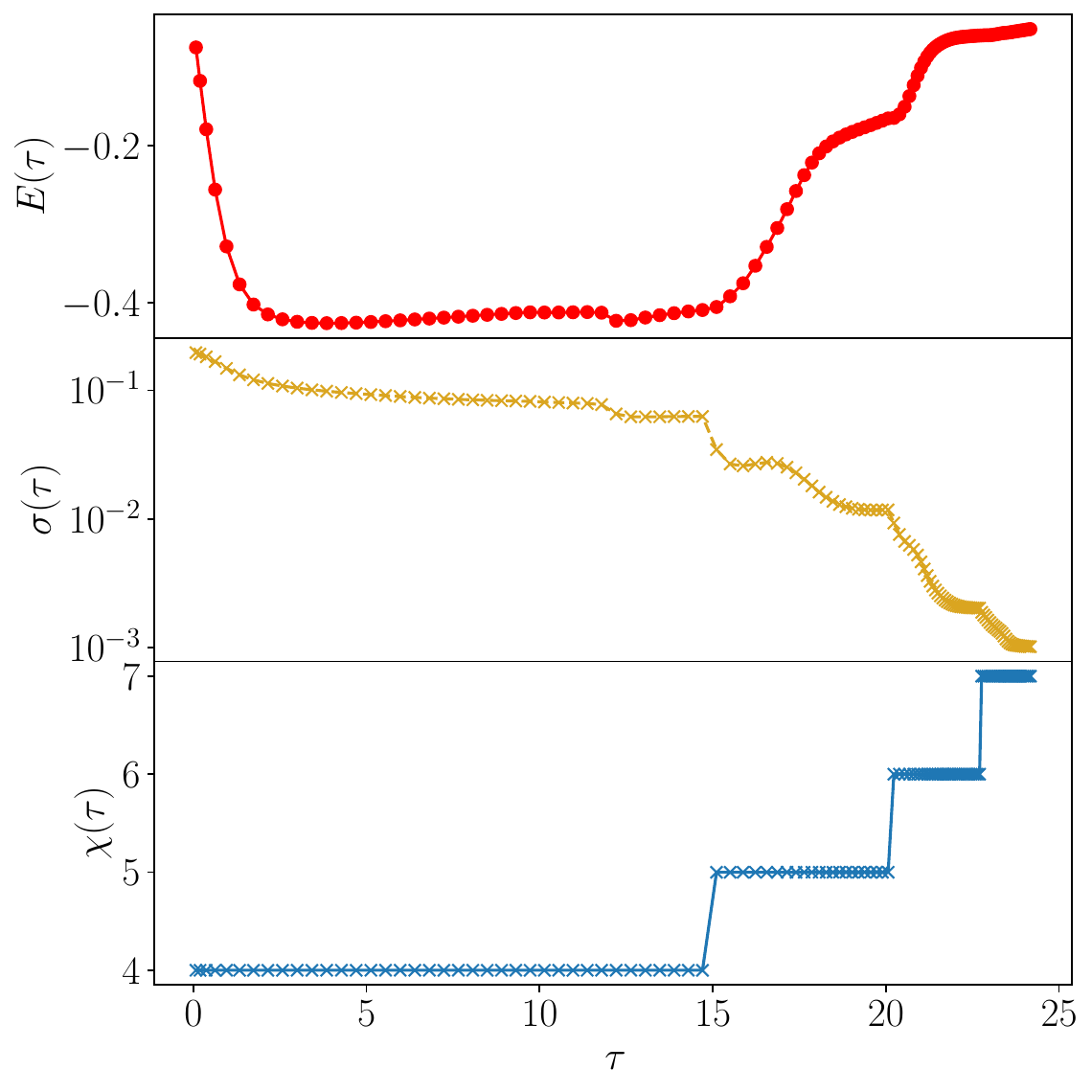}
    \includegraphics[width=0.4\linewidth]{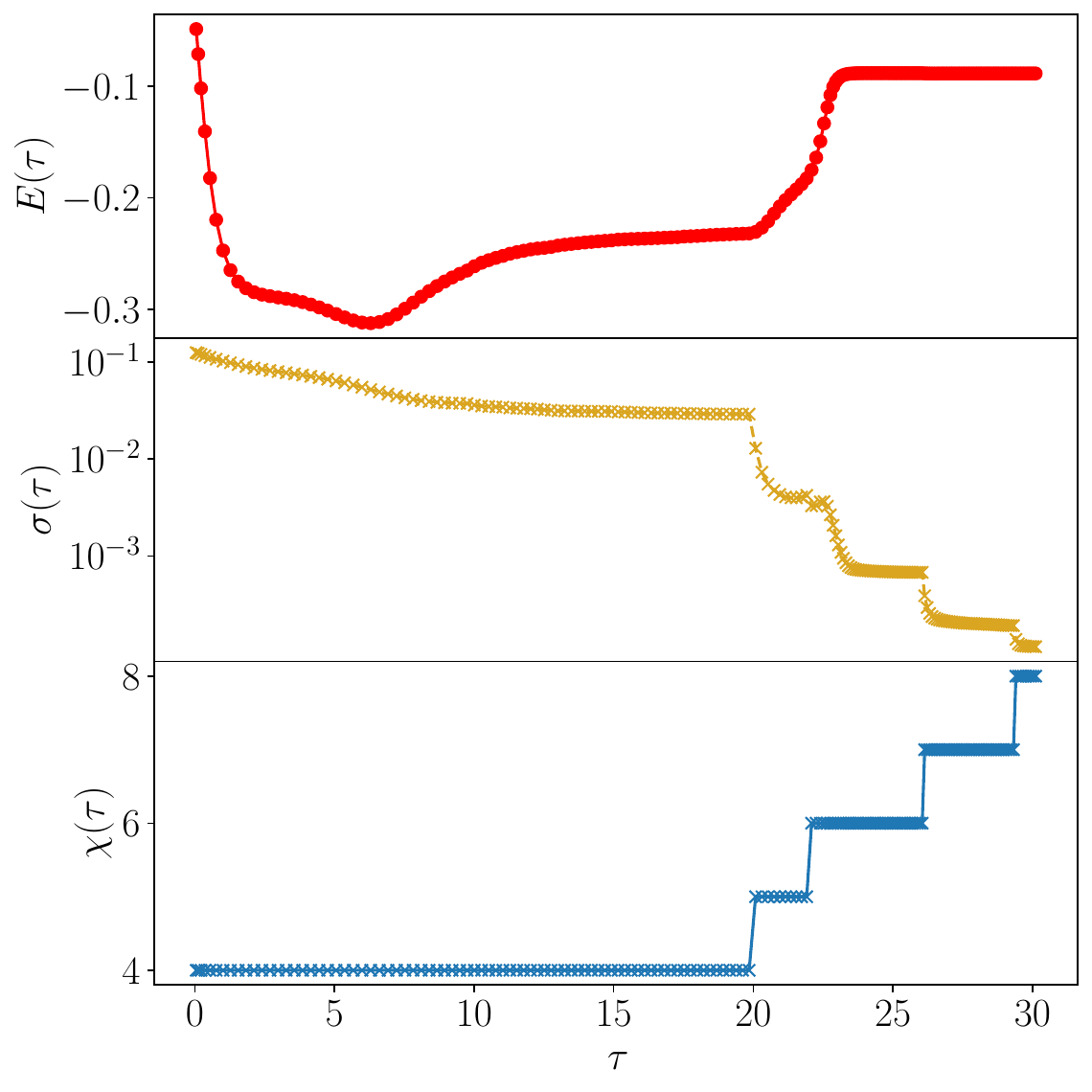}
    \includegraphics[width=0.4\linewidth]{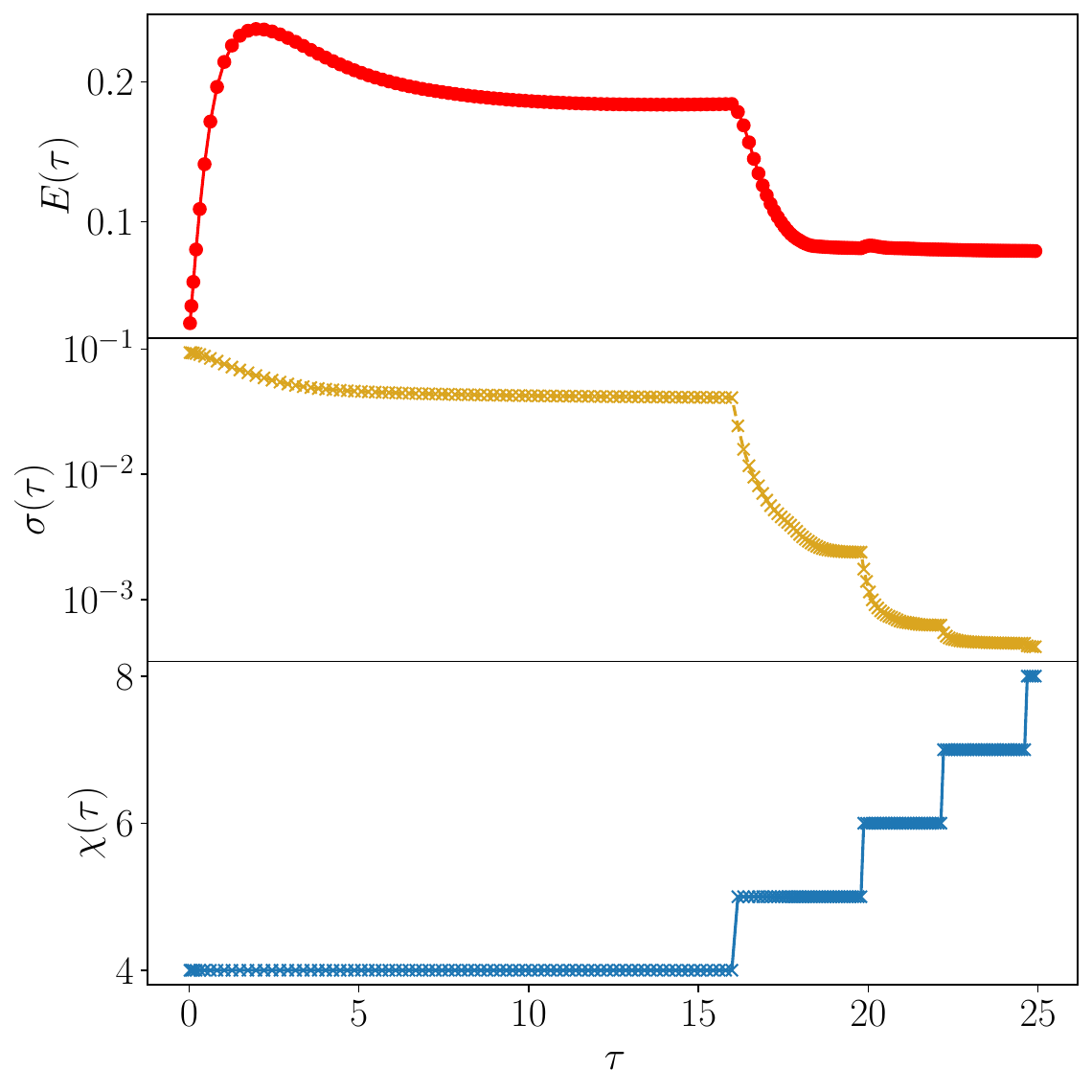}
    \includegraphics[width=0.4\linewidth]{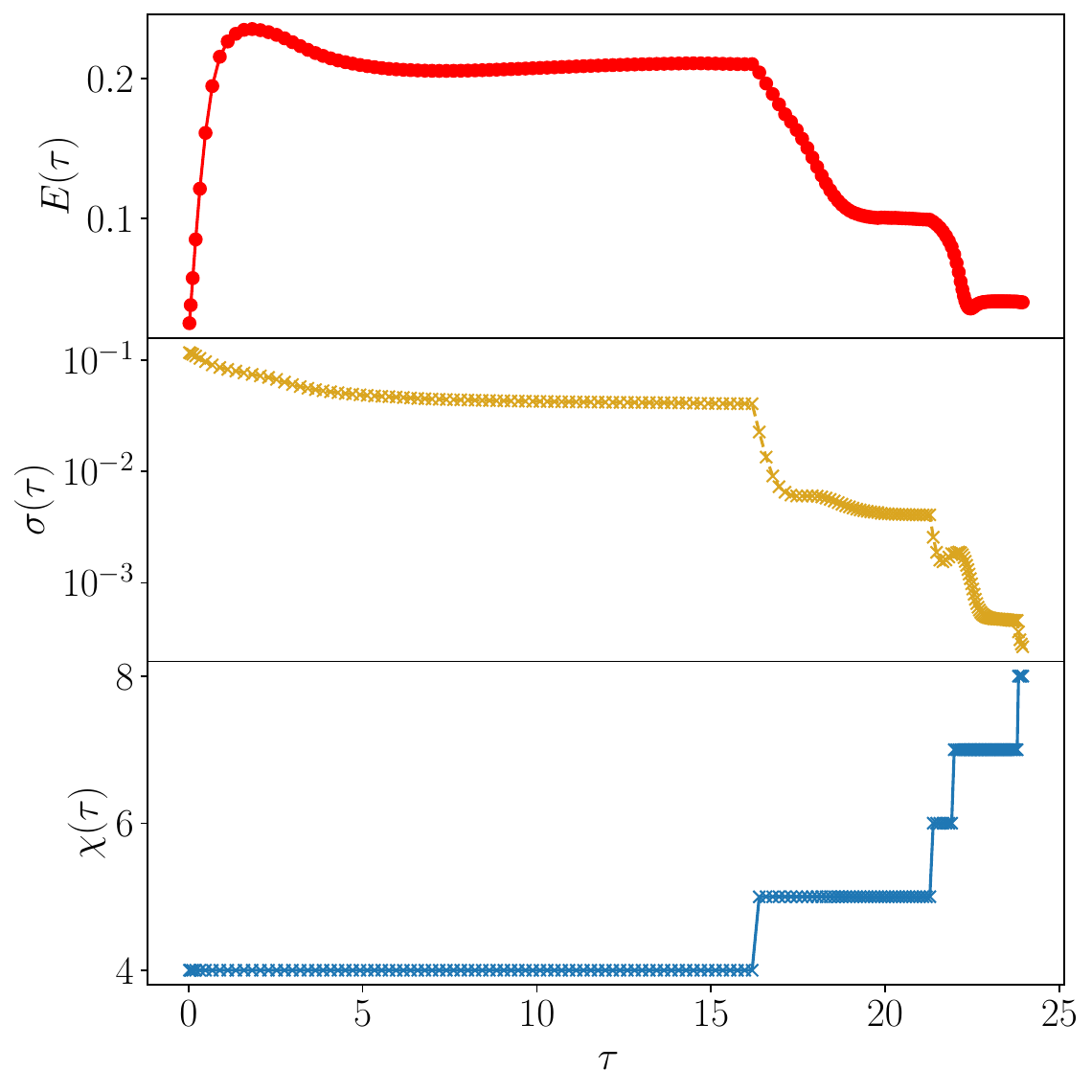}
    \includegraphics[width=0.4\linewidth]{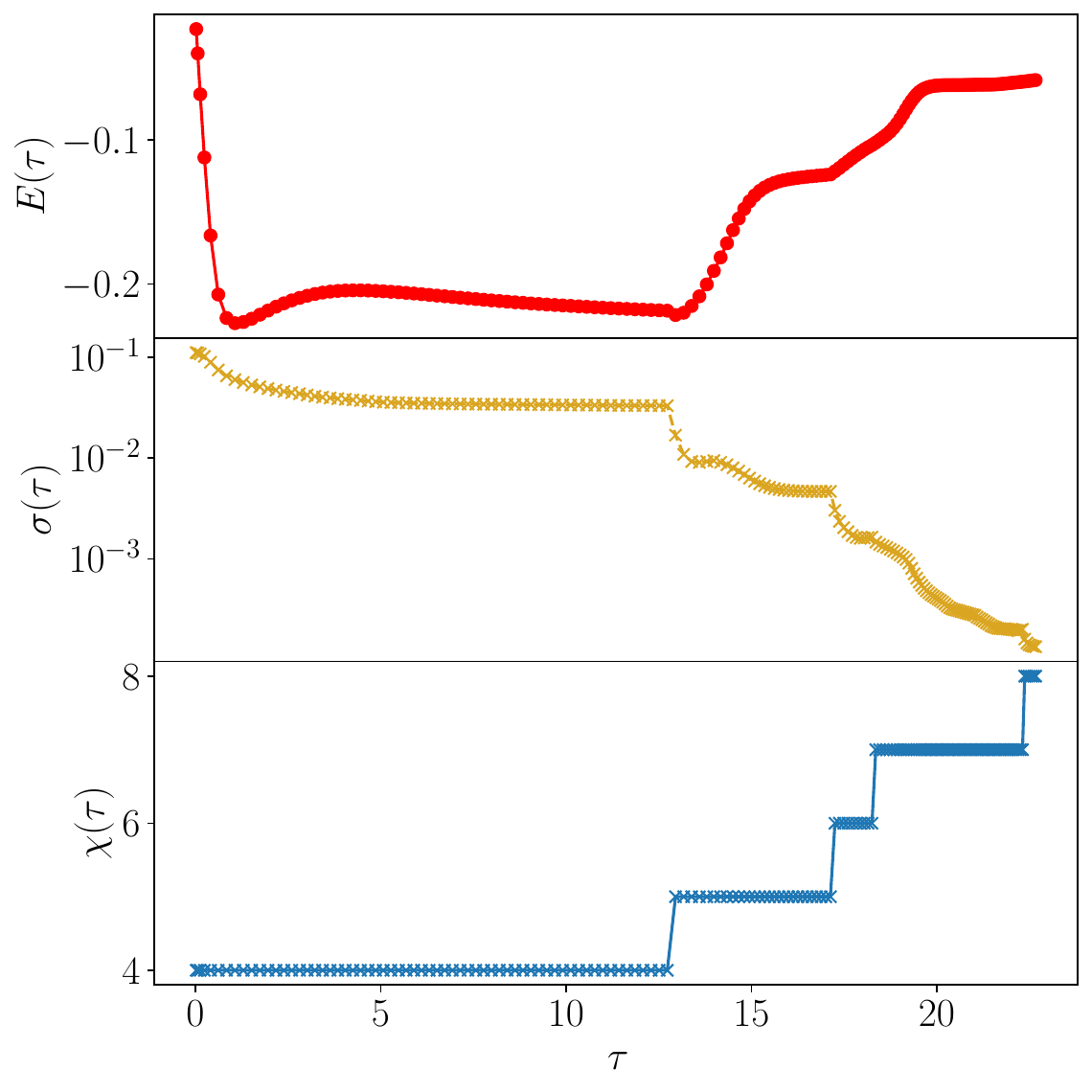}
    \caption{Additional representative results for $L=128$ and $W=6$. The combination of large system size and moderate disorder means that these computations are lengthy and did not reach our threshold criteria to be described as eigenstates, however they are nonetheless low-variance states which will asymptotically approach eigenstates as $\tau$ continues to increase.}
    \label{fig.further_results_128_6}
\end{figure}

\begin{figure}
    \centering
    \includegraphics[width=0.4\linewidth]{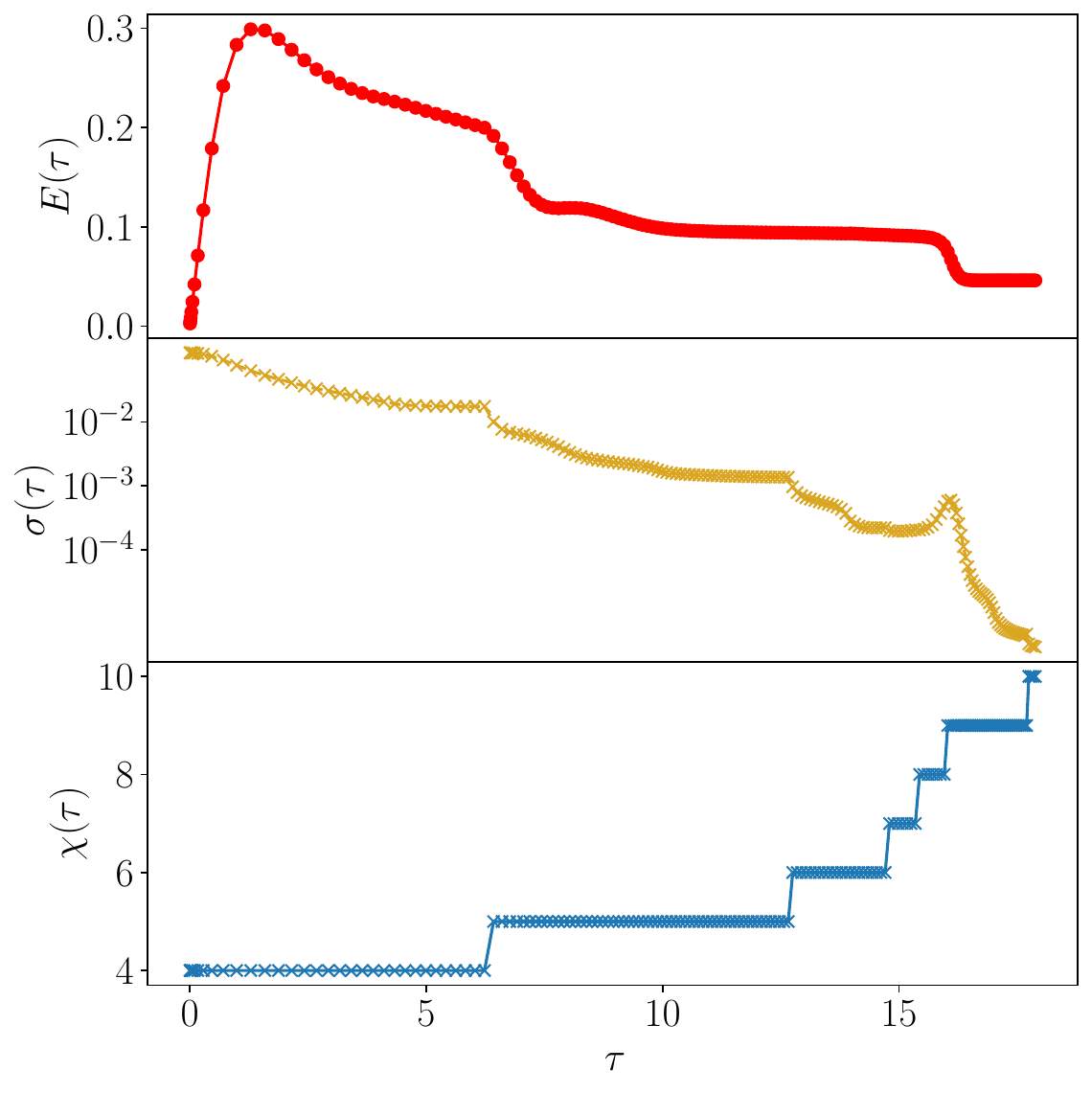}
    \includegraphics[width=0.4\linewidth]{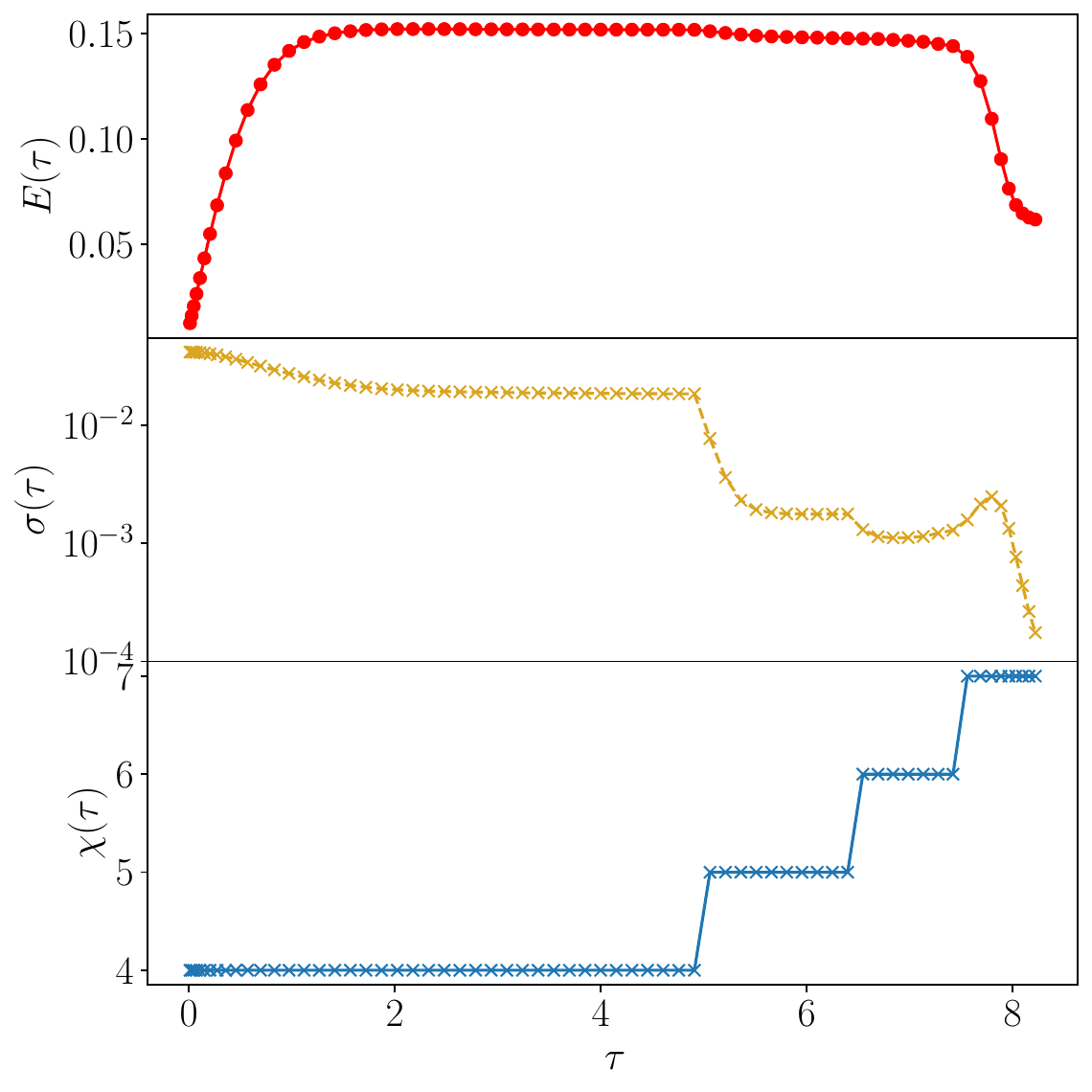}
    \caption{Additional representative results for $L=64$ and $W=6$, in addition to those shown in the main text.}
    \label{fig.further_results_64}
\end{figure}

\begin{figure}
    \centering
    \includegraphics[width=0.45\linewidth]{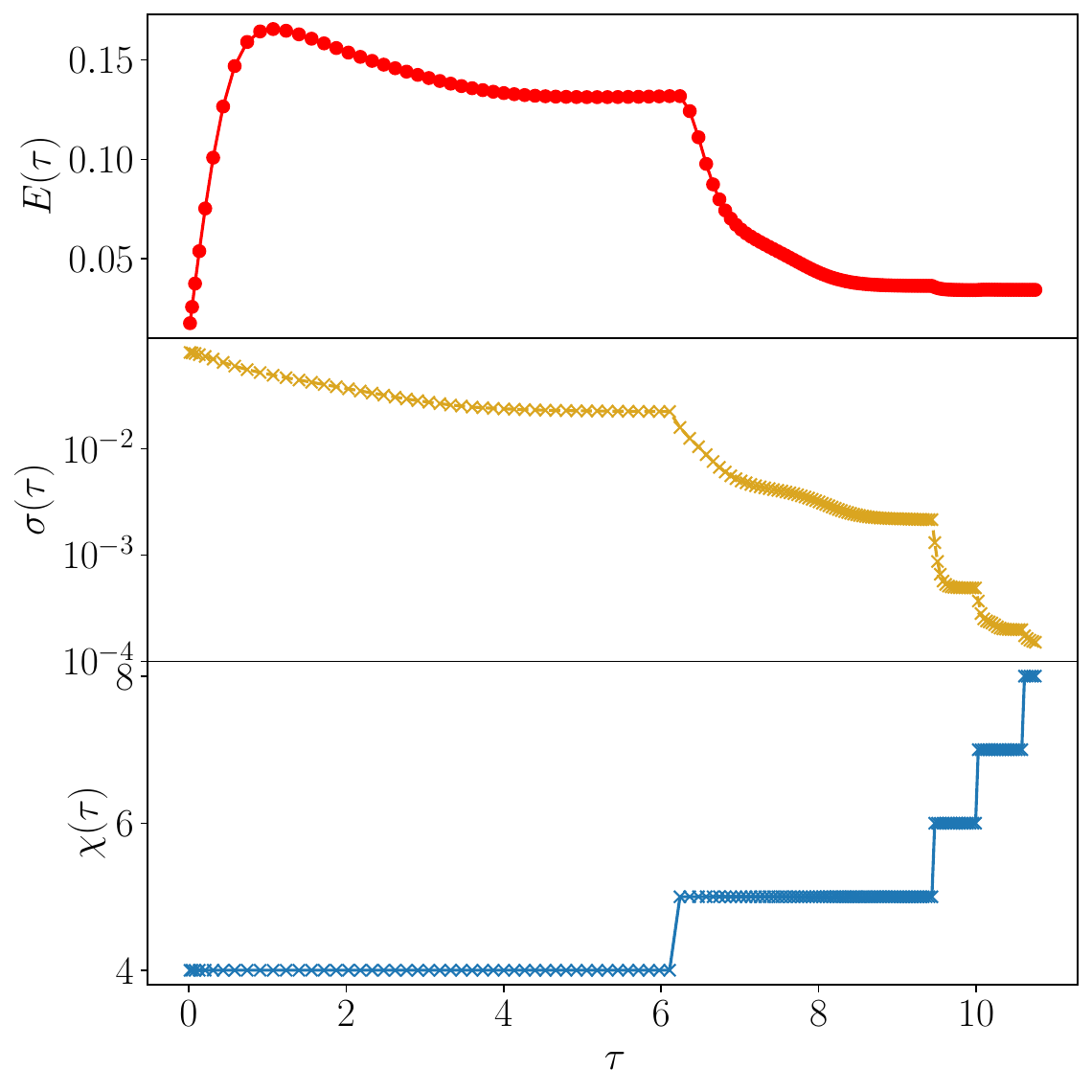}
    \includegraphics[width=0.45\linewidth]{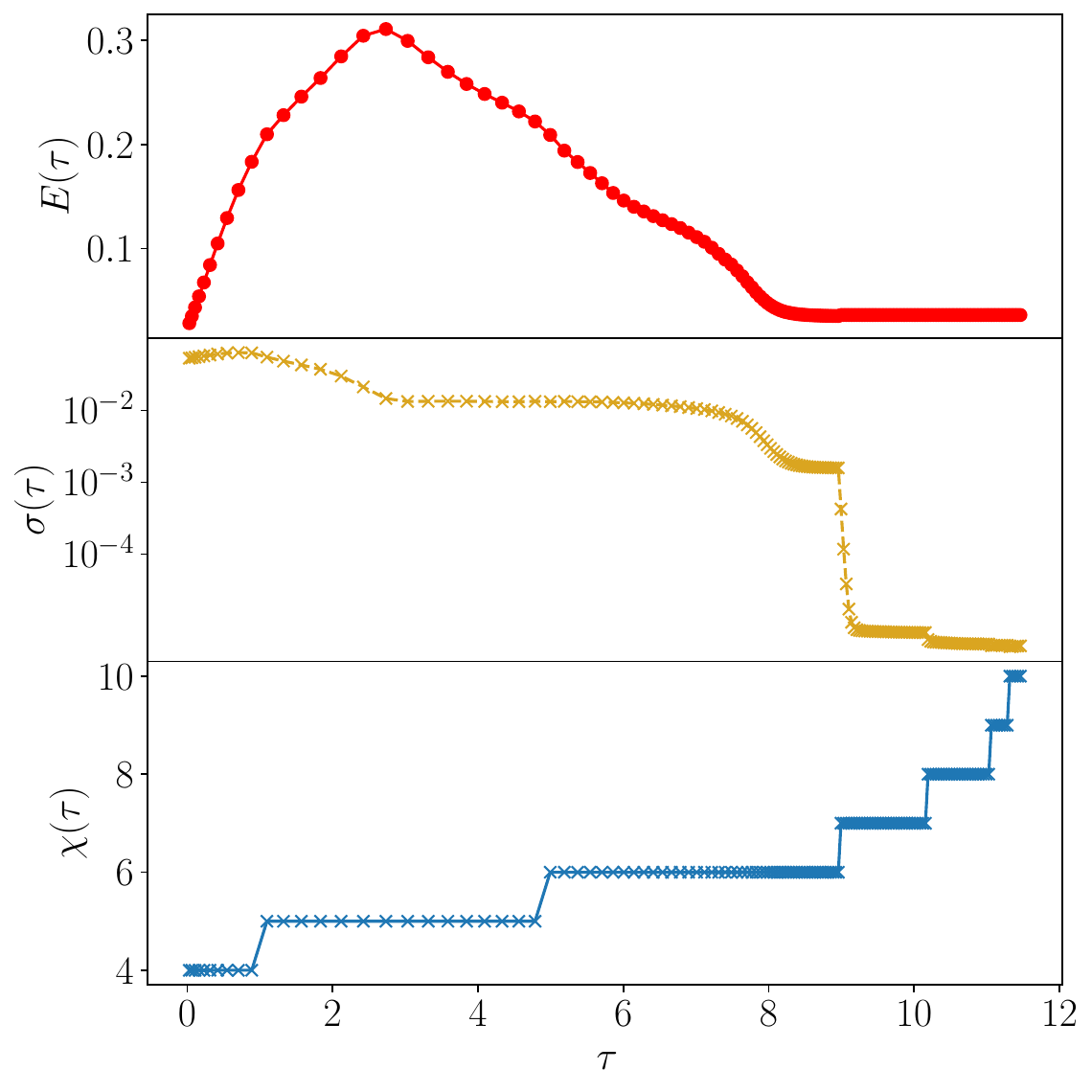}
    \includegraphics[width=0.45\linewidth]{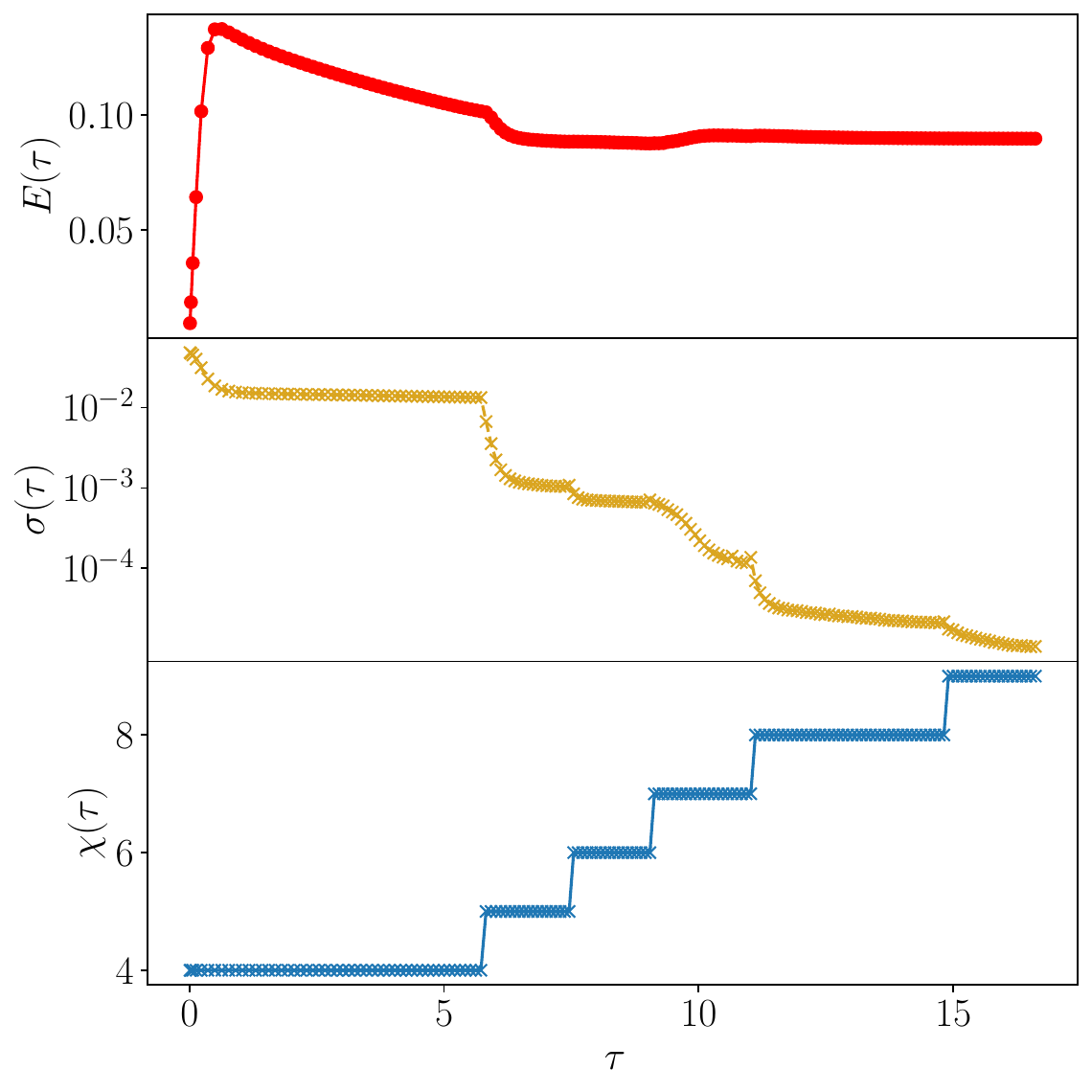}
    \includegraphics[width=0.45\linewidth]{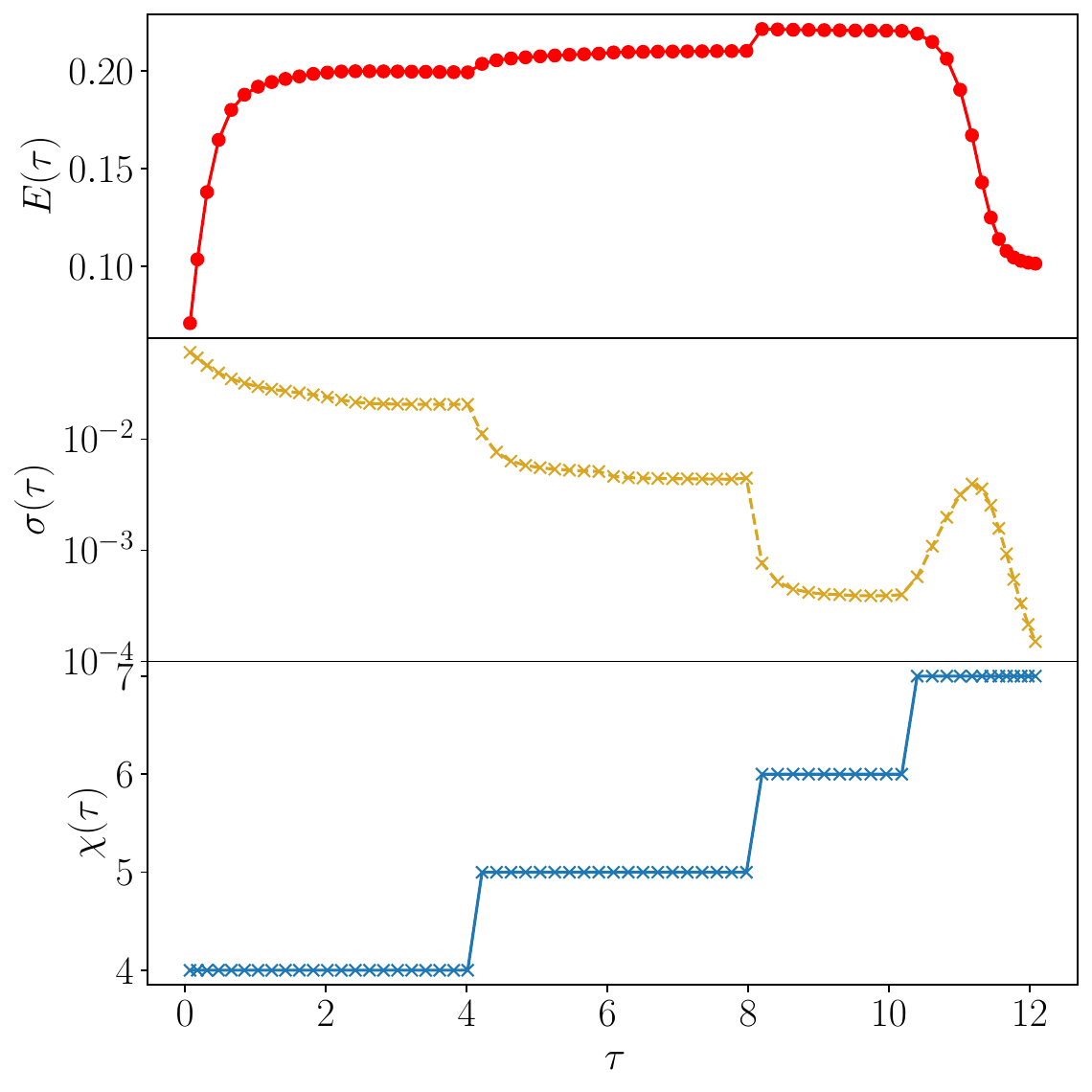}
    \caption{Additional representative results for $L=48$ and $W=6$.}
    \label{fig.further_results_48}
\end{figure}

In addition, the reader may notice that there are occasional regions in the evolution where there is a very high density of data points. This is because if the optimization algorithm fails at any point, we shrink the timestep until it is able to succeed and make progress. The regions where timesteps are clustered together represent challenging portions of the evolution where $\mathrm{d}\tau$ has been dynamically shrunk until the optimizer was able to converge, timesteps were allowed to increase again afterwards. This often occurs close to bond dimension increases, where the algorithm has reached the best possible state for a given bond dimension and can no longer optimize further.

\section{Entanglement Entropy}

\begin{figure}
    \centering
    \includegraphics[width=0.7\linewidth]{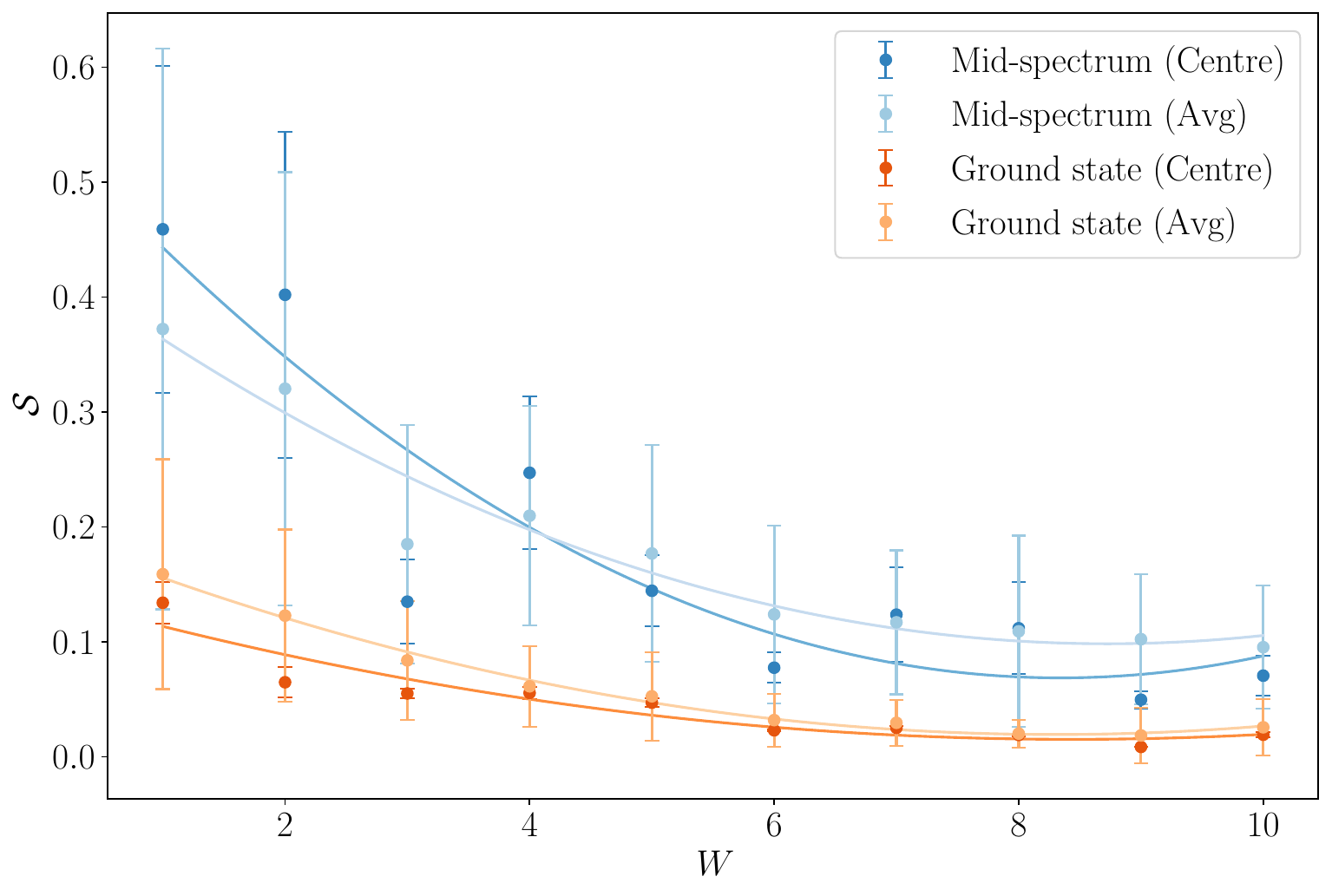}
    \caption{A comparison between the entanglement entropy on the central bond of the MPS versus the average over all bonds, for both the ground and mid-spectrum states. The error bars represent the standard deviation over disorder realizations. Solid lines are polynomial fits intended as guides to the eye.}
    \label{fig.sm_entanglement}
\end{figure}

In the main text, we showed the entanglement entropy averaged over all bonds. In Fig.~\ref{fig.sm_entanglement}, we show a comparison between the bond-averaged von Neumann entanglement entropy and the (perhaps more conventional) entanglement entropy across the central bond of the MPS, demonstrating that they behave qualitatively similarly. The bond-averaged entropy typically exhibits smoother behavior as a function of disorder strength, as one might expect from averaging only over a limited number of disorder realizations.

\section{Energy Gap}

While our main focus is on developing a technique able to prepare excited states, here we sketch an argument for why this method still gives a useful advantage for ground states. Given a Hamiltonian with $k$ eigenvalues denoted $E_k$, we can choose $\delta = E_n + \varepsilon$ to target the $n$-th eigenstate, where $\varepsilon < \textrm{min}(E_n-E_{n-1},E_{n+1}-E_n)$ is some small offset and we assume $n>0$. The eigenvalues of the shifted Hamiltonian $(\mathcal{H}-\delta)$ are given by $[...,E_{n-1}-E_n - \varepsilon, -\varepsilon, E_{n+1} - E_n - \varepsilon,...]$. After inversion, the eigenvalues are given by the reciprocal of the above values. The lowest eigenvalue will now be $-1/\varepsilon$, and the gap to the nearest eigenvalues above or below ($\pm$) this will be:
\begin{align}
    \Delta_{\textrm{eff}} &= \left| -\frac{1}{\varepsilon} - \frac{1}{E_{n \pm 1}-E_n - \varepsilon}\right| = \left| \frac{E_{n}-E_{n \pm 1}}{(E_{n \pm 1}-\delta)(E_{n}-\delta)}\right|.
\end{align}
For ground states ($n=0$), a similar procedure can be followed, resulting in a gap:
\begin{align}
    \Delta_{\textrm{eff}} &= \left| \frac{E_{1}-E_0}{(E_{1}-\delta)(E_{0}-\delta)}\right|.
\end{align}
For gapless systems where the energy difference between the ground and first excited states is $\Delta_0 \propto \exp(-L)$, if we target the ground state with $\delta = E_0 + \varepsilon$ then the ratio of the new energy gap to the old is given by:
\begin{align}
    \frac{\Delta_{\textrm{eff}}}{\Delta_0} \propto \left| \frac{1}{(\textrm{e}^{-L}-\varepsilon)\varepsilon} \right| \approx \frac{1}{\varepsilon^2} \gg 1.
\end{align}
Importantly, we do not require an exponentially small value of $\varepsilon$ for this ratio to be greater than one, and for the induced gap to be larger than the original microscopic energy gap.
This validates the use of the technique even for classical simulations: for gapless systems, it is possible to transform the problem into that of solving an equivalent Hamiltonian with a larger gap, widening the possibilities for classical methods to find the ground states of gapless systems, which would otherwise pose a problem for many variational techniques.